\documentclass[12pt,a4paper,final]{iopart}

\usepackage{iopams}
\usepackage{graphicx}
\usepackage{setstack}
\usepackage[breaklinks=true,colorlinks=true,linkcolor=blue,urlcolor=blue,citecolor=blue]{hyperref}
\usepackage{wasysym}
\usepackage{cite}

\newcommand{\mthr}{\mathrm}
\newcommand{\on}{\overline{n}}

\begin{document}

\title[Biomimetic linkers at fluid interfaces.]{Programmable interactions with biomimetic DNA linkers at fluid membranes and interfaces.}

\author{Bortolo Mognetti$^{1}$}
\address{$^1$Universit\'{e} libre de Bruxelles (ULB), Interdisciplinary Center for Nonlinear Phenomena and Complex Systems, Campus Plaine, CP
231, Blvd. du Triomphe, B-1050 Brussels, Belgium}
\ead{bmognett@ulb.ac.be}

\author{Pietro Cicuta$^{2}$}
\address{$^2$Biological and Soft Systems, Cavendish Laboratory, University of Cambridge, JJ Thomson Avenue, Cambridge CB3 0HE, United
Kingdom}
\ead{pc245@cam.ac.uk}

\author[cor1]{Lorenzo Di Michele$^{2}$}
\address{$^2$Biological and Soft Systems, Cavendish Laboratory, University of Cambridge, JJ Thomson Avenue, Cambridge CB3 0HE, United
Kingdom}
\ead{\mailto{ld389@cam.ac.uk}}

\begin{abstract}


At the heart of the structured architecture and complex dynamics of biological systems are specific and timely interactions operated by biomolecules. In many instances,  biomolecular agents are spatially confined to flexible lipid membranes where, among other functions, they control cell adhesion, motility and tissue formation.
Besides being central to several biological processes, \emph{multivalent interactions} mediated by reactive linkers confined to deformable substrates underpin the design of synthetic-biological platforms and advanced biomimetic materials. Here we review recent advances on the experimental study and theoretical modelling of a heterogeneous class of biomimetic systems in which synthetic linkers mediate multivalent interactions between fluid and deformable colloidal units, including lipid vesicles and emulsion droplets. Linkers are often prepared from synthetic DNA nanostructures, enabling full programmability of the thermodynamic and kinetic properties of their mutual interactions. The coupling of the statistical effects of multivalent interactions with substrate fluidity and deformability gives rise to a rich emerging phenomenology that, in the context of self-assembled soft materials, has been shown to produce exotic phase behaviour, stimuli-responsiveness, and kinetic programmability of the self-assembly process. Applications to (synthetic) biology will also be reviewed.

\end{abstract}

\vspace{2pc}
\submitto{\RPP}
\maketitle

\section{Introduction}
This review covers modern developments on the design, experimental study and theoretical modelling of artificial systems exploiting \emph{multivalent interactions} mediated by linkers at fluid membranes and interfaces.\\
Biological interfaces, typically lipid membranes, are decorated by a large variety of amphiphilic proteins that fulfil the most diverse range of functionalities including cell adhesion, motility, transport, environmental sensing and mechanosensing~\cite{alberts2015essential,Alberts}. In many scenarios, adhesive forces arise from specific attractive interactions between membrane-tethered receptors and ligands anchored to another cell or substrate. This is the case, for instance, of cell-adhesion molecules mediating tissue formation~\cite{Cavallaro:2011aa}, integrins anchoring cells to the extra-cellular matrix~\cite{Barczyk:2010aa}, and receptors inducing the active or passive endocytosis of particles~\cite{Doherty:2009aa}. The vast combinatorial freedom arising when a large number biomolecular agents interact simultaneously is already sufficient to produce a rich emerging phenomenology, which is made even more intriguing by the characteristic compliance of fluid biological interfaces.
The result is a complex picture in which several transport and molecular (de)complexation phenomena are strongly coupled and take place over a wide range of length-  and time-scales. In most cases, gaining a sound physical understanding of these phenomena by observing the biological systems is prohibitively challenging, owing to the extraordinary complexity of cells' active responses that often overshadow underlying fundamental mechanisms in unpredictable ways.\\
Artificial biomimetic systems, in which multivalent interactions are recreated by decorating the surface of compliant colloidal units or interfaces, offer the ideal playground to build such an understanding and, along the way, exploit it to produce tailored soft materials that could find application in several areas of advanced technology and healthcare.\\

Artificial liquid interfaces and surface films have been at the heart of colloid and interface science for over a century~\cite{lan92,sagis2011}, and synthetic lipid membranes are a classical area of activity in biophysics~\cite{BassereauBook}. Even just considering an interface decorated by a single-species of amphiphilic molecule, one has a huge space of complexity: there are 13 independent deformation modes, ranging from stretching, shearing and bending, to dilation and liquid-crystalline splay modes~\cite{lan92}. Each deformation mode correspond to different molecular mechanisms, they are all constantly excited as thermal fluctuations, and in terms of dynamics each of the modes has a corresponding spectrum of timescales. The modes can be coupled to each other, to external perturbations of the interface and to the fluid dynamics of the surroundings.
Bilayers, and multiple lipids, bring in further interface modes, and if surface tethered linkers are brought into the picture, then their interactions and dynamics, and those of resulting multivalent adhesive forces, become coupled to the underlying interface modes.\\
With artificial multivalent systems we have the unique opportunity to control this complexity, and arrange components and formulations so that only a much reduced set of degrees of freedom is at play. The simplest scenario is one of interacting solid interfaces decorated by fixed-anchor linkers, a class of systems that is still hugely relevant in the context of colloidal and nanoparticle self-assembly~\cite{Milam:2016aa,seeman-mirkin-review,lorenzo}. In this case, complexity arises from the fact that, despite being anchored to a fixed point on the surface, linkers possess a degree of conformational freedom so that various entropic and cooperative effects can already play a part~\cite{MognettiSoftMatt2012}.\\
The picture however gets much richer when the linkers are free to diffuse laterally, as is the case for fluid substrates such as supported lipid bilayers~\cite{Hook_Langmuir_2005,Boxer_PNAS_2007,VdMeulenJACS2013,shimobayashi2015direct}, emulsion droplets~\cite{Jasna_SoftMatter_2013} and lipid vesicles~\cite{Beales_JPCA_2007,parolini2014thermal}. Here, new phenomena arise from the interplay between enthalpic energy terms deriving from linker complexation, translational entropic effects associated to their redistribution, and excluded volume interactions in crowded conditions~\cite{DiMichelePRE2018}.\\
Finally, the full scale of complexity emerges with substrates that are both fluid and highly deformable, as for low-membrane-tension liposomes, where the free energy terms associated to linker conformation and lateral displacement complete with the wide variety of elastic deformation modes of the interface, as for example in~\cite{parolini2014thermal}. Clearly, situations in which substrates with different physical characteristics interact are both fundamentally intriguing and biologically relevant, as for the case of the multivalent interactions between fluid cell membranes and solid-like viruses or nanomedical probes~\cite{DiMichelePRE2018,Wel:2017aa}.\\

It would be overly simplistic to ascribe the rich phenomenology of biological-interface interactions only to the coupling between statistical multivalent effects and substrate deformability. In reality, membrane receptors are themselves highly sophisticated nanomachines with a vast range of structural features, mechanical responses, and many internal degrees of freedom. Cell-membrane receptors can form intra- or inter-membrane complexes, both dimeric and multimeric, the stability of which often depends on the presence of other molecular agents through allosteric mechanisms~\cite{Alberts,Cavallaro:2011aa,Barczyk:2010aa,Doherty:2009aa}.\\
Programming a similarly complex response with artificial linkers is however possible, if one relies on the tools of DNA nanotechnology~\cite{Seeman:2017aa}.
Indeed, the strict selectivity of DNA base-complementarity, along with the possibility of designing and (commercially) synthesising short DNA sequences, and chemically tagging them with a plethora of functional moieties, enables the construction of DNA nanostructures of near-arbitrary shape and dynamic behaviour~\cite{Seeman:2017aa}.\\
The toolkit of DNA nanotechnology has been systematically applied to engineer biomimetic linkers for synthetic multivalent systems, and this review will be mainly focusing on these implementations. Chemical functionalisations, often commercially available, enable coupling of DNA linkers to virtually any available substrate. The thermodynamic predictability of Watson-Crick base pairing~\cite{santalucia,santalucia2004thermodynamics} and other DNA-DNA complexation mechanisms such as Hoogteen interactions~\cite{Nikolova:2011aa} and G-quadruplexes~\cite{Lane:2008aa} enable control over the  free energy of interactions between the linkers, and thus the programmable formation of dimeric and multimeric complexes. Established mechanisms such as toehold-mediated strand displacement or exchange allow one to control the kinetics of DNA-bond formation and breakup, with kinetic rates that can be fine-tuned over several orders of magnitude~\cite{zhang2009control,srinivas2013biophysics}. Furthermore, thanks to chemical tags and DNA aptamers~\cite{Chen:2017aa} the kinetics and thermodynamics of DNA (de)complexation can be coupled to the presence and concentration of other molecular species, and allosteric mechanisms similar to those observed in cell receptors can be implemented~\cite{Ranallo:2017aa,Del-Grosso:2019aa,Engelen:2017aa}.\\

Thanks to the vast range of responses that can be programmed with synthetic linkers, it would be diminishing to consider linkers as mere  structural elements. Instead, we should regard interface-tethered nanostructures also as agents for sensing, signal transduction, and more generally information-processing. The community has grown around this concept, and recent years have seen a conceptual shift in the way artificial multivalent systems are perceived and applied, with an increasing number of examples in which synthetic linkers are coupled to live biological cells to perform \emph{in-situ} sensing~\cite{Salaita_NatComm_2014,You:2017aa} or drive tissue formation~\cite{Gartner_NMeth_2015}. Furthermore, the growing popularity of \emph{bottom-up synthetic biology}, aiming at constructing \emph{artificial cells}~\cite{Bartelt:2019aa}, provides a strong drive to attempt replicating life-like behaviours with ever increasing complexity such as communication~\cite{Joesaar:2019aa,Gines:2017aa,Kaufhold:2019aa} and motility~\cite{Bartelt:2018aa}.\\

In this review, we offer an overview on the milestone contributions and the state of art of biomimetic multivalent systems, covering their application to self-assembled soft materials and (synthetic) biology. Additionally, we provide a detailed summary of analytical and numerical methods developed to rationalise and predict the experimental observations, and unify them into a consistent framework for modelling multivalent interactions.\\
In Sec.~\ref{Sec2} we review common experimental implementations of artificial multivalent systems, including details on linker design and functionalisation strategies for various deformable substrates such as synthetic lipid bilayers, emulsion droplets and biological cells.
Section~\ref{Sec3} then summarises key experiments demonstrating control over the phase behaviour, structural responses, and self-assembly kinetics of compliant multivalent objects. Direct applications to biological cells are also discussed.\\
Sections~\ref{Sec4} and \ref{Sec5} are dedicated to modelling, with the former section outlining the statistical mechanical framework needed to derive multivalent-interaction free energy in all experimentally relevant scenarios, and the latter demonstrating the predictive power of these theoretical tools in designing the complex self-assembly behaviours and collective responses.\\
Finally, in Sec.~\ref{Sec6} we offer our view on the exciting future directions of the field, with particular emphasis on the applications of artificial multivalent systems to synthetic biological and healthcare technologies.\\

\section{Functional fluid interfaces.}\label{Sec2}
In this section, we review experimental realisations of artificial biomimetic systems in which forces between fluid interfaces are mediated by a large number of artificial surface-tethered linkers, resulting in multivalent interactions. We  dedicate particular emphasis to those realisations that make use of synthetic DNA nanostructures to construct  the linkers, as this approach has emerged as one of the most popular and all-around effective. In Sec.~\ref{Sec2Linkers} we describe general features of DNA linkers and DNA-mediated interactions. In Sec.~\ref{Sec2Membranes} we present methods for the functionalisation of lipid membranes and supported bilayers, before discussing the partitioning of linkers in different domains in phase-separated liposomes in Sec.~\ref{Sec2Separation}. In Secs~\ref{Sec2Droplets},~\ref{Sec2BilayerParticles} and~\ref{Sec2Cells} we describe functionalisation methods for oil-in-water droplets, bilayer-coated particles and biological cells, respectively.\\

\subsection{Synthetic DNA linkers and DNA-mediated multivalent interactions}\label{Sec2Linkers}
\begin{figure}[ht!]
\begin{center}
\includegraphics[width=8cm]{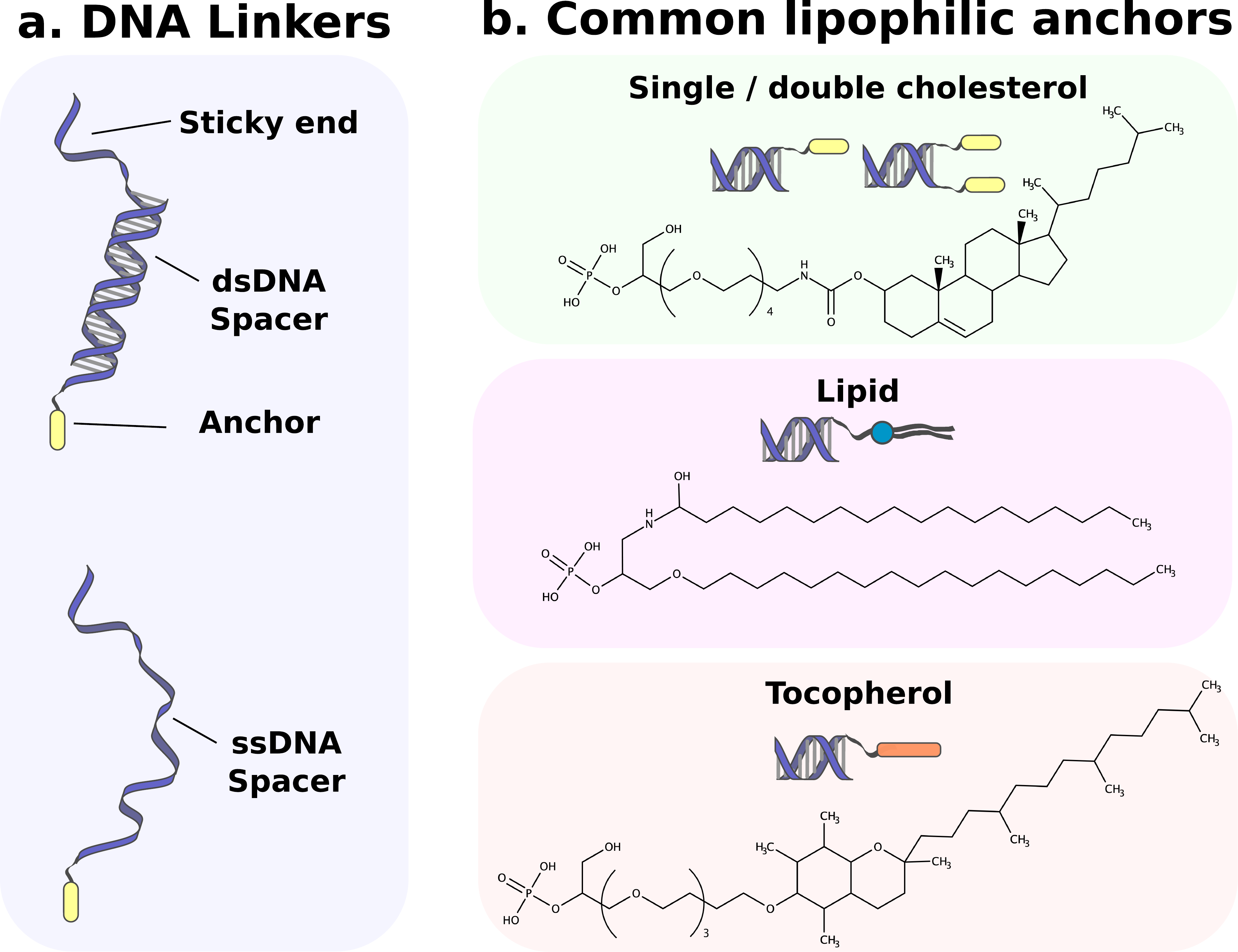}
\caption{\label{fig1} \textbf{Anatomy of DNA linkers and common hydrophobic anchors.} \textbf{a}, DNA linkers are composed by an anchoring moiety, a dsDNA (top) or ssDNA spacer, and a ssDNA sticky end mediating complexation. \textbf{b}, Hydrophobic anchors commonly adopted to confine DNA linkers to artificial lipid bilayers, emulsion droplets and biological-cell membranes. The phosphate groups on each chemical structure (left) indicates the point where they are coupled to DNA. The choleserol (top) and tocopherol (bottom) moieties are coupled to the phosphate though a tetra and a triethylene glycol spacer, respectively.}
\end{center}
\end{figure}

Since the pioneering proposals of Ned Seeman in the eighties~\cite{Seeman1982,Seeman:2017aa}, research in the field of \emph{DNA nanotechnology} proved the vast potential of synthetic nucleic acids as the building blocks for innovative nanostructured materials, owing to the unique selectivity and programmability of Watson-Crick base pairing but also to the remarkable robustness of nucleic acids and the possibility of functionalising them with a plethora of different chemical modifications. The latter enable coupling DNA nanostructures to substrates of near-arbitrary nature, and thus allow one exploit the programmability of DNA-DNA bonds to engineer multivalent interactions between particles at the nano- and colloidal scale.\\
We owe the introduction of DNA-mediated multivalent interactions to Chad Mirkin \emph{et al.}~\cite{mirkin} and Paul Alivisatos \emph{et al.}~\cite{alivisatos}, who first reported the functionalisation of gold nanoparticles with thiol-modified DNA linkers and their subsequent self-assembly mediated by DNA-bond formation. Surface-tethered DNA linkers typically feature three parts (Fig.~\ref{fig1}\textbf{a})~\cite{lorenzo}: \emph{i)} a chemical or physical~\emph{anchor} that confines the DNA construct to the substrate, reversibly or irreversibly; \emph{ii)} a \emph{spacer} prepared from single-stranded (ss) or double-stranded (ds) DNA, determining the length and polymer properties of the linker; \emph{iii)} a \emph{sticky end} consisting of a short segment of ssDNA designed to form  Watson-Crick bonds with other sticky ends, which ultimately mediates interactions. The choice of anchor depends on the substrate of interest and the functionalisation strategy, and as discussed in the following sections. The spacer plays a central role in determining the reach of the sticky end, and its flexibility and length significantly affect the linker-complexation free energy, as we discuss extensively in Sec.~\ref{Sec4:MobMob}. The base-sequence and length of the sticky end is the key design parameter, enabling the fine tuning of linker complexation free energy. The latter can be predicted from experiment-derived thermodynamic parameters~\cite{santalucia}, correcting for non-specific effects if required~\cite{DiMicheleJACS2014}.\\
The simple principle of DNA-mediated multivalent interactions resulted in a previously unthinkable control over nanoparticle and colloidal interactions, which has been demonstrated by producing a variety of self-assembled phases, spanning from crystalline superlattices~\cite{crystal-mirkin,crystal-mirkin2,auyeung2014dna,rogers-manoharan} to amorphous materials with unique local morphologies~\cite{varrato2012arrested,di2014aggregation,di2013multistep,amorphous1}. These advances have been reviewed in multiple instances~\cite{Milam:2016aa,seeman-mirkin-review,lorenzo}, and will thus not be covered here in further depth.\\
The aforementioned examples of DNA-mediated multivalent systems, despite the variety of particle materials and sizes used, share one key feature: the functionalised substrates are solid. This implies that the particles cannot be deformed under the action of the multivalent forces, but also that the linkers are anchored to a specific point on their surface and are incapable of undergoing lateral diffusion. The use of fluid and deformable substrates, reviewed here, unlocks a wide variety of complex phenomena that arise from changes in particle geometry, rearrangement of the mobile linkers, and the feedback between these two processes. These phenomena are reviewed in Sec.~\ref{Sec3} and theoretically rationalised in Secs~\ref{Sec4} and~\ref{Sec5}.

\subsection{Functionalised lipid membranes}\label{Sec2Membranes}
\begin{figure}[ht!]
\begin{center}
\includegraphics[width=12cm]{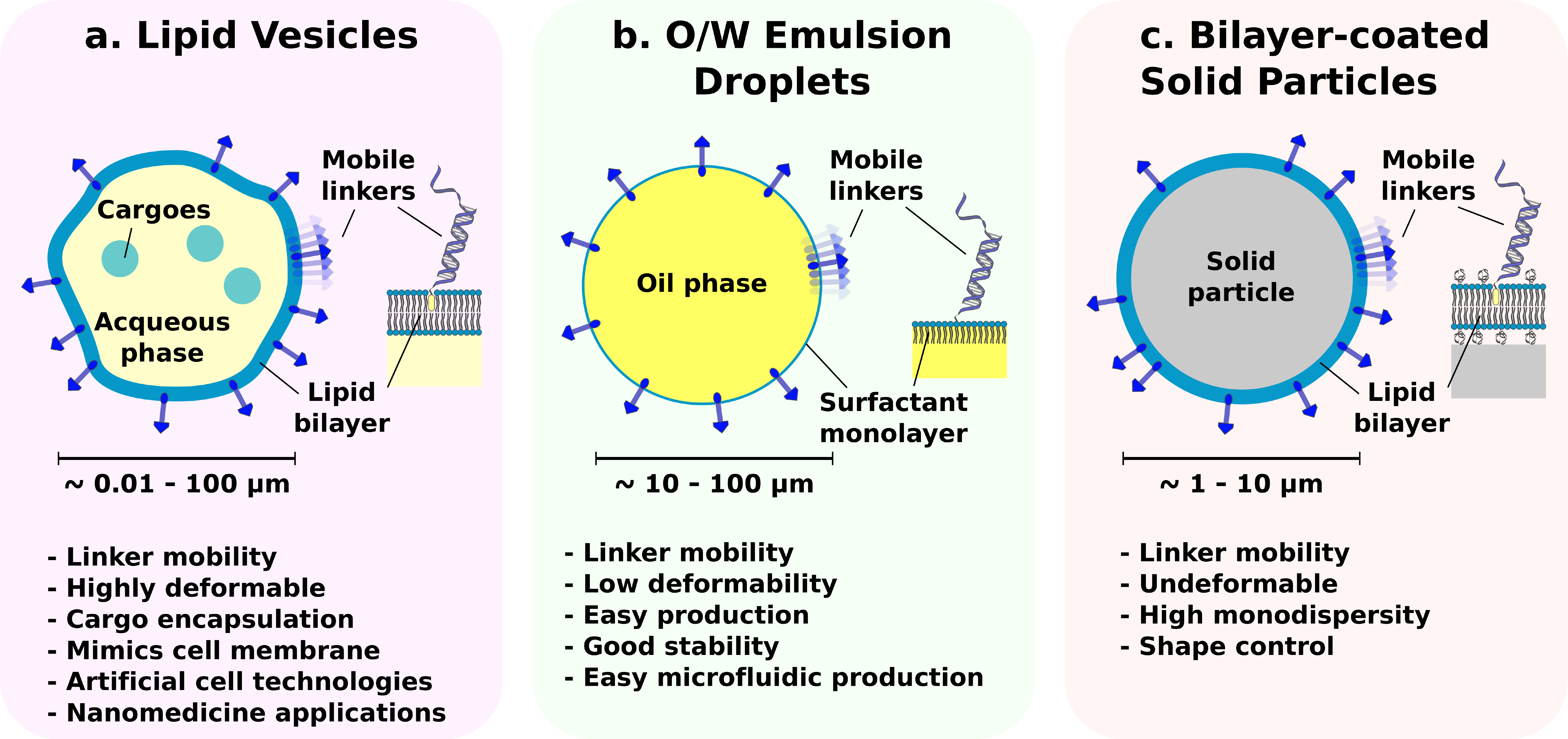}
\caption{\label{fig2} \textbf{Common types of fluid and/or compliant multivalent building blocks}. \textbf{a}, Unilamellar lipid vesicles. \textbf{b}, Oil-in-water emulsion droplets. \textbf{c}, Solid colloidal particles coated with a fluid lipid bilayer.}
\end{center}
\end{figure}

Synthetic lipid membranes exist in a variety of forms, from Unilamellar Vesicles which can be either Small (SUVs, $\sim10$\,nm), Large (LUVs, $\sim100-1000$\,nm), or Giant (GUVs, $\sim10$\,$\mu$m) (Fig.~\ref{fig2}\textbf{a}), to tubular structures, to flat Supported Lipid Bilayers (SLBs) deposited onto a solid surface. In all their flavours, artificial lipid membranes have been engineered and researched as simplified model systems to understand the biophysics of cellular membranes~\cite{BassereauBook,Chan:2007aa,Pick:2018aa}, to construct nano-therapeutic agents~\cite{BozzutoReview,Pick:2018aa}, and even to build complex multi-compartment systems that serve as the basis of entire artificial cells~\cite{Trantidou:2017aa}. In all these instances the ability to precisely control the interaction between synthetic lipid membranes, as it is enabled by carefully engineered multivalent functionalisation, is a highly desired feature. Indeed, artificial liposomes with precisely tailored multivalent interactions can provide insights into the (statistical) mechanics of cell-cell adhesion as mediated by membrane proteins, while at the same time they represent routes to program the interactions and mechanical response of artificial cells and tissues. Similarly, carefully engineered functionalisation of vesicle-based drug delivery vectors can lead to more effective cell-targeting strategies~\cite{Sercombe2015}.\\
Functionalisation of synthetic lipid bilayers with DNA linkers can be performed by utilising lipophilic anchors, which posses the capacity of spontaneously inserting within the hydrophobic core of the bilayer, forming a physical bond whose strength depends largely on the nature of the anchoring moiety (Fig.~\ref{fig1}\textbf{b}). Among others, this strategy has been explored and optimised by H\"{o}\"{o}k and coworkers~\cite{Hook_Langmuir_2004,Hook_JSB_2009,Hook_JACS_2004,Hook_Langmuir_2005}  and Boxer and coworkers~\cite{Boxer_JSB_2009,Boxer_Langmuir_2010,Boxer_Langmuir_2011,Boxer_Langmuir_2013,Boxer_FD_2013,Boxer_JACS_2005} for the case of SLBs as well as liposomes. Tested anchor moieties include cholesterol, lipids, tocopherol or carbon chains, as recently reviewed by Lopez and Liu~\cite{Lopez:2018aa}. Among these, cholesterol is most commonly used owing to the commercial availability of such chemical modification for DNA oligomers.  Although an individual cholesterol anchor has been demonstrated to provide relatively stable anchoring of DNA nanostructures, Pfeiffer \emph{et al.} proved that insertion is in fact reversible~\cite{Hook_Langmuir_2004}, with off rates  $\sim 5.8 \times 10^{-4}$\,s$^{-1}$. In turn, a double cholesterol anchor obtained by functionalising both ends of a DNA duplex, produces irreversible insertion~\cite{Hook_Langmuir_2004}. Cholesterol (or cholesteyl) moieties are typically conjugated \emph{via} a tetra ethylene-glycol (TEG) spacer~\cite{Lopez:2018aa,Bunge:2009aa}, as depicted in Fig.~\ref{fig1}\textbf{b} (top), although hydrophobic spacers such as  C6 or direct conjugation are also available. Similarly strong coupling with the membranes can be achieved using lipid anchors~\cite{Boxer_JACS_2005,Boxer_JSB_2009,Boxer_Langmuir_2010,Boxer_Langmuir_2011,Boxer_Langmuir_2013,Boxer_FD_2013},  however these are not easily accessible commercially (Fig.~\ref{fig1}\textbf{b}, centre). A recently popularised approach to lipid-DNA conjugation makes use of  $N_3$ - DBCO click chemistry~\cite{Mirkin_Small_2018}. Tocopherol anchors hold promises as good alternatives in view of their binding stability, biocompatibility, and commercial availability (Fig.~\ref{fig1}\textbf{b}, bottom)~\cite{Bunge:2007aa,Mirkin_JACS_2014}.\\
An appealing alternative to the aforementioned lipophilic anchors is the one introduced by Hern\'{a}ndez-Ainsa \emph{et al.}~\cite{Hernandez-Ainsa:2016aa}, who conjugated DNA oligonucleotides with azobenzene. This moiety exists in two conformations: a high-hydrophobicity trans state and a low-hydrophobicity cis state, and can be made to switch between the two by exposure to UV light of specific wavelengths. While the trans configuration has been shown to stably anchor the linkers to LUVs, transition to cis causes their detachment, offering a route for light-sensitive multivalent interactions~\cite{Hernandez-Ainsa:2016aa}.\\
Besides their use as membrane-tethered linkers, hydrophobised DNA  strands or constructs are utilised in several other contexts of DNA nanotechnology~\cite{Czogalla:2016aa}, \emph{e.g.} to produce tailorable micelle-like aggregates for cargo capture and delivery~\cite{Berti:2011aa,Liu:2010aa}, to aid self-assembly of macromolecular crystals~\cite{Brady:2019aa,Brady:2018aa,Brady:2017aa}, to membrane-anchor bulkier nanostructures based on DNA origami~\cite{Czogalla:2016aa,Franquelim:2018aa,Langecker:2014aa} and trans-membrane channels~\cite{Chidchob:2019aa,Burns:2013aa,Langecker:2014aa}, or to template the growth of nanoparticles~\cite{Trinh:2017aa}.\\
Multi-step functionalisation protocols relying on biotin-streptavidin labelling have also been developed, in which both lipids and DNA oligomers carry a biotin moiety, and biotin mediates bridging between the two~\cite{Hadorn_Langmuir_2013,Hadorn_PLOSONE_2010}.\\

\subsection{Partitioning of linkers in lipid domains}\label{Sec2Separation}
Lipid phase separation occurs in multicomponent membranes of biological or synthetic origin, typically in the presence of mixtures of saturated lipids, unsaturated lipids and sterols~\cite{SilviusJ.R.1982,Veatch:2003aa,Cicuta:2007aa,Honerkamp-Smith:2008aa}. Different membrane compositions and environmental conditions lead to coexisting lipid phases with a variety physical properties, including liquid ordered ($L_o$), liquid disordered ($L_d$) and gel-like phases~\cite{SilviusJ.R.1982,Veatch:2003aa}. Depending on the nature of the anchor used, DNA linkers have been demonstrated to strongly partition into different lipid phases, enabling patterning of the surface of liposomes~\cite{Beales_JPCB_2009,Beales_SoftMatter_2011,Emma_PNAS_2017,Bunge:2007aa}. In GUVs displaying $L_d$-$L_o$ phase separation, tocopherol-linked DNA oligomers have been demonstrated to preferentially partition within the liquid-disordered domains~\cite{Kurz:2006aa,Bunge:2007aa}. In contrast, Beales \emph{et al.} showed how the use of cholesterol anchors results in segregation within the $L_o$ phase~\cite{Beales_JPCB_2009,Beales_SoftMatter_2011}, and how the degree of partitioning is strongly dependent on the chemical composition of the membrane~\cite{Beales_SoftMatter_2011}.  Additionally, the degree of segregation of cholesterol-DNA constructs is stronger if a double anchor is utilised~\cite{Beales_JPCB_2009}. For membranes undergoing liquid-solid phase separation, cholesterol modified strands partition preferentially within the liquid domains~\cite{Beales_JPCB_2009}.\\
Partitioning of linkers within lipid domains ultimately produces anisotropy of the resulting multivalent interactions. For suitable lipid compositions, GUVs undergoing complete liquid-liquid phase separation acquire a Janus-like appearance, with roughly one hemisphere occupied by the $L_d$ phase and the other consisting of $L_o$ phase. In these conditions, attractive forces produced by $L_o$-segregating DNA linkers are highly directional, and the formation of size-limited GUV clusters is observed~\cite{Beales_SoftMatter_2011}.\\
The thermophoretic motion lipid domains has been exploited by Talbot~\emph{et al.} to direct the distribution of membrane-linked DNA cargoes in thermal gradients~\cite{Emma_PNAS_2017}.

\subsection{Functionalsied emulsion droplets}\label{Sec2Droplets}
Oil-in-water emulsion droplets represent an appealing alternative to liposomes as substrates for investigating multivalent interactions at fluid interfaces, both in the context of biomimicry~\cite{Jasna_PNAS_2012}, and in the fundamental study of soft-matter phenomena (Fig.~\ref{fig2}\textbf{b})~\cite{Jasna_SoftMatter_2013,Jasna_PRL_2018,zhang2018multivalent,zhang2017sequential,Hadorn_PNAS_2012,Hadorn_Langmuir_2016}.  With lipid vesicles, oil-in-water droplets share the mobility of surface tethered constructs but differ substantially in terms of mechanical properties: droplet elasticity is typically dominated by interfacial tension~\cite{Jasna_PNAS_2012}, while a variety of contributions including membrane bending, entropic effects associated to thermal fluctuations and the presence of excess surface area tend to play a major role in the deformation response of vesicles~\cite{DIMOVA2014}. Droplets offer advantages in terms of stability, robustness, and the possibility of easily generating large mono-disperse samples by means of microfluidic devices~\cite{Meissner:2017aa}.\\
Methods for anchoring molecular linkers to oil droplets have often involved the use of phospholipids whose head groups are functionalsied with PEG-biotin anchors~\cite{Jasna_SoftMatter_2013,Jasna_PRL_2018,zhang2017sequential,Hadorn_PNAS_2012,Hadorn_Langmuir_2016,Kraft_JPCM_2018}. Similar to the strategy mentioned above for DNA functionalisation of liposomes, streptavidin is then used to bridge between the lipid anchors and oligonucleotides~\cite{Jasna_SoftMatter_2013,Jasna_PRL_2018,zhang2017sequential,Hadorn_PNAS_2012,Hadorn_Langmuir_2016,Kraft_JPCM_2018} or more complex DNA nanostructures~\cite{zhang2018multivalent}, which then mediate multivalent interactions. However, streptavidin itself can be used as the linker molecule, producing near-irreversible bonds between droplets~\cite{Jasna_PNAS_2012}. As for the case of lipid bilayers, an appealing alternative consists in the use of copper-free click chemistry~\cite{Jasna_PRL_2018}.

\subsection{Solid particles with mobile linkers}\label{Sec2BilayerParticles}
Free lateral diffusion of linkers can be obtained also for solid colloidal particles, if the latter are coated by a fluid lipid bilayer \cite{TroutierAdvCollIntSci20071} to which the linkers can be anchored following the same strategies adopted for liposomes. In fact, the original protocol introduced by van der Meulen and Leunissen proceeds by first functionalising SUVs with DNA linkers featuring a double cholesterol anchor, and then incubating the latter with micron-scale silica particles~\cite{VdMeulenJACS2013}. The rupture of the SUVs against the silica surface results in the formation of a uniform, fluid, DNA-functionalised lipid bilayer. The functionalisation strategy has later been extensively applied and optimised by Kraft and coworkers~\cite{RinaldinSoftMatter2018,Wel:2017aa,Kraft_Nanoscale_2017}, who explored the effect of changing particle material (polystyrene, hematite, TMP) surface chemistry (carboxylated and amino-modified surfaces) and lipid-bilayer composition (presence and concentration of PEG-modified lipis)~\cite{RinaldinSoftMatter2018}.

\subsection{Functionalising biological cells with artificial linkers}\label{Sec2Cells}
Although the main focus of the present review is on synthetic multivalent systems, it is important to mention the usefulness of artificial (DNA) linkers in the context of cell biology and biotechnology, and the relevance of the functionalisation strategies we just discussed to cellular systems.
Gartner and coworkers, indeed, demonstrated how lipid modifications similar to those utilised to label liposomes and supported bilayers can anchor the synthetic DNA linkers to the plasma membrane of a variety of mammalian cells~\cite{Gartner_NMeth_2015,Gartner_CPCB_2016}, and a similar outcome was reported by You \emph{et al.} using lipid, tocopherol or cholesterol anchors~\cite{You:2017aa}.  Other labelling methods include direct conjugation of oligonucleotides modified by N-hydroxysuccinimidyl (NHS) ester to amine groups on the cell~\cite{Mathies_Lang_2012}, and the use of adhesive peptides connected to the DNA probes \emph{via} click chemistry~\cite{Salaita_NatComm_2014}.

\begin{figure}[ht!]
\begin{center}
\includegraphics[width=8cm]{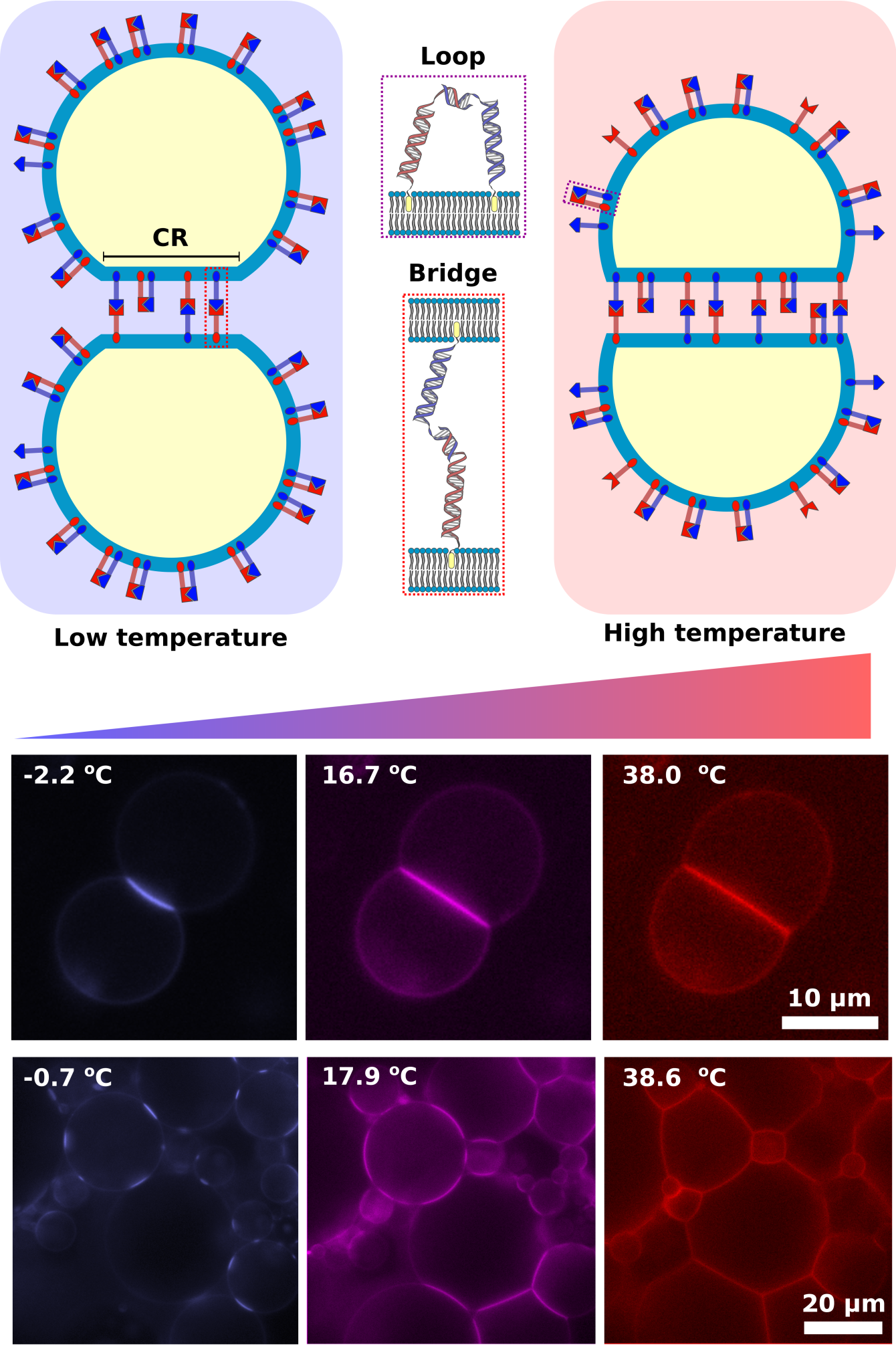}
\caption{\label{fig3} \textbf{Temperature-depedendent morphology of liposome dimers and tissue like materials~\cite{parolini2014thermal}}. GUVs are functionalised with linkers capable of forming both intra-liposome \emph{loops} and inter-liposome \emph{bridges}. The latter partition within the contact region (CR) of interacting vesicles, leading to adhesive forces. The liposome pairs undergo morphological changes upon tuning temperature, with the CR shrinking on cooling and expanding on heating, a process caused by the temperature-dependent excess area of the GUVs and enabled by linker diffusivity and relaxation of the bridge and loop distributions. Epifluorescence micrographs (reproduced from~\cite{parolini2014thermal}) demonstrate the structure change in GUV pairs (centre) and tissue-like materials (bottom). Micrographs are show in false colours to highlight the temperature trend.}
\end{center}
\end{figure}

\section{Controlling interactions between individual units and the self-assembly of tissue-like materials}\label{Sec3}
In this section we review a selection of experimental studies that demonstrated programmability, control, and understanding of the rich phenomenology achievable with synthetic multivalent systems, in which the substrates involved share features of deformability and reconfigurability. In Secs~\ref{Sec3Eq} and \ref{Sec3Kin} we focus on systems of multivalent liposomes and lipid membranes, with the former section discussing equilibrium and thermodynamic properties of the self-assembled constructs and materials, and the latter being dedicated to kinetic phenomena and programmability. We dedicate Sec.~\ref{Sec3Fusion} to those implementations where DNA linkers are designed to mediate membrane fusion, rather than simple adhesion. Section~\ref{Sec3EmulsAndSolid} concerns the self-assembly of less compliant fluid units, namely emulsion droplets and solid particles coated by a supported lipid bilayer. Section~\ref{Sec3Cells} discusses innovative biological applications in which complex DNA linkers are coupled to live cells. Throughout, we refer to theoretical models for multivalent interactions and, on occasions, coarse-grained molecular simulations implemented to rationalise the experimental observations. These are be reviewed in detail in Secs~\ref{Sec4} and \ref{Sec5}.

\subsection{Equilibrium properties of multivalent adhesion between lipid membranes}\label{Sec3Eq}
Functionalised synthetic liposomes and bilayer membranes have been the object of intense multidisciplinary research owing to their relevance to several fundamental and applicative contexts, including self-assembled soft materials~\cite{parolini2014thermal,Beales_JPCA_2007}, biophysical models for cell adhesion~\cite{BassereauBook,Chan:2007aa,Pick:2018aa},  \emph{in-vivo} delivery of therapeutic agents~\cite{BozzutoReview,Pick:2018aa}, and artificial-cellular systems~\cite{Trantidou:2017aa}. In all these situations, a precise control over the thermodynamic properties of the interactions and the structure of the resulting phases is an extremely desirable feature. Here we summarise key experimental results offering evidence and basic understanding of the unique phenomenology arising from the combination of multivalent interactions with lateral diffusivity of the linkers and deformability of the bilayers.\\
Early studies of DNA-controlled adhesion of liposomes were focused on demonstrating addressability and specificity of the interactions. Pioneering works from the groups of  Steven Boxer~\cite{Boxer_JSB_2009,Boxer_Langmuir_2010,Boxer_Langmuir_2011,Boxer_Langmuir_2013,Boxer_FD_2013,Boxer_JACS_2005,Boxer_PNAS_2007} and Fredrik H\"{o}\"{o}k~\cite{Hook_Langmuir_2004,Hook_JSB_2009,Hook_JACS_2004,Hook_Langmuir_2005} reported on specific adhesion of lipid vesicles of various sizes to SBLs, functionalised respectively with linkers featuring complementary sticky ends. DNA-functionalised liposomes tethered to SBLs retain the ability to diffuse laterally, and can interact with each other if functionalised with linkers orthogonal to those mediating the interactions with the SBL, making these systems ideal for studying the kinetics of vesicle-vesicle ``docking'' interactions~\cite{Boxer_PNAS_2007,Boxer_Langmuir_2010}. Boxer and coworkers demonstrated also the controlled rupture of GUVs adhering to SBLs, forming bilayer islands (referred to as ``patches'' by the authors) tethered to the underlying SBL by DNA linkers~\cite{Boxer_JSB_2009,Boxer_Langmuir_2011,Boxer_FD_2013}. The equilibrium distance between SLB and patch can be tuned by changing the length of the dsDNA spacers in the linkers, with longer spacers resulting in larger distances. Multi-storey patches can also be created. Remarkably, if linkers with spacers of different lengths are used on the same patch, the authors observe their segregation in domains rich with linkers of a given length~\cite{Boxer_FD_2013}. The observed phase separation is ascribed to the minimisation of the bending energy of the suspended membrane, as nearby tethers of different length induce local deformations that are minimised by lateral segregation~\cite{Boxer_FD_2013}.\\
Beales and Vanderlick investigated systematically the DNA-mediated self-assembly of both LUVs and GUVs suspended in solution, observing thermal reversibility and dependence of the presence and size of the aggregates on monovalent salt concentration and DNA surface coverage~\cite{Beales_JPCA_2007}. Later, the same authors performed a quantitative study of the thermal dissociation of DNA-functionalised LUVs and the hybridisation transition of the tethered DNA by means of dynamic light scattering (DLS) and UV-absorbance spectrophotometry~\cite{Beales_BJ_2009}. The authors focused on the effect of intermembrane interactions emerging if a small proportion of cationic lipids (DOPE) is included in the otherwise zwitterionic membranes (POPC), leading to a strengthening of the DNA-mediated attractive interaction due to non-specific bridging mediated by the polyanionic DNA~\cite{Beales_BJ_2009}.\\
Bachmann \emph{et al}. later reported on another quantitative study of the melting transition in DNA-functionalised LUVs, where the degree of association of the DNA linkers could be decoupled from the aggregation state of the liposomes~\cite{BachmannSoftMatter2016}. The former was  determined \emph{via} temperature dependent F\"{o}rster Resonance Energy Transfer (FRET) measurements, thanks to donor and acceptor fluorophores chemically conjugated to the complementary sticky ends, while the latter was assessed by Fourier analysis of microscopy images. The two measurements could be carried out simultaneously thanks to a fully automated epifluorescence microscopy setup. Note that different from earlier contributions, here all the LUVs in the systems were functionalised with both the mutually complementary linkers, such that the formation of intra-vesicle \emph{loops} would compete with that of inter-vesicle \emph{bridges} (Fig.~\ref{fig3}). While, owing to the fluidity of the membrane, loops can freely diffuse across the entire surface of the liposomes, bridges are confined to the Contact Region (CR) between interacting liposomes (Fig.~\ref{fig3}). The authors explored a wide range of surface densities of DNA-linkers and demonstrated that both the DNA melting temperature and the liposome dissociation temperature increase monotonically with linker concentration, an observation readily rationalisable with the greater attractive combinatorial contribution to the interaction free energy, as discussed in detail in Sec~\ref{Sec4:MobMob}. A saturation of both DNA melting temperature and LUV dissociation temperature was observed at high surface coverage, a behaviour that the authors ascribed to steric interactions between the DNA linkers. No LUV self-assembly was detected below a threshold DNA concentration, and in all cases where a dissociation transition was observed, it occurred at higher temperature compared to the melting of the DNA linkers. Finally, the authors determined the number of formed DNA bonds, including both loops and bridges, present at the dissociation temperature of the LUVs. Strikingly, this quantity was observed to increase with increasing surface coverage, hinting once again at the importance of steric interactions that might hinder bridge formation due to crowding effects in the CR.  Experimental observations on these systems were rationalised by means of coarse-grained Monte Carlo simulations that, however, neglected steric interactions. These are reviewed extensively in Sec.~\ref{Sec4:Elastic}.\\
Parolini \emph{et al.}~\cite{parolini2014thermal} performed a comprehensive fluorescence microscopy study of the static properties on individual pairs of GUVs interacting through bridge and loop-forming DNA tethers analogous to those of Ref.~\cite{BachmannSoftMatter2016} (Fig.~\ref{fig3}). Given the larger size of the vesicles, here the authors could determine the temperature-dependence of the geometry of the interacting pair, as well as the distribution of the fluorescently labelled DNA tethers. As previously reported~\cite{Beales_JPCA_2007}, owing to their deformability adhering GUVs were observed to give rise to a (usually) flat CR, where DNA linkers are preferentially recruited due to their tendency of forming bridges. Upon increasing temperature while remaining below the melting of the linkers and the dissociation of the GUVs, the authors observed a significant morphological change, with the CR expanding, and then reversibly returning to its original size upon cooling (Fig.~\ref{fig3}). This transition was ascribed to the change in the excess area of the GUV following the thermal expansion of the bilayer, which makes vesicles more ``deflated'' and thus easily deformable at higher temperatures, along with the rapid redistribution of the mobile linkers~\cite{parolini2014thermal}. Since bridges are necessarily confined to the contact region, the free energy of bridge formation depends on the area of the CR though an entropic confinement term (see Sec.~\ref{Sec4:MobMob}), producing a non-trivial temperature dependence on the fraction of linkers involved in different types of constructs, and hence in the overall GUV-adhesion free energy~\cite{parolini2014thermal}. As shrinkage of the CR leads to an increase in the centre-to-centre distance between the interacting GUVs, tissue-like materials formed by a large number of interacting GUVs posses a \emph{negative thermal expansion coefficient}: upon heating the tissue expands as large gaps open between neighbouring GUVs (Fig.~\ref{fig3}, bottom)~\cite{parolini2014thermal}.\\
Investigating a closely related system, Shimobayashi \emph{et al.}~\cite{shimobayashi2015direct} could quantitatively characterise the melting transition using a FRET assay similar to the one mentioned above for the case of LUVs~\cite{BachmannSoftMatter2016}. Here, as opposed to two pairs of interacting GUVs, the authors focused on the interaction of an individual GUV with a SLB, both functionalised with loop- and brigde-forming DNA tethers. In such configuration, the presence and size of the CR could be easily determined by means of confocal microscopy, which also enabled \emph{in situ} FRET measurements. Interestingly, the authors observed a higher melting temperature for DNA bonds within the CR as opposed to the surrounding Outer Region (OR). Confocal microscopy also allowed the authors to image the equatorial plane of the GUV, and record its thermal fluctuations. Flickering spectroscopy could then be used to determine the mechanical properties of the membrane from the power spectrum of its undulations, including the DNA-induced membrane tension~\cite{Pecreaux2004,shimobayashi2015direct,Rautu:2017aa}. The latter was surprisingly found to remain rather constant with changing temperature up to the point of GUV detachment, an evidence that was in agreement with theoretical predictions based on the multivalent models discussed in Sec.~\ref{Sec4}.\\
The same GUV-SLB configuration was employed by Amjad \emph{et al.} to investigate the effect of multivalency of the individual linkers~\cite{Amjad:2017aa}. Here, tetravalent streptavidin was used to connect dsDNA linkers anchored to the GUV and the SLB, whose sticky ends were replaced by biotin tags. Near-irreversible biotin streptavidin-bonds lead to the formation of constructs featuring one to four DNA linkers, with the proportion of the different valencies being dependent on the concentration ratio between biotinylated DNA linkers and streptavidin. Interestingly, a re-entrant adhesion transition was observed as a function of the streptavidin concentration. Too little streptavidin indeed would not produce a sufficient number of tethering complexes, while too much would saturate all available biotins such that all formed complexes are monovalent, and thus unable to form GUV-SLB bridges~\cite{Amjad:2017aa}.\\
Beales \emph{et al.}~\cite{Beales_SoftMatter_2011} exploited the preferential partitioning of cholesterol-anchored DNA linkers to the liquid-ordered phases of fully demixed GUVs to create ``Janus vesicles", in which the linkers are localised to one hemisphere. The authors demonstrated that the high directionality of the resulting interactions leads to the formation of finite-size clusters rather than extended networks, offering a route for controlling the morphology of the self-assembled aggregates~\cite{Beales_SoftMatter_2011}.\\
GUVs are routinely used to create compartments for artificial cells, and one of the most popular realisations of the latter consists in encapsulating cell extracts or reconstituted transcription/translation machinery capable of expressing proteins in the presence of the relevant (genomic) DNA templates~\cite{Garamella:2016aa}. Hadorn \emph{et al.} showed that these protein-expressing GUVs can be decorated with sticky DNA linkers and made to form artificial tissues~\cite{Hadorn_Langmuir_2013}.\\

\begin{figure}[ht!]
\begin{center}
\includegraphics[width=17cm]{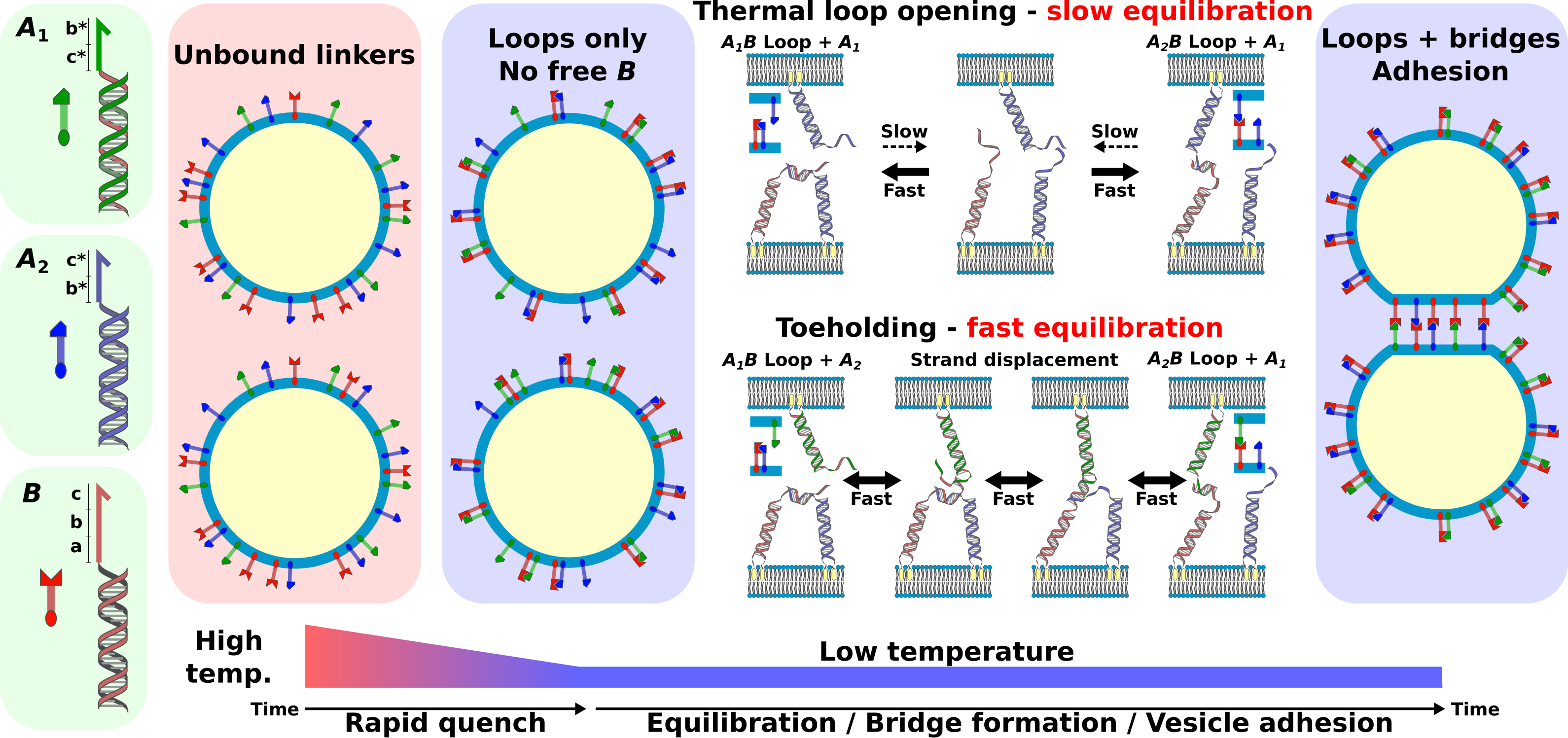}
\caption{\label{fig4} \textbf{Toehold-mediated strand exchange controls aggregation kinetics of DNA-functionalised liposomes~\cite{ParoliniACSNano2016}.} From left to right. LUVs are decorated with three types of DNA linkers, $A_1$, $A_2$ and $B$. $B$ can bind to $A_1$ and $A_2$, which can displace each other through toehold-mediated exchange so that $A_1B$ and $A_2B$ bonds can efficiently be swapped~\cite{zhang2009control}. DNA domains on the $B$ sticky end are marked as $a$, $b$ and $c$, and their complementary domains on $A_1$ and $A_2$ as $a*$, $b*$ and $c*$. Upon rapid cooling from above to below the DNA complexation temperature, kinetically favoured $A_1B$ or $A_2B$ loops form on the LUVs, sequestrating all available $B$ linkers and hindering LUV-LUV aggregation. The latter can thus occur only if, upon LUV-LUV collision, existing loops break and are replaced by bridges. This equilibration process can follow two routes: a slow process relying on spontaneous thermal opening of the loops and a fast toehold-mediated bond-swapping pathway. The prominence of each process, and thus the overall rate of aggregation can be regulated by tuning the surface concentration of the different linker species~\cite{ParoliniACSNano2016}.}
\end{center}
\end{figure}

\subsection{Controlling the kinetics of multivalent liposome adhesion}\label{Sec3Kin}
The equilibrium state of a multivalent system is characterised by a well defined configuration for the network of bonds linkers forming within the same object or across different ones, as thoroughly outlined in Sec.~\ref{Sec4}. For the case of functionalised lipid membranes or liposomes, owing to the mobility of the linkers and the deformability of the substrates, the equilibrium configuration also features non-trivial spatial distributions of the linkers and the different types of bonds they can form (\emph{e.g.} loops or bridges), as well as  specific 3D morphologies of the substrates, as reviewed in Sec. \ref{Sec3Eq}.\\
The different degrees of freedom at play, however, are found to relax towards equilibrium over timescales that can massively differ from each other, and change with experimental conditions by orders of magnitude, producing a complex and intriguing kinetic phenomenology. Key examples are the timescales associated with the 2D-diffusion of linkers on the membranes, the 3D diffusion of liposomes in bulk, and the rates of bond formation and breakup. In the following we  review experimental work where the influence of these processes on the kinetics of multivalent adhesion has been characterised, rationalised and programmed. Theoretical considerations and numerical methods developed to rationalise such experimental observations are reviewed in Sec.~\ref{Sec6:ReactionRates}.

Parolini \emph{et al.} exploited the interplay between relevant timescales, along with the programmability of binding-unbinding rates enabled by DNA nanotechnology, to devise a multivalent functionalisation scheme where the rate of liposome aggregation can be controlled isothermally over more than one order of magnitude~\cite{ParoliniACSNano2016}. As depicted in Fig.~\ref{fig4}, LUVs are decorated with three different DNA linkers with sticky ends labelled as $A_1$, $A_2$ and $B$, and designed in a way that $A_1$ and $A_2$ can bind to $B$ with similar free energy, while $A_1A_2$ bonds are forbidden. All constructs are present on all LUVs, such that both intra-vesicles loops and inter-vesicle bridges can form. However, the surface densities of the linkers are such that $[A_1] + [A_2]$ = $2[B]$, such that when all possible $A_1B$ and $A_2B$ bonds are formed, 50\% of all $A_1$ or $A_2$ linkers remain free. Upon rapid cooling, from above to below the association temperature of the DNA linkers, loop formation is kinetically favoured over bridge formation owing to the high surface density of the linkers, their relatively high lateral diffusivity, and the slow 3D diffusion of the LUVs. After an initial quench, therefore, nearly all available $B$ linkers are sequestered into loops, and only non-interacting $A_1$ and $A_2$ remain available, suppressing bridge formation and LUV-LUV aggregation~\cite{miriam-nature}. The loop-only configuration is however a kinetically arrested state, because owing to combinatorial entropy a fraction of the linkers must form bridges at equilibrium (Sec.~\ref{Sec4}). The system is therefore bound to relax towards a configuration in which bridges are present and LUVs form aggregates. The timescales of this relaxation are directly dependent on the breakup rate of formed loops, and as such can be tuned by changing temperature: if the system is kept well below the DNA melting temperature, loops are very stable and LUV aggregation is slower, while increasing temperature destabilises loops and speeds-up bridge formation. In their contribution, the authors demonstrate the validity of a different route to control the rate of loop breakup and bridge formation, and therefore the aggregation rate of LUVs, relying on \emph{toehold-mediated-exhange}~\cite{zhang2009control,srinivas2013biophysics}. As displayed in Fig.~\ref{fig4}, this is a process through which, in the attempt of forming a bridge, an $A_1$ ($A_2$) sticky end can transiently bind to a $A_2B$ ($A_1B$) loop though a short ssDNA domain exposed on $B$ (a \emph{toehold}), triggering the displacement of $A_2$ ($A_1$) through a \emph{branch-migration} process and producing a $A_1B$ ($A_2B$) bridge. The toehold assisted pathway massively reduces the free-energy barrier for loop-to-bridge swapping, making the relaxation significantly faster, particularly at low temperature where the toehold-independent pathway (relying on thermal breakup of the loops) is slowest~\cite{ParoliniACSNano2016}. The predictable rates of formation and breakup of DNA bonds, combined with our understanding of the statistical physics of multivalent interactions, enabled the prediction of the temperature dependence of bridge-formation rates in the presence and absence of toeholding~\cite{ParoliniACSNano2016}, as further discussed in Secs~\ref{Sec4:MobMob},~\ref{Sec4:Multimeric}, and~\ref{Sec6:ReactionRates}.\\
The combination of near-irreversible bond formation with surface mobility of the linkers can result in arrested adhesion processes also in the absence of loop/bridge competition, as recently demonstrated by Lanfranco~\emph{et al.}~\cite{LanfrancoLangmuir2019} in a system of DNA-functionalised gold nanoparticles (GNPs, $\diameter\sim20$\,nm) interacting with LUVs ($\diameter\sim200$\,nm) hosting the complementary sticky ends. As the first GNPs adhere to LUVs they recruit membrane-tethered linkers within the GNP-LUV CR, over timescales faster compared to the average time interval between subsequent GNP-LUV adhesion events. As a result, the surface of the LUV gets rapidly depleted of linkers, making it harder for late-coming GNPs to form stable bonds and eventually leading to kinetic arrest of the adhesion process, an effect that becomes more severe if the surface density of linkers on the LUVs is reduced~\cite{LanfrancoLangmuir2019}.  These conclusions have been backed by Ref.~\cite{LanfrancoLangmuir2019} using theory and simulations (see Sec.~\ref{Sec6:ReactionRates}). Interestingly, arrested adhesion processes due to slow bond-breakup rates have been reported by Lewis {\em et al.}~\cite{LewisLangmuir2019} while studying functionalised nanoparticles targeting supported bilayers. Although in this system linkers are not mobile, reduced lateral diffusivity of nanoparticles forming strong bonds with the surface hampers relaxation towards close-packed nanoparticle monolayers.

\subsection{Controlling the fusion of liposomes}\label{Sec3Fusion}
Besides mediating adhesion, membrane-anchored DNA linkers have been developed to induce the partial or complete fusion between liposomes, mimicking the functionality of   SNARE (soluble N-ethylmaleimide-sensitive factor attachment protein receptors) proteins that regulate the fusion of biological membranes~\cite{Stengel:2007aa,Stengel:2008aa,Chan:2008aa,Lengerich:2013aa,Rawle:2016aa,Flavier:2017aa,Chan:2009aa}. In adhesion-inducing linkers, the double-stranded DNA spacers provide steric repulsion between the interacting membranes, typically keeping them a few nanometers apart~\cite{Boxer_JSB_2009,Boxer_Langmuir_2011,Boxer_FD_2013,parolini2014thermal}. Linkers designed to induce fusion do not feature repulsive spacers and the sticky ends are designed to zip up from their tip to their anchoring point, bringing the two membranes very close to each other. Extreme proximity then triggers fusion of the liposomes. In some cases, fusion is limited to the outer leaflet of the membranes, producing a mixing of the lipids but keeping the lumen of the two vesicles separated. In other cases both leaflets fuse, leading to complete merging of the liposomes and content mixing. Generally, outer-leaflet fusion can be obtained with nearly 100\% efficiency while content mixing occurs less readily, with efficiencies of a few percents~\cite{Stengel:2007aa,Stengel:2008aa,Chan:2008aa,Lengerich:2013aa,Rawle:2016aa,Flavier:2017aa,Chan:2009aa}. DNA-mediated vesicles fusion has been demonstrated using different types of hydrophobic anchors, including cholesterol~\cite{Stengel:2007aa,Stengel:2008aa} and various lipids~\cite{Chan:2008aa,Lengerich:2013aa,Rawle:2016aa,Flavier:2017aa,Chan:2009aa}.

\begin{figure}[ht!]
\begin{center}
\includegraphics[width=8cm]{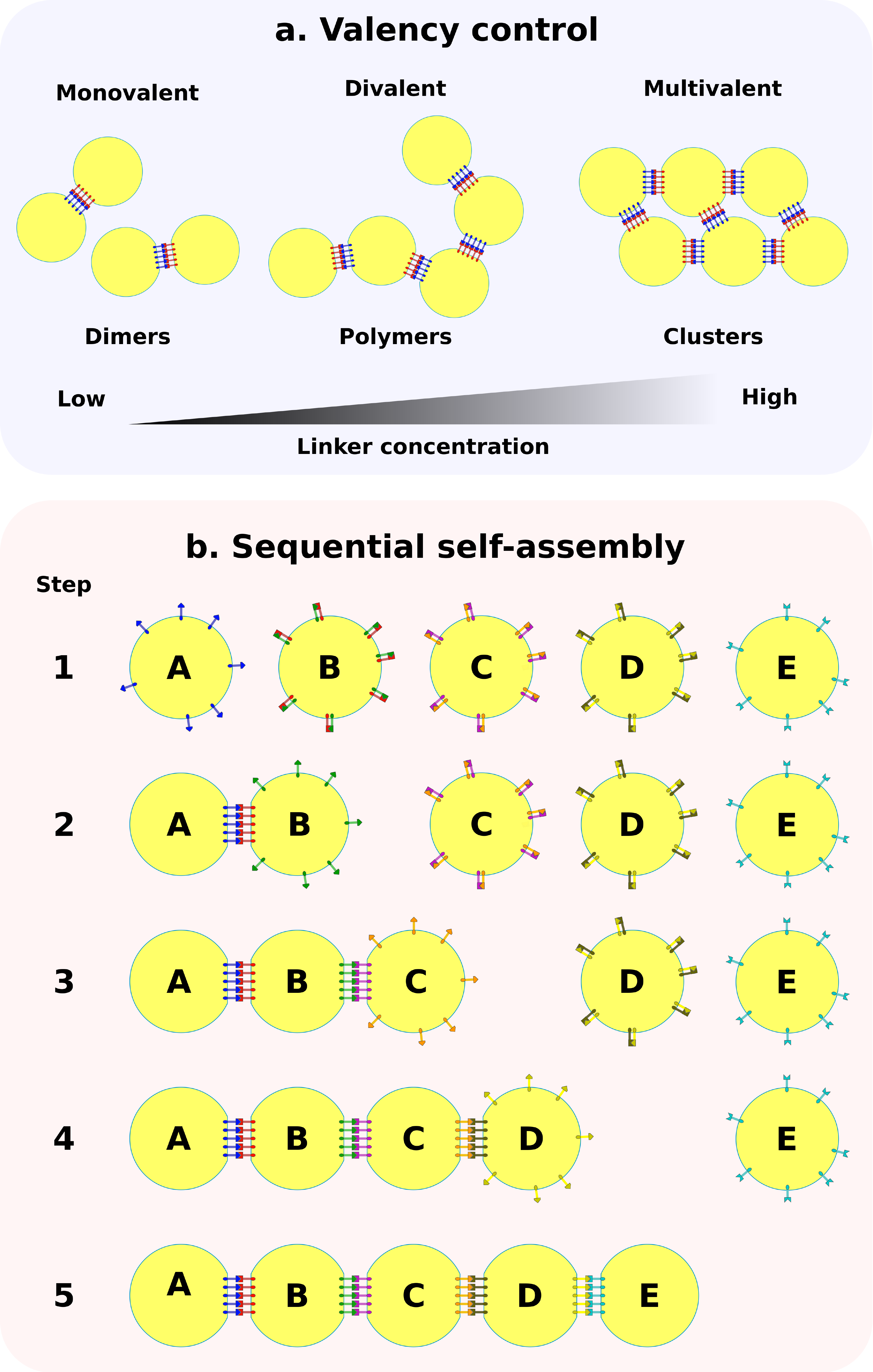}
\caption{\label{fig5} \textbf{Valency control and sequential self-assembly in DNA-functionalised emulsion droplets}. \textbf{a}, Mechanism of valency control in emulsion droplets demonstrated in Ref.~\cite{Jasna_SoftMatter_2013} and subsequently adopted in Refs~\cite{Jasna_PRL_2018} and~\cite{zhang2017sequential}. At low linker surface coverage only one adhesion patch per droplet can be stabilised, and dimers are formed. Upon increasing linker density droplets become divalent, forming polymers, and then multivalent, forming clusters. \textbf{b}, Mechanisms to achieve sequential self-assembly of droplet polymers devised in Ref.~\cite{zhang2017sequential}. Droplets B-D have initially stable loops, and are thus unable to interact with any other element. Through a toeholding reaction~\cite{zhang2009control,srinivas2013biophysics}, linkers on the initiator droplet A can form bridges with one of the linkers on B. The second linker on B is thus freed and becomes available to trigger, again through toeholding, bridge formation with C. The chain reaction progresses until droplet D forms stable bonds with E.}
\end{center}
\end{figure}

\subsection{Self-assembly of emulsion droplets and solid particles with mobile linkers}\label{Sec3EmulsAndSolid}
Owing to their robustness and facile preparation, linker-functionalised emulsion droplets have been extensively employed to investigate the rich phenomenology of multivalent interactions associated with linker mobility and the recruitment of bonds in the CR~\cite{Jasna_SoftMatter_2013,Jasna_PRL_2018,zhang2018multivalent,zhang2017sequential,Hadorn_PNAS_2012,Hadorn_Langmuir_2016}.\\
Feng~\emph{et al.} demonstrated control over the valency of DNA-functionalised oil-in-water droplets, obtained by tuning the surface density of the linkers (Fig.~\ref{fig5}\textbf{a})~\cite{Jasna_SoftMatter_2013}. Upon adhesion of two droplets functionalised with complementary sticky ends, the bridge-forming linkers are found to strongly recruit within the formed CRs, until no free linkers are left in the OR or the CR becomes saturated due to steric repulsion between the bridges. In the latter scenario, enough linkers may be available to form a second droplet-droplet bond, and perhaps a third... until all linkers are confined to CRs and none is available to form further connections. As a result, at low DNA coverage the droplets behave as monovalent and tend to form dimers, while upon increasing  linker density chain-forming divalent droplets and cluster-forming multivalent ones appear~\cite{Jasna_SoftMatter_2013}. The response of the system proposed in Ref.~\cite{Jasna_SoftMatter_2013} arises from a synergistic combination of the many-body and kinetic effects typical of multivalent interactions mediated by mobile linkers. These are extensively reviewed in Secs~\ref{Sec6:Many-body} and~\ref{Sec6:ReactionRates}, respectively. As reviewed in Sec.~\ref{Sec6:CollectiveBehav}, Angioletti-Uberti {\em et al.} used theory and computer simulations to devise a different mechanism leading to valency control, that relies programming equilibrium interactions rather than on exploiting kinetic effects~\cite{A-UbertiPRL2014}. Specifically, the authors exploit a combination of many-body effects (Sec~\ref{Sec6:Many-body}) and the possibility of tuning particle--particle steric repulsion using inert linkers. \\
Recently, McMullen~\emph{et al.} utilised a system similar to that of Ref.~\cite{Jasna_SoftMatter_2013} to study the polymer properties of chains formed by divalent droplets~\cite{Jasna_PRL_2018}. Owing to the mobility of the DNA-linkers and the resulting adhesion patches, the droplet-polymers behave as 2D freely-jointed chains, and their end-to-end distance and diffusivity are well described by Flory and Zimm models, respectively~\cite{Jasna_PRL_2018}.\\
Using a self-protected interaction scheme, Zhang~\emph{et al.} demonstrated sequential activation of droplet interactions and self-assembly (Fig.~\ref{fig5}\textbf{b})~\cite{zhang2017sequential}. Here, linkers on most droplets are initially closed in irreversible loops, similarly to what demonstrated by Parolini \emph{et al.} with LUVs. One ``initiator'' droplet (A in Fig.~\ref{fig5}\textbf{b}) features linkers that can break up the loops on another particle following a toeholding reaction~\cite{zhang2009control,srinivas2013biophysics}. Of the two linkers initially forming the loop, one remains bound into a bridge with the initiator droplet, leading to adhesion, while the latter is freed up and can act as initiator for adhesion with a third droplet~\cite{zhang2017sequential}. The process can then repeat multiple times, leading to the sequential self-assembly of droplet polymers~\cite{zhang2017sequential}. \\
The same collaboration led by Paul Chaikin, later demonstrated a more direct approach to control droplet valency and binding selectivity, where the number and chemistry of the adhesion patches can be precisely prescribed~\cite{zhang2018multivalent}. Rather than relying on simple DNA linkers, here droplets were decorated by large DNA-origami rafts~\cite{Rothemund:2006aa}, anchored to the oil interface by means of multiple biotin-streptavidin bonds. Each ``functional" raft features a number of linkers that can mediate droplet-droplet adhesion, but also a second set of sticky ends with affinity for another family of ``shepherding" rafts, which bridge together the functional rafts forming patches independent of droplet-droplet interactions. Besides controlling the number of patches and thus the valency of the droplets, this technique enables control of the sequence of the linkers on each patch and thus their selective binding to specific patches on different droplets. Exploiting these ``multi-flavoured'' units, the authors could produce complex architectures such as alternated-droplet co-polymers or hereto-tetramers~\cite{zhang2018multivalent}.\\
Multivalent droplets have also been utilised to produce and study three-dimensional tissue like materials~\cite{Hadorn_PNAS_2012,Hadorn_Langmuir_2016,Jasna_PNAS_2012}, \emph{e.g.} to investigate the effect of external applied pressure on the local morphology of the packings~\cite{Jasna_PNAS_2012}.\\
Once again we stress how the functionalities achieved in the systems of Refs~\cite{zhang2017sequential,zhang2018multivalent}, such as the ability to activate linkers at different steps of the aggregation process, rely on the mobility of the linkers. In Sec.~\ref{Sec6:CollectiveBehav}, we present the theoretical design of three different systems \cite{A-UbertiPRL2014,HalversonJCP2016,JanaArxiv2018} that also owe their peculiar properties to the fluidity of the interfaces.\\
Similar to emulsion droplets, solid particles covered by a fluid lipid bilayer provide a robust tool for studying the effects of linker motility, with the additional advantage of near-perfect particle monodispersity and the possibility of utilising particles of different shapes~\cite{VdMeulenJACS2013,Kraft_Nanoscale_2017,RinaldinSoftMatter2018,Wel:2017aa}. In their original contribution van der Meulen and Leunissen demonstrated how linker mobility enables the formation of two-dimensional colloidal crystals, a challenging task to achieve with fixed linkers~\cite{VdMeulenJACS2013}. Indeed, the unimpeded pivoting of colloidal junctions enable the relaxation of the aggregates towards a close-packed hexagonal lattice maximising particle-particle contacts~\cite{VdMeulenJACS2013}.\\
The properties of ``colloidal joints'' formed from bilayer-coated particles, including their range of motion and flexibility, were then systematically characterised by Chakraborty \emph{et al.}~\cite{Kraft_Nanoscale_2017}. By studying the interactions between cubic and spherical particles, the authors also demonstrated the effect of surface curvature. Spherical particles display a preference towards adhering to the flat face of the cube, where a larger CR can be formed. In turn, adhesion to the edge or the vertex of the cube significantly reduces the size of the CR, leading to a stronger confinement of the linkers and thus to a free energy penalty~\cite{Kraft_Nanoscale_2017}.

\subsection{Applications to biological cells}\label{Sec3Cells}
Gartner and coworkers demonstrated the applicability of DNA-mediated multivalent interactions to drive the assembly of biological cell aggregates, known as organoids~\cite{Gartner_NMeth_2015}. The authors could target the adhesion of DNA-functionalised cells to specific areas on a surface decorated by complementary linkers, and then proceed with the synthesis of micro-tissues by depositing further cells in a layer-by-layer fashion~\cite{Gartner_NMeth_2015}. The sequential nature of the developed strategy enabled control over the composition and spatial heterogeneity of the microtissues, as well as their size and shape~\cite{Gartner_NMeth_2015}.\\
Besides artificially prescribing cell-cell interactions, more complex surface-tethered DNA nanostructures, in many cases mimicking the functionalities of cell-membrane receptors, have been developed for gaining insights in various aspects of cell behaviour and biochemistry. For instance, Zhang \emph{et al.} produced DNA linkers capable of probing molecular forces experienced by integrins, mechanotransducing cell-membrane receptors~\cite{Salaita_NatComm_2014,Barczyk:2010aa}. The authors designed DNA linkers that can connect integrins to solid substrates and feature a hairpin motif that can unfold when an above-threshold tension is applied, producing a fluorescence signal in the process. By changing the length and GC content of the hairpin, the authors could tune the activation force of the probe and thus determine the tension experienced by key force-transducing sensor in live cells, as well as the map out a tension distribution by means of optical microscopy~\cite{Salaita_NatComm_2014}.\\
Synthetic DNA constructs have also been applied by You \emph{et al.} to probe the encounter rate of of molecules diffusing on the plasma membrane of live cells~\cite{You:2017aa}. The collision rate of membrane receptors plays a key role in the kinetics of signalling pathways that rely on receptor-receptor complexation, and is influenced by parameters such as lipid composition and poorly understood biophysical phenomena such as the segregation of receptors in specific lipid domains~\cite{You:2017aa,Sezgin:2017aa}. The simple sensor introduced by the authors is based on two membrane-anchored constructs, one carrying a fluorophore and the second a quencher, which bind to each other following a toehold-mediated strand displacement reaction, the rate of which depends on the frequency of the encounters between the two constructs~\cite{zhang2009control,srinivas2013biophysics}. Dimerisation of the constructs produces quenching of the fluorophore, such that the progression of the reaction could be monitored by microscopy or flow cytometry, providing a simple route to estimate the rate of encounter~\cite{You:2017aa}. The authors determined that constructs tethered \emph{via} lipophilic anchors known to partition in the same lipid domains collide more frequently than those partitioning in different domains, and could directly probe the encounter rate of membrane proteins after having labelled them with the constructs \emph{via} DNA aptamers~\cite{You:2017aa}.

\begin{figure}[ht!]
\begin{center}
\includegraphics[width=16.cm]{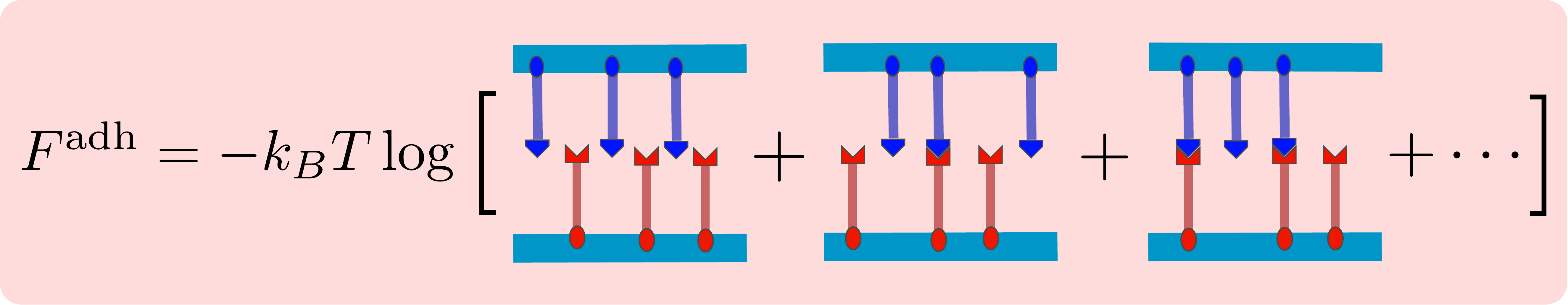}
\caption{{\bf Ensemble averages and adhesion free energy.} The multivalent adhesion free energy between interfaces interacting through complementary linkers can be computed from the ensemble average over all possible microstates featuring a given configuration of bonds and free linkers. \label{FigMod0}}
\end{center}
\end{figure}

\section{Modelling multivalent interactions between fluid interfaces}\label{Sec4}
\begin{figure}[ht!]
\begin{center}
\includegraphics[width=16.cm]{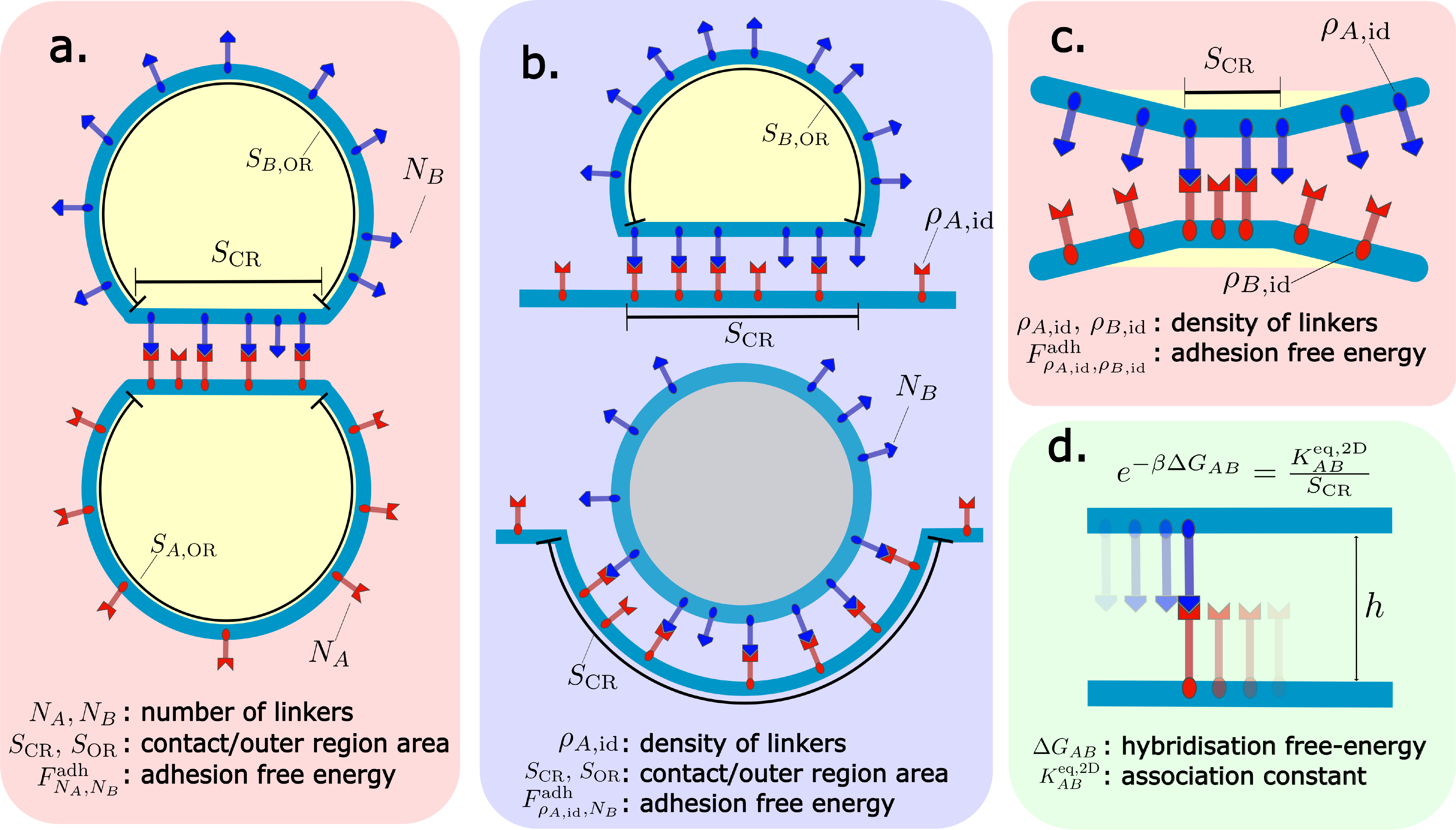}
\caption{
{\bf System parameters controlling the adhesion free energy between interfaces functionalised by complementary linkers.} Interactions between, {\bf a}, two finite-size particles, {\bf b}, a finite particle in contact with an infinite plane, and, {\bf c}, two infinite planes are mediated by the formation of reversible bridges in the contact region (CR). $N_A$ and $N_B$ are the total numbers of linkers {\em per} particle while $\rho_{A,\mathrm{id}}$ is the areal density of linkers bound to the plane. Each interface is partitioned into a contact region (CR) with area $S_\mathrm{CR}$ and an outer region (OR). $S_{A,\mathrm{OR}}$ and  $S_{B,\mathrm{OR}}$ are the surfaces of the ORs of the two particles. The adhesion free energy is not affected by the deformability of the interfaces to the extent that a well defined CR is formed. For that being the case, at least one of the two interfaces should be compliant. The adhesion free energy between bilayer--coated solid particles is given in Sec.~\ref{Sec6:Many-body} (see Fig.~\ref{Fig:Sec6Fig05}{\bf b}). {\bf d}, The reactions between complementary linkers in the CR are controlled by the hybridisation free energy ($\Delta G_{AB}$) or the association constant of the surface-bound linkers ($K^\mathrm{eq,2D}_{AB}$).}\label{Fig:Sec4Fig1}
\end{center}
\end{figure}

Most of the experimental systems presented in Sec.~\ref{Sec3} operate in reversible conditions, in which linkers continuously bind and unbind even after reaching an equilibrium steady state.
An equilibrium statistical-mechanical approach can thus be pursued to theoretically investigate multivalent linker-mediated interactions. Following from the pioneering work of Bell and collaborators~\cite{BellBJourn1984}, this approach aims at deriving \emph{multivalent free energies}, here indicated as $F^\mthr{adh}$, which quantify the strength of the adhesion between interfaces linked by reversible bonds~\cite{BellBJourn1984, CoombsBJourn2004,KitovJACS2003,LicataPRL2008,Martinez-Veracoechea_PNAS_2011,VarillyJCP2012,XuShawBJour2016}. Rather than considering the internal energy of any specific combination of bonds, the statistical approach requires that $F^\mthr{adh}$ is derived through Boltzmann averages over all possible binding configurations, as graphically depicted in Fig.~\ref{FigMod0}. As a result, $F^\mthr{adh}$ explicitly accounts for entropic contributions, including combinatorial terms or avidity, which play an important role in multivalent systems~\cite{MammenAngCIE1998}. A key condition for applying  free energies derived following this approach to the description of self-assembly or biological processes is that the population of linkers must always remain in thermal equilibrium. This requirement is fulfilled if the complexation and decomplexation rates are much faster than any of the processes that involve reconfiguration or diffusion of the substrates hosting the linkers and the linker themselves~\cite{DemboPRSL1998,HAMMER1987475}.
Under these assumptions, the expressions of $F^\mthr{adh}$  are usually tractable, and do not depend on system-specific kinetic parameters.\\
In this section we review recent theoretical advances that provided portable expressions of the multivalent adhesive free energy $F^\mthr{adh}$ for a variety of systems, as summarised in Fig.~\ref{Fig:Sec4Fig1} and discussed for their experimental implementation in Sec~\ref{Sec3}. Previous reviews have covered the mathematical framework supporting multivalent theories~\cite{A-UbertiPCCP2016} and have highlighted some of their most interesting applications, such as the design of selective nano-medical vectors~\cite{Curk7210,A-UbertiNPJ2017,LanfrancoChapter2019}. Differently from earlier contributions, here we specifically address the case of fluid and deformable interfaces functionalised by linkers that can undergo lateral diffusion. In these systems peculiar terms contribute to $F^\mthr{adh}$, such as the entropic costs of confining linker complexes to the contact region between two surfaces.
Moreover, we highlight the challenges of accounting for the deformability of the interfaces, which often requires the development of coarse-grained numerical simulation methods. We attempt to provide a unified and generalised framework, which prevents us from adopting the nomenclature or following the exact mathematical derivation of any specific literature contribution.  However, relevant works from ourselves and others are referred to whenever we make use of concepts and results proposed therein.\\

The remainder of this section is laid out as follows.  In Sec.~\ref{Sec4:MobMob}, we introduce the formalism required to derive $F^\mthr{adh}$, and study the simplest scenario of interfaces functionalised by ``ideal'' mobile linkers, which can only form bridges and do not experience steric interactions. In Sec.~\ref{Sec4:CompCon}, we relax one of the approximations used in Sec.~\ref{Sec4:MobMob}, and study the more realistic scenario in which linkers experience compression from the approaching interface. In Sec.~\ref{Sec4:MobileFixed}, we consider systems in which one of the two interfaces features linkers with fixed tethering points, incapable of undergoing diffusion. In Sec.~\ref{Sec4:ComplexLinkages}, we analyse cases in which linkers can form different types of competing complexes, including loop-bridge systems (Sec.~\ref{Sec4:Loop-bridges}) and systems featuring complexes made by more than two linkers (Sec.~\ref{Sec4:Multimeric}). Sec.~\ref{Sec4:NonSel} considers the effects of non-specific (steric) interactions between linkers. In Sec.~\ref{Sec4:Elastic} the effects of interface elastic deformation are discussed.  Finally, in Sec.~\ref{Sec4:FutureDirections} we highlight what in our opinion are the limitations of the current modelling approaches, which should be addressed in the future. After acquiring the formalism and key results presented in Sec.~\ref{Sec4:MobMob}, the reader should be able to follow each section independently from the others.

\subsection{Ideal mobile linkers}\label{Sec4:MobMob}
\begin{figure}[ht!]
\begin{center}
\includegraphics[width=10.cm,angle=0]{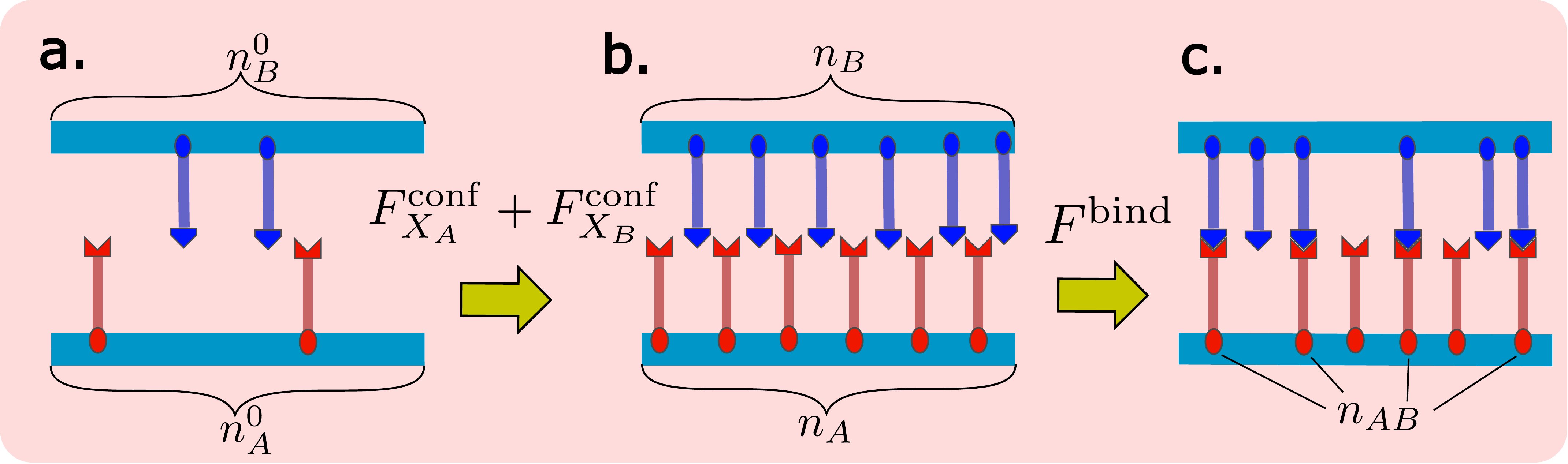}
\caption{  {\bf Two step calculation of the adhesion free energy $F^\mathrm{adh}_{X_A,X_B}$.}  {\bf a}, $F^\mathrm{adh}_{X_A,X_B}$ is calculated starting from a reference configuration in which the interfaces are far from each other with the densities of linkers in the CR ($n^0_A/S_\mathrm{CR}$ and $n^0_B/S_\mathrm{CR}$) equal to the ones in the ORs (see Fig.~\ref{Fig:Sec4Fig1}). {\bf b}, The first two contributions to $F^\mathrm{adh}_{X_A,X_B}$ arise from the entropic costs of confining a given amount of linkers ($n_A$ and $n_B$) in the CR.  Such confining costs are peculiar to systems with mobile linkers and are given by Eqs~\ref{Eq:Sec4:ZconfId}, \ref{Eq:Sec4:FreeSP} in the case of infinite interfaces ($X_A=\rho_{A,\mathrm{id}}$ or $X_B=\rho_{B,\mathrm{id}}$) and Eqs~\ref{Eq:Sec4:ZconfN}, \ref{Eq:Sec4:FreeSP} in the case of finite-size particles ($X_A=N_A$ or $X_B=N_B$). $F^\mathrm{conf}$ also accounts for eventual configurational restrictions experienced by the linkers in the CR (see Sec.~\ref{Sec4:CompCon}) or non-specific linker-linker interactions (see Sec.~\ref{Sec4:NonSel}). {\bf c}, $F^\mathrm{bind}$ (see Eq.~\ref{Eq:Sec4:Fbind}) is the contribution to $F^\mathrm{adh}_{X_A,X_B}$ resulting from the formation of a certain number of bonds from the linkers present in the CR, and accounts for the hybridisation free energy of forming each bond ($\Delta G_{AB}$) and the combinatorial contribution (avidity). In Sec.~\ref{Sec4:ComplexLinkages} we generalise the expression of $F^\mathrm{bind}$ to systems featuring multiple types of bonds (Figs.~\ref{fig3}{\bf a}, \ref{Fig:Sec4Fig3}{\bf a} and Sec.~\ref{Sec4:Loop-bridges}) and bonds made by more than two linkers (Figs.~\ref{fig4}, \ref{Fig:Sec4Fig3}{\bf b} and Sec.~\ref{Sec4:Multimeric}). }\label{Fig:Sec4Fig2}
\end{center}
\end{figure}

In this section we derive the adhesion free-energy $F^\mthr{adh}$ for the case of fluid interfaces decorated by ideal mobile linkers.  We consider the different scenarios depicted in Fig.~\ref{Fig:Sec4Fig1}, in which the two interfaces are both finite-size particles  (Fig.~\ref{Fig:Sec4Fig1}{\bf a}), as derived in Ref.~\cite{parolini2014thermal}, in which one finite particle interacts with an infinite plane (Fig.~\ref{Fig:Sec4Fig1}{\bf b}), as studied in~Ref.~\cite{shimobayashi2015direct}, or in which two infinite planes interact through a finite-size contact region (Fig.~\ref{Fig:Sec4Fig1}{\bf c}). The scheme in Fig.~\ref{Fig:Sec4Fig1}{\bf a} is applicable to the interaction of two deformable objects of similar size, such as liposomes, droplets, or cells, while the  one of Fig.~\ref{Fig:Sec4Fig1}{\bf b} describes well the interaction of a liposome with a large SLB or that of a cell with a small object, such as a nanoparticle or a virus. We consider the geometry of Fig.~\ref{Fig:Sec4Fig1}{\bf c} for completeness, and show that the other two cases converge to this scenario in the limit of weak adhesion. We neglect non-specific interactions, such as steric and electrostatic effects, which are then introduced in Secs~\ref{Sec4:CompCon} and \ref{Sec4:NonSel}.\\
 As shown in Fig.~\ref{Fig:Sec4Fig1}, each interacting particle or plane is functionalised by one of two types of complementary linkers, labelled as $A$ and $B$, so that only bridges can be formed. Since each interface only hosts one type of linker, in this section labels $A$ and $B$ identify one of the two interfaces univocally. This is no longer true in Sec.~\ref{Sec4:ComplexLinkages}, where we consider the case of interfaces decorated by multiple types of linkers. We divide each interface into a contact region (CR) in which proximity to the other interface enables the formation of bridges, and an outer region (OR), where bridging is not permitted. We define the area of the contact region as $S_\mthr{CR}$ and, for finite-size particles, those of the ORs as $S_{A,\mthr{OR}}$ and $S_{B,\mthr{OR}}$. Accordingly, the total areas of the particles are defined as $S_{A,\mthr{Tot}}=S_{A,\mthr{OR}}+S_\mthr{CR}$ and $S_{B,\mthr{Tot}}=S_{B,\mthr{OR}}+S_\mthr{CR}$. Finite-size particles carry a fixed number of linkers $N_A$ or $N_B$. For the case of infinite planes, we mimic the infinite OR using a grand-canonical ensemble, so that an arbitrary number of linkers can be recruited in the CR without reducing their concentration in the OR~\cite{DiMichelePRE2018}. With this approach, the number of linkers $A$ in the CR is modulated by a chemical potential $\mu_A$ that, in the ideal case, is related to the surface density of linkers in the OR reservoir by $\rho_{A,\mthr{id}} \propto \exp [-\beta \mu_A]$. In the non-ideal case, if non-specific linker-linker interactions are present, $\rho_{A,\mthr{id}}$ does not correspond to the real density of linkers in the OR~\cite{DiMichelePRE2018}. Similar definitions apply to type-$B$ linkers.\\
In this section, we assume that the distance $h$ between the interfaces in the CR is sufficiently large that the configurational space available to unbound linkers in the CR and the OR are the same. The reliability of this hypothesis was tested in Ref.~\cite{parolini2014thermal}. We study the general case in Sec.~\ref{Sec4:CompCon}, where we clarify how compression of the linkers in the CR produces repulsive steric interactions between the interfaces~\cite{melting-theory1,melting-theory2,LeunissenJCP2011}.\\
To distinguish the different scenarios under consideration, we label the adhesive free energy as $F^\mthr{adh}_{X_A,X_B}$, where labels $X_A$ and $X_B$ are equal to $N_A$ / $N_B$ for finite particles, and $\rho_{A,\mthr{id}}$ / $\rho_{B,\mthr{id}}$ for infinite interfaces (Fig.~\ref{Fig:Sec4Fig1}). For example, the free energy of interaction between two finite-size particles is  $F^\mthr{adh}_{N_A,N_B}$ (Fig.~\ref{Fig:Sec4Fig1}\textbf{a}), while that between an $B$-type particle and a $A$-type infinite plane is $F^\mthr{adh}_{\rho_{A,\mthr{id},N_B}}$ (Fig.~\ref{Fig:Sec4Fig1}\textbf{b}).\\
In general, $F^\mthr{adh}_{X_A,X_B}$ can be expressed as
\begin{equation}\label{Eq:Sec4:FreeDef}
F^\mthr{adh}_{X_A,X_B} = -k_B T  \log \sum_{n_A,n_B,n_{AB}} Z^\mthr{adh}_{X_A,X_B}(n_A, n_B, n_{AB}),
\end{equation}
where the partition function $Z^\mthr{adh}_{X_A,X_B}(n_A, n_B, n_{AB})$ describes all microstates featuring $n_{AB}$ bonds and a total number of $A$ and $B$ linkers in the CR equal to $n_A$ and $n_B$, and is then summed up over all possible values of  $n_A$, $n_B$ and $n_{AB}$.\\
We calculate $Z^\mthr{adh}_{X_A,X_B}(n_A, n_B, n_{AB})$  following the
two--step pathway depicted in Fig.~\ref{Fig:Sec4Fig2}. The contribution associated to the first step describes the relative statistical weight of confining $n_A$ and $n_B$ linkers in the CR, as compared to a reference state in which the interfaces are non-interacting (Fig.~\ref{Fig:Sec4Fig2}\textbf{a}-\textbf{b}). For ideal linkers, this term is the product of two partition functions $ Z^\mthr{conf}_{X_A}(n_A) \cdot Z^\mthr{conf}_{X_B}(n_B)$. The general case of non-ideal linkers is covered in Sec.~\ref{Sec4:NonSel} and Ref.~\cite{DiMichelePRE2018}. The second contribution, $Z^\mthr{bind}(n_A,n_B,n_{AB})$, describes the probability of forming $n_{AB}$ bonds starting from the $n_A$ and $n_B$ linkers present in the CR (Fig.~\ref{Fig:Sec4Fig2}\textbf{b}-\textbf{c}).
Note that from the configuration of Fig.~\ref{Fig:Sec4Fig2}\textbf{b} to that of Fig.~\ref{Fig:Sec4Fig2}\textbf{c} the total numbers of $A$ and $B$ linkers in the CR remain constant.\\
We can thus extract $Z^\mthr{adh}_{X_A,X_B}(n_A, n_B, n_{AB})$ as
\begin{equation}
Z^\mthr{adh}_{X_A,X_B}(n_A, n_B, n_{AB}) = Z^\mthr{conf}_{X_A}(n_A)\,Z^\mthr{conf}_{X_B}(n_B)\,Z^\mthr{bind}(n_A, n_B, n_{AB}),
\end{equation}
which combined with Eq.~\ref{Eq:Sec4:FreeDef} renders
\begin{eqnarray}
&&F^\mthr{adh}_{X_A,X_B} = \nonumber \\
    && = -  k_B T \log \left[\sum_{n_A,n_B} Z^\mthr{conf}_{X_A}(n_A) Z^\mthr{conf}_{X_B}(n_B) \sum_{n_{AB}} Z^\mthr{bind}(n_A, n_B, n_{AB})
\right]. \label{Eq:Sec4:Free}
\end{eqnarray}
For the case of finite particles, the sums over $n_A$ and $n_B$ in Eqs~\ref{Eq:Sec4:FreeDef} and \ref{Eq:Sec4:Free} are limited by $N_A$ and $N_B$. For infinite planes the number of linkers in the CR is not restricted, so these sums proceed to infinity. In all cases  $n_{AB}$ is limited by the minimum value between $n_A$ and $n_B$.\\
Boltzmann weighting guarantees that the sums in Eqs~\ref{Eq:Sec4:FreeDef} and \ref{Eq:Sec4:Free} are dominated by the terms corresponding to the most probable choices of $n_A$, $n_B$ and $n_{AB}$, labelled as $\on_A$, $\on_B$ and $\on_{AB}$. A saddle--point approximation thus allows us to neglect all but these dominant terms, obtaining $F^\mthr{adh}_{X_A,X_B} \approx -k_B T \log  Z^\mthr{adh}_{X_A,X_B}(\on_A, \on_B, \on_{AB})$. Eq.~\ref{Eq:Sec4:Free} then simplifies as follows
\begin{eqnarray}
F^\mthr{adh}_{X_A,X_B}
= F^\mthr{conf}_{X_A}(\on_A)+ F^\mthr{conf}_{X_B}(\on_B) +   F^\mthr{bind}(\on_A,\on_B,\on_{AB})&&
\label{Eq:Sec4:FreeSP}
\\
F^\mthr{conf}_{X}(\on) = -k_B T \log Z^\mthr{conf}_X(\on) &&
\nonumber \\
F^\mthr{bind}(\on_A,\on_B,\on_{AB}) = -k_B T \log Z^\mthr{bind}(\on_A,\on_B,\on_{AB}) &&
\nonumber
\end{eqnarray}
where in the second equation $X=X_A$ or $X=X_B$ and, accordingly, $\on=\on_A$ or $\on=\on_B$. We calculate $\on_A$, $\on_B$, and $\on_{AB}$ by determining the stationary point of $Z^\mthr{adh}_{X_A,X_B}$, satisfying the following set of coupled equations
\begin{eqnarray}
{\mathrm{d}  Z^\mthr{adh}_{X_A,X_B}\over \mathrm{d} n_A}\Big{|}_{\{\on_A, \on_B, \on_{AB}\}} = {\mathrm{d}  Z^\mthr{adh}_{X_A,X_B} \over \mathrm{d} n_B}\Big{|}_{\{\on_A, \on_B, \on_{AB}\}}={\mathrm{d}  Z^\mthr{adh}_{X_A,X_B} \over \mathrm{d} n_{AB}}\Big{|}_{\{\on_A, \on_B, \on_{AB}\}}=0 \, .
\nonumber \\
\label{Eq:Sec4:ChemEq}
\end{eqnarray}
The strategy highlighted in Fig.~\ref{Fig:Sec4Fig2} of decoupling configurational from binding contributions to $F^\mathrm{adh}$ provides a modular and versatile route to describe a variety different systems, as shown in this and in the following sections. In the remainder of this section we first discuss the derivation of $F^\mthr{conf}$ and  $F^\mthr{bind}$ and then provide expressions of $F^\mthr{adh}$ for the systems of Fig.~\ref{Fig:Sec4Fig2} (Eqs~\ref{Eq:Sec4:FNN}, \ref{Eq:Sec4:FrhoN}, and \ref{Eq:Sec4:Frhorho}). To enable the derivation of $F^\mthr{bind}$, we will also review the calculation of the  hybridisation free energy of linkers with mobile tethering points.
\\

{\bf Free energy of confining linkers in the CR.} Here we derive expressions for $Z^\mthr{conf}_{X_A} (n_A)$, and thus $F^\mthr{conf}_{X_A}(\on_A)$. Similar definitions follow for $X_B$. $Z^\mthr{conf}_{\rho_{A,\mthr{id}}}$ and $Z^\mthr{conf}_{N_A}$ are, respectively, a Poissonian and a binomial distribution that read as follows
\begin{eqnarray}
Z^\mthr{conf}_{\rho_{A,\mthr{id}}}(n_A) &=& {1\over Z^\mthr{ref}_{\rho_{A,\mthr{id}}}}{\left( \rho_{A,\mthr{id}} S_\mthr{CR}\right)^{n_A}\over n_A! } = {\exp[-\rho_{A,\mthr{id}} S_\mthr{CR}]\over n_A!} \left( \rho_{A,\mthr{id}} S_\mthr{CR}\right)^{n_A}
\label{Eq:Sec4:ZconfId}
\\
Z^\mthr{conf}_{N_A} (n_A) &=&{1\over Z^\mthr{ref}_{N_A} N_A!} {N_A\choose n_A}\left( S_\mthr{CR} \right)^{n_A} \left( S_{A,\mthr{OR}} \right)^{N_A-n_A}
\nonumber \\
&=& {N_A\choose n_A} \left( S_\mthr{CR} \over S_{A,\mthr{Tot}} \right)^{n_A} \left( S_{A,\mthr{OR}} \over  S_{A,\mthr{Tot}}\right)^{N_A-n_A} \, .
\label{Eq:Sec4:ZconfN}
\end{eqnarray}
In Eqs~\ref{Eq:Sec4:ZconfId} and~\ref{Eq:Sec4:ZconfN}, $Z^\mthr{ref}_{\rho_{A,\mthr{id}}}$ and $Z^\mthr{ref}_{N_A}$ are the reference-state partition functions associated to isolated planes and finite particles, respectively, and are derived below (Eqs~\ref{Eq:Sec4:ZrefPart} and~\ref{Eq:Sec4:ZrefPlane}).
It is straightforward  to show that in the limit of very large $N_A$, $S_{A,\mthr{Tot}}$, and $S_{A,\mthr{OR}}$, with $N/S_{A,\mthr{OR}}\to \rho_{A,\mthr{id}}$ and $S_{A,\mthr{Tot}}/S_{A,\mthr{OR}}\to 1$, $Z^\mthr{conf}_{N_A} (n_A)$ converges towards $ Z^\mthr{conf}_{\rho_{A,\mthr{id}}}(n_A)$. $F^\mthr{conf}_{X_A}(\on_A)$ follows from the combination of Eq~\ref{Eq:Sec4:FreeSP} with Eqs~\ref{Eq:Sec4:ZconfId} and~\ref{Eq:Sec4:ZconfN}:
\begin{eqnarray}
{F^\mthr{conf}_{\rho_{A,\mthr{id}}} \over k_B T}&=& \on_A \log {\on_A \over S_\mthr{CR} \rho_{A,\mthr{id}} } - \on_A + S_\mthr{CR} \rho_{A,\mthr{id}},
\label{Eq:Sec4:FconfId}
\\
{F^\mthr{conf}_{N_A} \over k_B T} &=& \on_A \log {\on_A \over N_A } + (N_A-\on_A) \log\left(1-{\on_A\over N_A} \right) - \on_A \log {S_\mthr{CR} \over S_{A,\mthr{Tot}}}+
\nonumber \\
&& -(N-\on_A) \log {S_{A,\mthr{OR}} \over S_{A,\mthr{Tot}}},
\label{Eq:Sec4:FconfN}
\end{eqnarray}
where Eq.~\ref{Eq:Sec4:FconfId} describes the case of infinite interfaces and Eq.~\ref{Eq:Sec4:FconfN} that of finite particles. Note that explicit expressions for $F^\mthr{conf}_{X_A}$ require  those for the most probable number of linkers in the CR, $\on_{A}$, which can only be calculated after deriving $Z^\mthr{bind}$ as shown below.\\

{\bf Reference state.}
 In the reference state, the densities of linkers in the CR and the OR are equal, $n^0_A/S_\mthr{CR}=(N_A- n^0_A)/S_{A,\mthr{OR}}$ if $X_A=N_A$ and $n^0_A/S_\mthr{CR}=\rho_{A,\mthr{id}}$ if $X_A=\rho_{A,\mthr{id}}$, where  $n^0_A$ and $n^0_B$ are the number of $A$ and $B$ linkers in the CR and no bonds are formed. The partition functions of the interfaces in the reference state can then be written as
\begin{eqnarray}
Z^\mthr{ref}_{N_A} &=&{1\over N_A!} (S_{A,\mthr{Tot}})^{N_A}, \label{Eq:Sec4:ZrefPart}\\
 Z^\mthr{ref}_{\rho_{A,\mthr{id}}} &=& \sum_{n=0}^\infty {1\over n!} \left( \rho_{A,\mthr{id}} S_\mthr{CR} \right)^{n} = \exp[\rho_{A,\mthr{id}} S_\mthr{CR}].\label{Eq:Sec4:ZrefPlane}
\end{eqnarray}
Similar definitions apply for $Z^\mthr{ref}_{X_B}$.\\

{\bf Complexation free energy of individual tethered linkers.}
As the starting point for deriving the linker-binding contributions to the partition function and the adhesive free energy, namely $Z^\mthr{bind}$ and $F^\mthr{bind}$, we quantify the complexation free energy $\Delta G_{AB}(h)$ of a pair of $A$ and $B$ mobile linkers confined to the CR, with interfaces placed at a distance $h$.
$\Delta G_{AB}(h)$ can be divided in two contributions
\begin{equation}\label{Sec3EqDeltaG}
\Delta G_{AB}(h) = \Delta G^0_{AB} + \Delta G^\mthr{conf}_{AB}(h).
\end{equation}
In Eq.~\ref{Sec3EqDeltaG}, $\Delta G^0_{AB}$  is the complexation free energy of the linkers free in solution, which depends on their chemical details and is usually determined experimentally or \emph{via} atomistic simulations. When considering DNA oligomers $\Delta G^0_{AB}$ is calculated using the nearest-neighbour thermodynamic model~\cite{santalucia,santalucia2004thermodynamics,zadeh2011nupack,markham2005dinamelt,WangZhangNatComm2016}, possibly corrected for the presence of dangling DNA tails~\cite{DiMicheleJACS2014,WangZhangNatComm2016}.\\
$\Delta G^\mthr{conf}_{AB}(h)$ describes the additional constraints that surface-tethered linkers experience whenever they form a complex, as compared to the same complex formed by untethered linkers~\cite{melting-theory1,melting-theory2}. First, we derive $\Delta G^\mthr{conf}_{AB}$ for the case of linkers with fixed tethering points placed at a relative distance ${\bf r}$, a scenario in which $\Delta G^\mthr{conf}_{AB}\equiv\Delta G^\mthr{conf}_{AB}(h,{\bf r})$. For the simple case of rigid linkers of length $L$ and point-like reactive regions, free to pivot around their tethering point, $\Delta G^\mthr{conf}_{AB}({\bf r},h)$  can be computed analytically. This simple geometry is often adopted to describe typical DNA linkers, with rigid dsDNA spacers of length $L$ and comparatively shorter sticky ends (see Sec.~\ref{Sec2Linkers} and Fig.~\ref{fig1}\textbf{a}, top). Prior to bond formation, the point-like sticky ends can describe the surfaces of two hemispheres, while after bridging they are confined to a circle or an arc~\cite{melting-theory1,melting-theory2,LeunissenJCP2011}. The free energy term encoding for this reduction in configurational freedom reads as~\cite{MognettiSoftMatt2012}
\begin{eqnarray}
\Delta G^\mthr{conf}_{AB}({\bf r},h) =  -k_\mathrm{B}T \log \left({1-\alpha({\bf r},h) \over 2 \pi \rho_0 L^2 |{\bf r}|}\right),
\label{Eq:DGABrh}
\end{eqnarray}
where $\rho_0$ is the reference concentration used to define $\Delta G^0_{AB}$, \emph{e.g.} $\rho_0=1\,$M~\cite{santalucia, santalucia2004thermodynamics,zadeh2011nupack,markham2005dinamelt}, and $\alpha({\bf r},h)$ is the fraction of configurational space available to paired linkers that is excluded by the interfaces ($0\leq \alpha \leq 1$). Ref.~\cite{MognettiSoftMatt2012} reports the general expression of $\alpha({\bf r},h)$, also considering cases in which the two linkers have different lengths. For non-ideal and flexible linkers, $\Delta G^\mthr{conf}_{AB}({\bf r},h)$ is usually calculated using Monte Carlo methods~\cite{MladekSoftMatter2013, MognettiPNAS2012,VarillyJCP2012,DeGernierJCP2014}.\\
To derive an expression for $\Delta G_{AB}^\mathrm{conf}(h)$ valid for the case in which linkers $A$ and/or $B$ are mobile, the corresponding fixed-linker expression (\emph{e.g.} Eq.~\ref{Eq:DGABrh}) must be averaged to sample all the possible configurations of the tethering points
\begin{eqnarray}
\Delta G_{AB}^\mathrm{conf}(h) = -k_\mathrm{B}T \log \left({1\over S_\mthr{CR}} \int \mthr{d}^2 {\bf r}_{||} \exp[-\beta\Delta G_{AB}^\mathrm{conf}({\bf r},h)]\right)\, ,
\label{Eq:DGABh}
\end{eqnarray}
where ${\bf r}_{||}$ is the distance between the tethering points of $A$ and $B$ projected onto the CR plane.\\
In the simple scenario of mobile rigid linkers of length $L$ and inter-plane distance $h=L$, by combining Eqs~\ref{Eq:DGABrh} and \ref{Eq:DGABh} we find $\Delta G^\mthr{conf}_{AB} = - k_\mathrm{B}T \log \left(\rho_0 L S_\mthr{CR}\right)$, which using Eq.~\ref{Sec3EqDeltaG} gives~\cite{parolini2014thermal}.
\begin{equation}\label{Sec3EqDeltaGNatComm}
\Delta G_{AB}=\Delta G^0_{AB} - k_\mathrm{B}T \log \left(\rho_0 L S_\mthr{CR}\right).
\end{equation}
Ref.~\cite{parolini2014thermal} also reports the general expressions of $\Delta G_{AB}(h)$ for mobile rigid linkers as a function of $h$.\\
In general, it is convenient to describe the stability of $AB$ bonds in terms of  an equilibrium constant, defined as the ratio of the areal concentrations of the dimerised and free linkers and related to $\Delta G_{AB}(h)$ by~\cite{DiMichelePRE2018}:
\begin{eqnarray}
 K^\mthr{eq,2D}_{AB} = {\on_{AB} S_\mthr{CR} \over (\on_A-\on_{AB}) (\on_B-\on_{AB})} = S_\mthr{CR} \cdot \exp[-\beta \Delta G_{AB}(h)] \, .
 \label{Eq:Sec4:K2D}
 \end{eqnarray}
For the case of rigid linkers of length $L$, with $h=L$, combining Eqs \ref{Sec3EqDeltaGNatComm} and \ref{Eq:Sec4:K2D} produces $K^\mthr{eq,2D}_{AB} = \exp[-\beta \Delta G^0_{AB}] / (\rho_0 L)$, offering a remarkably simple relation between the 2D equilibrium constant and that of free linkers in solution $K^\mthr{eq,3D}_{AB}$: $K^\mthr{eq,2D}_{AB} = K^\mthr{eq,3D}_{AB}/L$. This result, which is peculiar of rigid linkers, follows from the fact that the reactive spots (\emph{e.g.} the sticky ends) tipping rigid linkers freely diffusing on a flat surface are uniformly distributed within the layer of thickness $L$, resulting in local bulk concentrations of paired and unpaired reactive spots equal to $\rho_{A}=\on_A/(L \, S_\mthr{CR})$,  $\rho_{B}=\on_B/(L \, S_\mthr{CR})$, and $\rho_{AB}=\on_{AB}/(L \, S_\mthr{CR})$. The chemical-equilibrium definition of $K^\mthr{eq,3D}_{AB}$ then proves the aforementioned relation with $K^\mthr{eq,3D}_{AB}$
\begin{eqnarray}
K^\mthr{eq,3D}_{AB} = {\rho_{AB} \over \rho_A \rho_B} = {\on_{AB} S_\mthr{CR} L \over (\on_A-\on_{AB}) (\on_B-\on_{AB}) } = K^\mthr{eq,2D}_{AB} L.
\end{eqnarray}
The analogy between the formation of inter--particle bonds and bulk chemical reactions has been previously evoked to motivate the use of local chemical equilibrium approaches to calculate interactions between DNA functionalised particles~\cite{RogersPNAS2011,MognettiPNAS2012,RogersAnswer,VarillyJCP2012}.
\\
Referring to equilibrium constants instead of free energies enables the direct comparison between theory and experiments measuring the fractions of formed complexes. Moreover, the use of experimentally-derived equilibrium constant can compensate for relevant factors neglected by Eq.~\ref{Eq:DGABrh}. For instance, Refs~\cite{hu2013binding,XuWeiklJCP2015} have shown how to correct the definition of $K^\mthr{eq,2D}_{AB}$ to effectively account for the roughness of the bilayers supporting the linkers.\\

 {\bf Free energy of forming $n_{AB}$ bonds.}
 We now consider the configuration depicted in Fig.~\ref{Fig:Sec4Fig2}\textbf{b}, featuring $n_A$ and $n_B$ mobile linkers confined to the CR, and calculate the partition function $Z^\mthr{bind}(n_A,n_B,n_{AB})$ describing the statistical weight of forming $n_{AB}$ bonds (Eq.~\ref{Eq:Sec4:Free}). The binding contribution to the adhesion free energy $F^\mthr{bind}(\on_A,\on_B,\on_{AB})$ will be then derived through the saddle-point approximation (Eq.~\ref{Eq:Sec4:FreeSP}). $Z^\mthr{bind}(n_A,n_B,n_{AB})$ reads as follows
 \begin{eqnarray}
 Z^\mthr{bind}(n_A,n_B,n_{AB}) &=& {n_A \choose n_{AB}} {n_B \choose n_{AB}} n_{AB}!  \left(  K^\mthr{eq,2D}_{AB} \over S_\mthr{CR} \right)^{n_{AB}}
 \nonumber \\
 &=&  {n_A! n_B!  \over n_{AB}! (n_A-n_{AB})! (n_B- n_{AB})!}  \left(  K^\mthr{eq,2D}_{AB} \over S_\mthr{CR} \right)^{n_{AB}},
 \label{Eq:Sec4:Zbind}
 \end{eqnarray}
 The first term on the right-hand side of Eq.~\ref{Eq:Sec4:Zbind} enumerates the ways in which $n_{AB}$ bonds can be formed starting from $n_A$ and $n_B$ linkers~\cite{Martinez-Veracoechea_PNAS_2011}, while the second term is the Boltzmann weight accounting for the complexation free energy of $n_{AB}$ bonds, expressed in terms of $K^\mthr{eq,2D}_{AB}$ using Eq.~\ref{Eq:Sec4:K2D}. \\
Equation~\ref{Eq:Sec4:Zbind}, along with the saddle-point approximation in Eq.~\ref{Eq:Sec4:ChemEq}, then allows us to compute $F^\mthr{bind}(\on_A, \on_B, \on_{AB})$ as
\begin{eqnarray}
{F^\mthr{bind}(\on_A, \on_B, \on_{AB})\over k_B T} &=& \on_A \log {\on_A - \on_{AB} \over \on_A} + \on_B \log{\on_B - \on_{AB} \over \on_B} +  \on_{AB}  \, .
\label{Eq:Sec4:Fbind}
\end{eqnarray}
Note that, similarly to the case of $F^\mthr{conf}$, also the explicit expression for $F^\mthr{bind}$ depends on the most probable number of linkers in the CR ($\on_{A}$ and $\on_{B}$), as well as on the most probable number of bonds $\on_{AB}$, which are derived in the following paragraph. Angioletti-Uberti~{\em et al.}~\cite{A-UbertiJCP2013} first derived Eq.~\ref{Eq:Sec4:Fbind} and showed how it can be immediately generalised to systems featuring different types of bonds (\emph{e.g.}~those studied in Sec.~\ref{Sec4:Loop-bridges}). Equation~\ref{Eq:Sec4:Fbind} is applicable to systems with mobile linkers~\cite{VarillyJCP2012,A-UbertiJCP2013} and systems with fixed tethering point at a sufficiently high coating density~\cite{VarillyJCP2012,TitoJCP2016}.\\

{\bf Most probable number of linkers in the CR and bonds formed}.
Here we derive the most probable numbers $\on_{A}$ and $\on_B$ of  $A$ and $B$ linkers present in the CR, as well as that of the formed $AB$ bonds ($\on_{AB}$). This is done by combining the saddle-point approximation in Eq.~\ref{Eq:Sec4:FreeSP} with the associated equilibrium conditions described in Eq.~\ref{Eq:Sec4:ChemEq} and the expressions for $Z^\mthr{conf}_{\rho_{A,\mthr{id}}}$, $Z^\mthr{conf}_{N_A}$, and $Z^\mthr{bind}$ (Eqs~\ref{Eq:Sec4:ZconfId}, \ref{Eq:Sec4:ZconfN} and~\ref{Eq:Sec4:Zbind}, respectively).\\
By combining Eq.~\ref{Eq:Sec4:ZconfN} with the first equality in Eqs~\ref{Eq:Sec4:ChemEq} we obtain the following expression associating $\on_A$ to $\on_{AB}$ for the case of finite particles
\begin{eqnarray}
{\mthr{d} Z^\mthr{adh}_{N_A,X_B } \over \mthr{d} n_A}\Big{|}_{\{\on_A, \on_B, \on_{AB}\}}=0 & \quad \leftrightarrow \quad & {\on_A-\on_{AB} \over N_A-\on_A}={S_\mthr{CR} \over S_{\mthr{OR},A}}.
\label{Eq:Sec4:onA-N}
\end{eqnarray}
Likewise, the use of Eq.~\ref{Eq:Sec4:ZconfId} leads to an analogous expression for the case of infinite interfaces
\begin{eqnarray}
{\mthr{d} Z^\mthr{adh}_{\rho_{A,\mthr{id}},X_B } \over \mthr{d} n_A}\Big{|}_{\{\on_A, \on_B, \on_{AB}\}}=0 & \quad \leftrightarrow \quad & \on_A=\on_{AB}+\rho_{A,\mthr{id}} S_\mthr{CR}
\label{Eq:Sec4:onA-rho}.
\end{eqnarray}
Identical expressions follow for $\on_B$. Note that Eqs~\ref{Eq:Sec4:onA-rho} and~\ref{Eq:Sec4:onA-N} state that at equilibrium the density of unbound linkers in the CR (equal to $(\on_A-\on_{AB})/S_\mthr{CR}$) is identical to that in the OR. This condition is only verified under the assumption of ideal linkers relevant to this section, \emph{i.e.} if the configurational freedom of linkers in the CR is not limited by excluded-volume interactions with other linkers, discussed in Sec.~\ref{Sec4:NonSel}, or the facing surface, discussed in Sec.~\ref{Sec4:CompCon}.\\
The combination of Eq.~\ref{Eq:Sec4:Zbind} with the last equality in  Eqs~\ref{Eq:Sec4:ChemEq} leads to another expression associating $\on_A$ to $\on_{AB}$, valid for both finite particles and infinite planes, and independent from those in Eq.~\ref{Eq:Sec4:onA-N} and~\ref{Eq:Sec4:onA-rho}
\begin{eqnarray}
{\mthr{d} Z^\mthr{adh}_{X_A,X_B} \over \mthr{d} n_A}\Big{|}_{\{\on_A, \on_B, \on_{AB}\}}=0 & \quad \leftrightarrow \quad & { \on_{AB} \over (\on_A - \on_{AB})(\on_B-\on_{AB}) } = { K^\mthr{eq,2D}_{AB} \over S_\mathrm{CR}}.
\label{Eq:Sec4:onAB}
\end{eqnarray}
The final expressions for $\on_A$, $\on_B$, and $\on_{AB}$ are then by obtained combining Eq.~\ref{Eq:Sec4:onAB} with the two conditions imposed by Eq.~\ref{Eq:Sec4:onA-N} and/or Eq.~\ref{Eq:Sec4:onA-rho} for both $A$ and $B$ linkers.  These enable explicit analytical expressions for $F^\mthr{bind}$, $F^\mthr{conf}$, and thus $F^\mthr{adh}$ as described in the following paragraph.\\

 {\bf Adhesion free energies.} We are now in the condition of calculating the adhesion free energies $F^\mthr{adh}_{X_A,X_B}$ for the different scenarios considered in Fig.~\ref{Fig:Sec4Fig1}.\\
 For systems in which two finite-size particles interact, as shown in Fig.~\ref{Fig:Sec4Fig1}\textbf{a}, $F^\mthr{adh}_{N_A,N_B}$ is calculated by combining the binding free energy $F^\mthr{bind}$ (Eq.~\ref{Eq:Sec4:Fbind}) with the two confining terms for the $A$ and $B$ linkers (Eq.~\ref{Eq:Sec4:FconfN}). Using Eqs~\ref{Eq:Sec4:onA-N} and~\ref{Eq:Sec4:onAB} we can simplify the final expression as follows
 \begin{eqnarray}
 \boxed{
{ F^\mthr{adh}_{N_A,N_B} \over k_B T } =  N_A \log { N_A - \on_{AB}  \over N_A }
 +  N_B \log {N_B - \on_{AB} \over N_B  } + \on_{AB}  \, .
 }
 \label{Eq:Sec4:FNN}
 \end{eqnarray}
 Not surprisingly, $F^\mthr{adh}_{N_A,N_B}$ has the same functional form of $F^\mthr{bind}$.
 This result follows from Ref.~\cite{A-UbertiJCP2013} stating that, in general, adhesive free energies in multivalent systems are proportional to the total number of bonds ($\on_{AB}$ in Eq.~\ref{Eq:Sec4:FNN}), augmented,  for each type of linker, by contributions proportional to the logarithm of the fraction of free linkers ($(N_A-\on_{AB})/N_A$ and $(N_B-\on_{AB})/N_B$ in Eq.~\ref{Eq:Sec4:FNN}). $F^\mthr{adh}_{N_A,N_B}$ has been used in Ref.~\cite{parolini2014thermal} to study system of interacting GUV (see Sec.~\ref{Sec3Eq}).
 \\
 We now discuss the adhesion free energy of a finite-size particle in contact with an infinite interface, relevant for the case of a liposomes interacting with a much larger SLB, or of a nanoparticle interacting with a cell (Fig.~\ref{Fig:Sec4Fig1}\textbf{b}). In this case the binding free energy (Eq.~\ref{Eq:Sec4:Fbind}) is augmented by $F^\mthr{conf}_{\rho_{A,\mthr{id}}}$ (Eq.~\ref{Eq:Sec4:FconfId}) and $F^\mthr{conf}_{N_B}$ (Eq.~\ref{Eq:Sec4:FconfN}). Using Eqs~\ref{Eq:Sec4:onA-N}, \ref{Eq:Sec4:onA-rho}, and~\ref{Eq:Sec4:onAB} the final expression becomes
\begin{eqnarray}
\boxed{
{F^\mthr{adh}_{\rho_{A,\mthr{id}},N_B} \over
k_B T}= N_B \log {N_B - \on_{AB} \over N_B  } =-N_B \log\left( 1+ {S_\mthr{CR} \rho_{A,\mthr{id}} K^\mthr{eq,2D}_{AB}\over S_{B,\mthr{Tot}} } \right)\, .
}
\label{Eq:Sec4:FrhoN}
\end{eqnarray}
 Equation~\ref{Eq:Sec4:FrhoN} has been used in Ref.~\cite{shimobayashi2015direct} to study systems of GUVs interacting with SLBs (see Sec.~\ref{Sec3Eq}). Similar expressions were derived in Refs~\cite{Curk7210,PhysRevE.96.012408}.\\
 Finally, it is instructive to consider the adhesion between two infinite interfaces interacting through a finite-size CR (Fig.~\ref{Fig:Sec4Fig1}\textbf{c}). Adding $F^\mthr{conf}_{\rho_{A,\mthr{id}}}+F^\mthr{conf}_{\rho_{B,\mthr{id}}}$ to $F^\mthr{bind}$ leads to
  \begin{eqnarray}
  \boxed{
 { F^\mthr{adh}_{\rho_{A,\mthr{id}},\rho_{B,\mthr{id}}} \over k_B T } = -  \on_{AB}\, .
 }
 \label{Eq:Sec4:Frhorho}
 \end{eqnarray}
 As largely discussed in the past all multivalent free energies including Eqs~\ref{Eq:Sec4:FNN} and~\ref{Eq:Sec4:FrhoN} tend to the expression in Eq.~\ref{Eq:Sec4:Frhorho} in the weak-binding regime, \emph{i.e.} in the limit of small $\on_{AB}/N_A$ and $\on_{AB}/N_B$~\cite{melting-theory2,melting-theory1, MognettiSoftMatt2012,RogersPNAS2011,MognettiPNAS2012,RogersAnswer,VarillyJCP2012}. A  connection between weak-binding approximations and systems in which the numbers of linkers are Poisson-distributed (see Eq.~\ref{Eq:Sec4:ZconfId}) has recently been studied by Curk~{\em et al.}~\cite{CurkMolPhys2018}.

\subsection{Compression of the linkers in the CR}\label{Sec4:CompCon}

In this section, we study how changing the distance $h$ between the interacting surfaces in the CR  influences the adhesive free energy (Fig.~\ref{Fig:Sec4Fig1}\textbf{d}). As previously done for systems of colloidal particles functionalized by DNA linkers~\cite{melting-theory2,melting-theory1,MognettiSoftMatt2012, RogersPNAS2011,MognettiPNAS2012,RogersAnswer,VarillyJCP2012}, we show that the reduction of configurational space available to linkers in the CR upon decreasing $h$ produces an entropic repulsions between the interfaces.\\
We define the configurational volumes of free $A$ and $B$ linkers in the CR as $\chi_A(h)$ and $\chi_B(h)$, as relative to the same quantities calculated for $h=\infty$. When considering solely steric interactions between the linkers and the interfaces we have $\chi_A, \, \chi_B \le 1$.\\
To calculate $F^\mthr{adh}$, we follow the same steps presented in Sec.~\ref{Sec4:MobMob}.\\
In the present case, the confining partition functions (Eqs~\ref{Eq:Sec4:ZconfId} and \ref{Eq:Sec4:ZconfN}) are multiplied by a factor equal to $\chi_A(h)^{n_A}$ and $\chi_B(h)^{n_B}$. Accordingly, as compared to Eqs~\ref{Eq:Sec4:FconfId} and~\ref{Eq:Sec4:FconfN}, $F^\mthr{conf}_{X_A}$ and $F^\mthr{conf}_{X_B}$ include additional terms $-k_BT n_A\log \chi_A(h)$ and $-k_BT n_B\log \chi_B(h)$, respectively.\\
The expression of $Z^\mthr{bind}$ changes only through the equilibrium constant $K^\mthr{eq,2D}_{AB}(h)$, which depends on $h$ in a model-dependent manner related to the polymeric properties of the linkers. Reference~\cite{parolini2014thermal} (Supporting Information) reports analytic expressions of $K^\mthr{eq,2D}_{AB}(h)$ for rigid linkers. For flexible linkers, $K^\mthr{eq,2D}_{AB}(h)$ can be calculated using Monte Carlo simulations~\cite{VarillyJCP2012,DeGernierJCP2014,A-UbertiPCCP2016}.\\
The chemical-equilibrium relations in Eqs~\ref{Eq:Sec4:onA-N}, \ref{Eq:Sec4:onA-rho} are modified by a factor $\chi_A(h)$ multiplying  $S_\mthr{CR}$, while Eq.~\ref{Eq:Sec4:onAB} remains unchanged.\\
Finally, we arrive at the following expression of the particle-particle adhesion free energy
  \begin{eqnarray}
 {F^\mthr{adh}_{N_A,N_B}(h) \over k_B T} &&=  N_A \log \left( { N_A - \on_{A}(h)  \over N_A } {S_{A,\mthr{Tot}} \over S_{A,\mthr{OR}} } \right) +  \nonumber \\
  && + N_B \log \left( { N_B - \on_{B}(h)  \over N_B } {S_{B,\mthr{Tot}} \over S_{B,\mthr{OR}} } \right) + \on_{AB}(h)
 \label{Eq:Sec4:FNNh}
 \end{eqnarray}
 where, as before, $\on_A(h)$, $\on_B(h)$, and $\on_{AB}(h)$ are the total numbers of (bound or unbound) $A$ and $B$ linkers and $AB$ bonds in the CR at a given $h$. In the absence of compression ($h \rightarrow \infty$) Eq.~\ref{Eq:Sec4:FNNh} converges to Eq.~\ref{Eq:Sec4:FNN}. This can be demonstrated by observing that the term $(N_A-\on_A)/S_{A,\mathrm{OR}}$ appearing in the first logarithm of Eq.~\ref{Eq:Sec4:FNNh} is the density of linkers in the OR, and  that in the absence of compression such density is equal to the density of unbound linkers $(N_A-\on_{AB})/S_{A,\mathrm{Tot}}$ (see Eq.~\ref{Eq:Sec4:onA-N}).\\
Reference~\cite{parolini2014thermal} derived an expression of the multivalent free energy equivalent to Eq.~\ref{Eq:Sec4:FNNh} for a more general system also featuring intra--particle bonds (see Sec.~\ref{Sec4:Loop-bridges}). As compared to the results presented in Ref.~\cite{parolini2014thermal} (see Eq.~18 in the Supplementary Materials of~\cite{parolini2014thermal}), here we have offset Eq.~\ref{Eq:Sec4:FNNh} by a constant value that so that  $F^\mthr{adh}_{N_A,N_B}\rightarrow 0 $ for $h\to \infty$ ($\on_{AB} \rightarrow 0$). Reference~\cite{parolini2014thermal} also shows how for rigid linkers $F^\mthr{adh}_{N_A,N_B}$ features a sharp minimum for a value of $h$ comparable with the length of the linkers $L$.
 \\
The particle--plane and plane--plane multivalent free energies at a given $h$ read as follows, and can be compared to the ideal $h$-independent expressions derived in Sec.~\ref{Sec4:MobMob} (Eqs~\ref{Eq:Sec4:FrhoN}, \ref{Eq:Sec4:Frhorho})
\begin{eqnarray}
{ F^\mthr{adh}_{\rho_{A,\mthr{id}},N_B}(h) \over
k_B T } =N_B  \log  \left( { N_B - \on_{B}(h)  \over N_B } {S_{B,\mthr{Tot}} \over S_{B,\mthr{OR}} } \right) +\rho_{A,\mthr{id}}S_\mthr{CR}\left[ 1-\chi_A(h) \right], \label{Eq:Sec4:FrhoNh}
\\
{F^\mthr{adh}_{\rho_{A,\mthr{id}},\rho_{B,\mthr{id}}}(h) \over k_B T } =
\rho_{A,\mthr{id}}S_\mthr{CR}\left[ 1-\chi_A(h) \right] + \rho_{B,\mthr{id}}S_\mthr{CR}\left[ 1-\chi_B(h) \right] -\on_{AB}(h).
\label{Eq:Sec4:Frhorhoh}
\end{eqnarray}
Equation~\ref{Eq:Sec4:FrhoNh} can also be derived from Eq.~\ref{Eq:Sec4:FNNh} by taking the limit $S_{A,\mthr{OR}}\to \infty$, $N_A \to \infty$ with  constant $S_\mthr{CR}$ and $N_A/S_{A,\mthr{Tot}}=N_A/S_{A,\mthr{OR}}=\rho_{A,\mthr{id}}$. To derive Eq.~\ref{Eq:Sec4:Frhorhoh}, an analogous limit is considered also for the quantities relative to interface $B$.

\subsection{Interactions between fluid interfaces and solid particles with fixed ligands}\label{Sec4:MobileFixed}
In this section we determine the multivalent interaction free energy between a particle decorated by $N_A$ linkers of type $A$, uniformly distributed with fixed tethering points, and either a particle or an infinite interface featuring mobile $B$-type linkers. This scenario is relevant for the interaction between biological-cell membranes and solid-like particles, including viruses, parasites and artificial nanoparticles. The results reviewed here are therefore useful to unravel ubiquitous biological processes such as cell invasion and endocytosis~\cite{McMahon2011,Nabi2003,Mercer2010,Boulant2015}, as well as to aid the rational design of nanomedical probes for targeted intracellular delivery of drugs or genetic material~\cite{Zhang2015,Bareford2007,Duzgunes1999,Deshpande2013,Akinc2013}.\\
As compared to the calculations of Sec.~\ref{Sec4:MobMob}, in this case the number of $A$-type linkers in the CR is constant for a given $S_\mthr{CR}$, and equal to $n_{A}=N_A S_\mthr{CR}/S_{A,\mthr{Tot}}$. Moreover, the adhesion free energy does not include the entropic cost of confining $A$ linkers in the CR. The expressions for the most probable number of $B$ linkers in the contact region, $\on_B$, and the number of bonds, $\on_{AB}$, can be derived from Eqs~\ref{Eq:Sec4:onA-rho}, \ref{Eq:Sec4:onA-N} and~\ref{Eq:Sec4:onAB}.
\\
For the particle-particle scenario, the adhesion free energy can be expressed as
\begin{eqnarray}
F^\mthr{adh,fixed}_{N_B} &=& F^\mthr{bind}(N_A S_\mthr{CR}/S_{A,\mthr{Tot}},\on_B,\on_{AB}) + F^\mthr{conf}_{N_B}(\on_B),
\end{eqnarray}
which explicitly reads
\begin{eqnarray}
\boxed{
{F^\mthr{adh,fixed}_{N_B} \over k_B T}= { S_\mthr{CR} N_A \over S_{A,\mthr{Tot}} } \log { S_\mthr{CR} N_A / S_{A,\mthr{Tot}}  -\on_{AB} \over S_\mthr{CR} N_A / S_{A,\mthr{Tot}} } + N_B \log {N_B -\on_{AB} \over N_B}
+ \on_{AB}.
}
\nonumber \\
\label{Eq:Sec4:fixed-NA}
\end{eqnarray}
Instead, for the case of a solid-like particle interacting with an infinite fluid interface we obtain:
\begin{eqnarray}
F^\mthr{adh,fixed}_{\rho_{B,\mthr{id}}} &=& F^\mthr{bind}(N_A S_\mthr{CR}/S_{A,\mthr{Tot}},\on_B,\on_{AB}) + F^\mthr{conf}_{\rho_{B,\mthr{id}}} (\on_B)\label{Eq:Sec4:fixed-rho-imp}
 \\
&=& k_B T  { S_\mthr{CR} N_A \over S_{A,\mthr{Tot}} } \log \left( { S_\mthr{CR} N_A / S_{A,\mthr{Tot}}  -\on_{AB} \over S_\mthr{CR} N_A / S_{A,\mthr{Tot}} } \right),
\nonumber
\end{eqnarray}
which results in
\begin{eqnarray}
\boxed{
{ F^\mthr{adh,fixed}_{\rho_{B,\mthr{id}}} \over k_B T}
= -{ S_\mthr{CR} N_A \over S_{A,\mthr{Tot}} } \log (1+\rho_{B,\mthr{id}} K^\mthr{eq,2D}_{AB}).
}
\label{Eq:Sec4:fixed-rho}
\end{eqnarray}
To derive Eq.~\ref{Eq:Sec4:fixed-rho} from Eq.~\ref{Eq:Sec4:fixed-rho-imp}, we used $\on_{AB} = S_\mthr{CR} N_A \rho_{B,\mthr{id}} K^\mthr{eq,2D}_{AB}/ [S_{\mthr{Tot},A}(1+\rho_{B,\mthr{id}} K^\mthr{eq,2D}_{AB})]$, valid for the case relevant here in which type $B$ linkers are freely diffusing on an infinite interface. Equation~\ref{Eq:Sec4:fixed-rho} was originally derive in Ref.~\cite{DiMichelePRE2018} to study endocytosis.

\begin{figure}[ht!]
\begin{center}
\includegraphics[width=10.cm,angle=0]{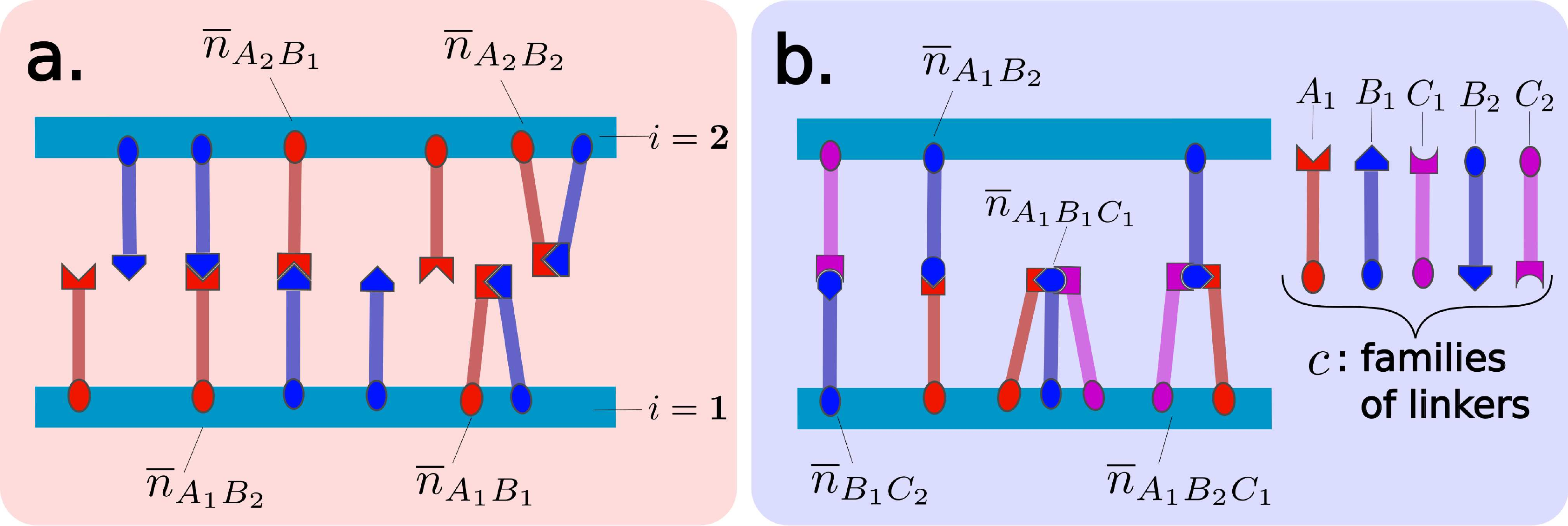}
\caption{{\bf Systems featuring multiple types of linker complexes.} {\bf a},  Loops and bridges appear when coating each interface with two types of complementary linkers as in the system presented in Fig.~\ref{fig3}. The interfaces are labelled as, $i=1,\, 2$. {\bf b}, The presence of $c$ families of linkers (with $c>2$) can lead to the fomation of multimeric complexes.  Note that linkers tethered to different interfaces belong to different families. Complexes made by three linkers appear in the system presented in Fig.~\ref{fig4} as transient states between loops and and bridges \cite{ParoliniACSNano2016}. }\label{Fig:Sec4Fig3}
\end{center}
\end{figure}

\subsection{Multiple types of bonds}\label{Sec4:ComplexLinkages}
As demonstrated in Sec.~\ref{Sec3}, functionalities and responsive behaviours often arise from using linkers capable of forming multiple types of (competing) complexes. In this section we apply the formalism introduced in the Sec.~\ref{Sec4:MobMob} to tackle the experimentally-relevant scenario in which the formation of cross-surface bridges competes with that of intra-surface loops (Sec.~\ref{Sec4:Loop-bridges}), and the most general case in which linkers can form multimeric complexes of arbitrary size (Sec.~\ref{Sec4:Multimeric}).

\subsubsection{Competition between bridge and loop formation.}\label{Sec4:Loop-bridges}
 In this section we discuss the case in which both interacting interfaces are functionalised by two types of mutually complementary linkers, which can thus form either bridges, if a bond forms across different interfaces, or loops, if the bond occurs within the same interface (see Fig.~\ref{Fig:Sec4Fig3}{\bf a}). This configuration is relevant to a number of experimental studies on equilibrium liposome-liposome~\cite{parolini2014thermal,BachmannSoftMatter2016} and liposome-SLB interactions~\cite{shimobayashi2015direct}, discussed in Sec.~\ref{Sec3Eq}.
We denote by $A_1$, $B_1$ and $A_2$, $B_2$ the linkers of the two complementary types ($A$ and $B$) anchored to the first and second interface ($1$ and $2$). Importantly, linkers of the same type tethered to different interfaces are treated as belonging to different families when calculating adhesion free energies (see Fig.~\ref{Fig:Sec4Fig3}{\bf b}). This is required given that the configurational part of the linker-linker complexation free energies ($\Delta G^\mthr{conf}_{AB}$ in Sec.~\ref{Sec4:MobMob}) depends on the properties and position of the interfaces supporting the linkers. For instance, in a system where two finite-size particles interact, the configurational free energy of forming loops is dependent on the area of each particle, so that $\Delta G_{A_1 B_1}\neq \Delta G_{A_2 B_2}$ if the two particles have different area.\\ 
The first step in calculating the adhesion free energy consists in identifying the list of different complexes featured by the system; in the present case two types of bridges, $A_1B_2$ and $A_2B_1$, and two types of loops, $A_1B_1$ and $A_2B_2$. However, we further distinguish the case in which loops are contained in the OR rather than the CR. This choice enables the derivation of a more general expression of $F^\mthr{adh}$ for cases in which linkers and complexes are compressed when found in the CR (Sec.~\ref{Sec4:CompCon}) or in the presence of non-specific interactions (Sec.~\ref{Sec4:NonSel}). Consequently, we regard the system as featuring six types of complexes, two of which are bridges and four of which are loops. We then set off to derive the most likely number of each complex: $\on_{A_1B_2}(h)$ and $\on_{A_2B_1}(h)$ for the two types of bridges and $\on^\mthr{CR}_{A_1B_1}(h)$, $\on^\mthr{OR}_{A_1B_1}(h)$, $\on^\mthr{CR}_{A_2B_2}(h)$, $\on^\mthr{OR}_{A_2B_2}(h)$ for the four types of loops. Note that the numbers of loops in the ORs are also a function of the distance between the interfaces, $h$, given that smaller values of $h$ reduce the number of linkers in the CR. For loops in the OR of infinitely extended interfaces, reservoir densities $\rho^\mthr{OR}_{A_1B_1}$  and  $\rho^\mthr{OR}_{A_2B_2}$ replace $\on^\mthr{OR}_{A_1B_1}(h)$ and $\on^\mthr{OR}_{A_2B_2}(h)$. \\
The most probable number of each type of complex can be derived by imposing stationary conditions on the partition function $Z^\mthr{adh}$, as done in Eq.~\ref{Eq:Sec4:ChemEq}. References~\cite{parolini2014thermal} and~\cite{shimobayashi2015direct} follow this procedure for the case of two interacting particles and a particle interacting with an infinite interface, respectively. However, here we want to stress how using the results of Sec.~\ref{Sec4:MobMob} one can avoid the explicit derivation of $Z^\mthr{adh}$, which is in fact rather cumbersome if multiple types of complexes are involved. Instead one can directly impose chemical equilibrium conditions as reported in  Sec.~\ref{Sec4:MobMob}. For instance, by generalising Eq.~\ref{Eq:Sec4:onAB}, we can calculate the number of loops in the CR and the OR of a particle using the following expressions
\begin{eqnarray}
\on^\mthr{CR}_{A_1B_1}(h) &=&\on^\mthr{free,CR}_{A_1}(h) \on^\mthr{free,CR}_{B_1}(h) { K^\mthr{eq,2D}_{A_1 B_1}(h) \over S_\mthr{CR} }
\nonumber \\
 \on^\mthr{OR}_{A_1B_1}(h) &=& \on^\mthr{free,OR}_{A_1}(h) \on^\mthr{free,OR}_{B_1}(h) { K^\mthr{eq,2D}_{A_1 B_1}(\infty) \over S_\mthr{OR} },
\label{Eq:Sec4:ChemLB1}
\end{eqnarray}
where we define the number of free linkers as $\on^\mthr{free,CR}_{A_1}(h) = \on_{A_1}(h)-\on^\mthr{CR}_{A_1B_1}(h)-\on^\mthr{CR}_{A_1B_2}(h)$ and
$\on^\mthr{free,OR}_{A_1}(h) = N_A-\on_{A_1}(h)-\on^\mthr{OR}_{A_1 B_1}(h)$.
After deriving expressions analogous to Eqs~\ref{Eq:Sec4:ChemLB1}
for all the complexes involved, one can calculate the adhesion free energy using the portable expressions for $F^\mthr{bind}$ and $F^\mthr{conf}$ derived in the previous section. Using the ``lb''  superscript to label the adhesion free energy in the presence of loop and bridge formation, for the scenario of two finite-size interacting particles, we can write:
\begin{eqnarray}
F^\mthr{adh,lb}_{N_{A_1}N_{A_2}N_{B_1}N_{B_2}} =\sum_{i=1,2}\left[ F^\mthr{conf}_{N_{A_i}}(\on_{A_i}(h))+F^\mthr{conf}_{N_{B_i}}(\on_{B_i}(h)) \right] + F^\mthr{bind}.
\label{Eq:Sec4:adhlb1}
\end{eqnarray}
 The expression for $F^\mthr{conf}_{N_A}$ in Eq.~\ref{Eq:Sec4:adhlb1} can be taken from Eq.~\ref{Eq:Sec4:FconfN}, modified as described in Sec.~\ref{Sec4:CompCon} to account for compression of the linkers in the CR, while $F^\mthr{bind}$ can be calculated following the general recipe of Ref.~\cite{A-UbertiJCP2013} as presented in Sec.~\ref{Sec4:MobMob}
\begin{eqnarray}
{F^\mthr{bind} \over k_B T} &=& \sum_{Y=A,B}\sum_{i=1,2} \left[
(N_{Y_i}-\on_{Y_i}(h)) \log {\on^\mthr{free,OR}_{Y_1}(h)\over N_{Y_i}-\on_{Y_i}(h)} + \on_{Y_i}(h) \log {\on^\mthr{free,CR}_{Y_1}(h) \over \on_{Y_i}(h) } \right],
\nonumber \\
&& \qquad +\on^\mthr{bridges}(h) + \on^\mthr{loops}(h)
\end{eqnarray}
where $Y$ identifies the type of molecule ($Y=A$ and $Y=B$) and $\on^\mthr{bridges}(h)$ and $\on^\mthr{loops}(h)$ are the total number of bridges and loops (\emph{e.g.} $\on^\mthr{bridges}(h)=\on_{A_1B_2}(h)+\on_{A_2B_1}(h)$).\\
Assuming that the two particles are identical in size and for the total number of each type of linker they host ($S_{A,\mthr{OR}}=S_{B,\mthr{OR}}=S_\mthr{OR}$, $N_{A_1}=N_{B_1}=N_{A_2}=N_{B_2}=N$), Eq.~\ref{Eq:Sec4:adhlb1} renders
\begin{eqnarray}
\boxed{
{F^\mthr{adh,lb}_{N,N} \over k_B T} = 4 N \log \left({ \on^\mthr{free,OR}(h)\over N} {S_\mthr{Tot} \over S_\mthr{OR}}\right) + \on^\mthr{bridges}(h) +  \on^\mthr{loops}(h),
}
\label{Eq:Sec4:NComm}
\end{eqnarray}
where $\on^\mthr{free,OR}(h)$ is the total number of free linkers of a given type in the OR and $\on^\mthr{free,OR}(h)/S_\mthr{OR}=\on^\mthr{free,CR}(h)/(\chi S_\mthr{CR})$. Equation~\ref{Eq:Sec4:NComm} was derived in Ref.~\cite{parolini2014thermal} to study GUV-GUV interactions, as discussed in Sec.~\ref{Sec3Eq}. Importantly, $F^\mthr{adh,lb}_{N,N}$ is not zero at large $h$ given the presence of intra--particle bonds. Therefore the adhesion free energy is usually offset by its value at $h\to\infty$.
\\
A similar procedure leads to the adhesion free energy of a particle interacting with an infinite interface
\begin{eqnarray}
F^\mthr{adh,lb}_{N_{A_1} \rho_{A_2,\mthr{id}} N_{B_1} \rho_{B_2,\mthr{id}}} =\sum_{Y=A,B}\left[ F^\mthr{conf}_{N_{Y_1}}(\on_{Y_1}(h))+F^\mthr{conf}_{\rho_{{Y_2},\mthr{id}}}(\on_{Y_2}(h)) \right] + F^\mthr{bind},
\label{Eq:Sec4:adhlb2}
 \end{eqnarray}
 where $F^\mthr{conf}_{\rho_{{Y_2},\mthr{id}}}$ is given by Eq.~\ref{Eq:Sec4:FconfId}. If $N=N_{A_1}=N_{B_1}$ and $=\rho_{{A_2},\mthr{id}}=\rho_{{B_2},\mthr{id}}=\rho_\mthr{id}$ we find
\begin{eqnarray}
\boxed{
{F^\mthr{adh,lb}_{N,\rho_\mthr{id}} \over k_B T } =  2 N \log \left( {\on^\mthr{free,OR}(h) \over N} {S_\mthr{Tot} \over S_\mthr{OR}}\right) + \on^\mthr{loops}(h) + 2 \rho_\mthr{id} S_\mthr{CR} [1-\chi(h)],
}
\label{Eq:Sec4:PCCP}
\end{eqnarray}
where in this case $\on^\mthr{loops}$ is the total number of loops on the particle.
Equation~\ref{Eq:Sec4:PCCP} was derived in Ref.~\cite{shimobayashi2015direct} without accounting for compression ($\chi(h)=1$), and with the argument of the logarithm term expressed as $(N-\on_{A_1B_1}-\on_{A_1B_2})/N$.

\subsubsection{Multimeric bonds}\label{Sec4:Multimeric}
 In this section, we detail the calculation of the multivalent free energies in systems where complexes featuring more than two linkers can form.  Multimeric bonds are for instance found in implementations relying on toehold-mediated-strand displacement (Secs~\ref{Sec3Kin} and ~\ref{Sec3EmulsAndSolid}) and are likely to become more common in future applications aiming at mimicking the ability of cell-bound proteins to process and transduce chemical signals.
\\
Here we detail the derivation of the binding free energy $F^\mthr{bind}$ (see Fig.~\ref{Fig:Sec4Fig2}); the overall adhesive free energy can then be derived by adding confining terms ($F^\mthr{conf}$) as done in the previous sections.
\\
Employing the notation of  Ref.~\cite{DiMicheleJCP2016}, we consider  $c$ families of linkers tethered to the CR between two interfaces (see Fig.~\ref{Fig:Sec4Fig3}{\bf b}). As done in~\ref{Sec4:Loop-bridges}, also in this case we regard identical linkers tethered to different interfaces as belonging to different types (families). $N_\alpha$ (with $\alpha=1,\cdots c$) is the total number of linkers of type $\alpha$. We define by $g_\alpha(X)$ the number of linkers of type $\alpha$ entering a complex of type $X$, while $m(X)$ is the total number of linkers making up $X$. Any given linker complex $X$ is specified by the list of the linkers forming it \{$ \alpha_1,\, \alpha_2,\cdots \alpha_m$\}, with $\alpha_i\in [1,c]$. We label with \{$X$\} the ensemble of bonds displayed by the system. $\on_X$ is the number of linker complexes of type $X$ at equilibrium, and is calculated using chemical equilibrium equations
\begin{eqnarray}
\on_X = { K^\mthr{eq,2D}_X \over {S_\mathrm{CR}}^{m(X)-1}} \prod_{i=1}^{m(X)} \on^\mthr{free}_{\alpha_i}
\qquad \qquad
\on^\mthr{free}_{\alpha} = N_\alpha - \sum_{X\in \mthr{\{} X \mthr{\}} } g_\alpha(X) \on_X \, \, ,
\label{Eq:Sec4:ChemEqMulti}
\end{eqnarray}
where $\on^\mthr{free}_{\alpha}$ is the number of free linkers of type $\alpha$. A saddle-point approximation of the binding partition function $Z^\mthr{bind}$ (Eq.~\ref{Eq:Sec4:Zbind}) adapted to the presence of multimeric bonds allows writing the binding free energy as (see Ref.~\cite{DiMicheleJCP2016} for the derivation)
\begin{eqnarray}
\boxed{
{F^\mthr{bind} \over k_B T} =
\sum_{\alpha=1}^c N_\alpha \log { \on^\mthr{free}_\alpha \over N_\alpha } + \sum_{X\in \mthr{\{} X \mthr{\}}} \left(m(X)-1\right) \on_X
}
\label{Eq:Sec4:FbindMulti}
\end{eqnarray}
In the weak binding regime, here defined as the limit in which $\on^\mthr{free}_\alpha/N_\alpha \to 1$,  Eq.~\ref{Eq:Sec4:FbindMulti} becomes
\begin{eqnarray}
{F^\mthr{bind} \over k_B T} =- \sum_{X\in \mthr{\{} X \mthr{\}}} \on_X.
\label{Eq:Sec4:FbindMultiWB}
\end{eqnarray}
To derive Eq.~\ref{Eq:Sec4:FbindMultiWB} from Eqs~\ref{Eq:Sec4:ChemEqMulti} and~\ref{Eq:Sec4:FbindMulti} we noted that the formation of a single bond of type $X$ depletes the total number of linkers by $m(X)$.

\subsection{Non specific interactions between linkers}\label{Sec4:NonSel}
Despite the fact that non-specific (\emph{e.g.}~electrostatic or steric)  interaction between linkers are sometimes accounted for in the derivation of the linker-linker complexation free energy~\cite{MognettiPNAS2012,VarillyJCP2012,DeGernierJCP2014} (see Sec.~\ref{Sec4:MobMob}), so far we have neglected non-specific interactions between linkers that do not belong to the same complex. Steric interactions between fixed linkers are negligible at low coating densities, as typically found in systems using bulky streptavidin-biotin complexes to anchor DNA linkers to the surface of colloids. In these cases, steric interactions only contribute with a fraction of $k_B T$ {\em per} linker~\cite{LeunissenJCP2011}. Non-specific interactions only become relevant for high-density linker coatings for both micro~\cite{KimCrockerLang2006,WangPineNatComm2015} and nano sized
particles~\cite{MladekPRL2012,NykypanchukNatMat2008}.\\
The situation is significantly different for the case of mobile linkers~\cite{DiMichelePRE2018}. Indeed, given the possibility of the linkers to accumulate in the CR~\cite{SackmannSoftMatter2014,SmithPNAS2008}, large linker densities can locally emerge even if the average coating density is small, and the resulting steric interactions can drastically affect the adhesion free energy.\\
Reference~\cite{DiMichelePRE2018} examined the effect of excluded volume interactions in the context of passive endocytosis, by studying the multivalent adhesion between a rigid particle, representing a virus, parasite or nanomedical probe, and a deformable infinite interface representing a cell membrane. The authors considered the cases in which linkers on the invading particle are either fixed or mobile, while those on the membrane were always considered as mobile. Critically for the discussion at hand, excluded-volume interactions of different magnitude were introduced between mobile linkers, which were modelled as hard disks at a given areal packing fraction $\eta$. The multivalent adhesion free energy $F^\mthr{adh}$ could then be decomposed in its ideal terms ($F^\mthr{bind}$ and $F^\mthr{conf}$ discussed in the previous sections) and a non-ideal contribution $F^\mthr{exc}(\eta)$ accounting for steric effects. In passive endocytosis, the dependence of $F^\mthr{adh}$ on the contact area $S_\mthr{CR}$, along with the elastic costs for membrane deformation, determine the tendency for the membrane to wrap around the particle. If the particle features mobile, non-sterically interacting linkers then one finds $F^\mthr{adh} \propto - \log (S_\mthr{CR}/S_\mthr{Tot})$ (Eq.~\ref{Eq:Sec4:FrhoN}), whereas for particles with fixed linkers $F^\mthr{adh}$ decreases linearly with $S_\mthr{CR}$ ($F^\mthr{adh}\propto -S_\mthr{CR}/S_\mthr{Tot}$, see Eq.~\ref{Eq:Sec4:fixed-rho}). The steadier decrease in $F^\mthr{adh}$ with $S_\mthr{CR}$ results in a greater tendency for the membrane to wrap around the particle for the case of fixed linkers, whereas ideal mobile linkers tend to stabilise partially-wrapped states~\cite{DiMichelePRE2018}. Importantly, the authors of Ref.~\cite{DiMichelePRE2018} observed that, for non-ideal mobile linkers, increasing $\eta$ produces a crossover between the logarithmic behaviour of $F^\mthr{adh}(S_\mthr{CR})$ observed for ideal mobile linkers, and the linear trend of fixed linkers. Therefore, although steric interactions weaken the adhesion, they also facilitate endocytosis by hampering partially wrapped states that dominate systems with mobile ideal linkers~\cite{DiMichelePRE2018}. Interestingly, for $S_\mthr{CR}/S_\mthr{Tot} \approx 1$, Ref.~\cite{DiMichelePRE2018} reported a strengthening of the adhesion at high $\eta$ because linker complexation minimises excluded volume interactions.

Reference~\cite{DiMichelePRE2018} used virial expansions~\cite{hill1986introduction} to estimate $F^\mthr{adh}$. However, the resulting expression of $F^\mthr{adh}$ is not portable and has limited validity~\cite{DiMichelePRE2018}. In particular, linker packing fractions greater than 5\% are sufficient to produce significant discrepancies assessed using Monte Carlo simulations (see below). Such discrepancies become important at small values of the CR because of higher local concentrations of the linkers~\cite{DiMichelePRE2018}.
Monte Carlo simulations can instead be implemented to reliably estimate the adhesion free energy through thermodynamic integration. By sampling the number of bonds $\langle n_{AB} \rangle_{K^\mthr{eq,2D}}$ at different equilibrium constants, we find~\cite{LeunissenJCP2011,VarillyJCP2012,DiMichelePRE2018}
\begin{eqnarray}
\boxed{
F^\mthr{adh} = -k_B T\int_0^{K^\mthr{eq,2D}} \mthr{d} \overline K^\mthr{eq,2D} { \langle n_{AB} \rangle_{\overline K^\mthr{eq,2D}} \over \overline K^\mthr{eq,2D}} + F^\mthr{adh}_{K^\mthr{eq,2D}=0},\label{Eq:Sec4:TI}
}
\end{eqnarray}
where $F^\mthr{adh}_{K^\mthr{eq,2D}=0}$ is the free energy of a system where bridge formation is forbidden. The expression in Eq.~\ref{Eq:Sec4:TI} can be obtained by deriving the partition function of the system (also including non--ideal terms) with respect to $\Delta G_{AB}$, and realising that $\beta \mthr{d} \Delta G_{AB}= -\mthr{d} K^\mthr{eq,2D}/ K^\mthr{eq,2D}$. In the most general case, the integral in Eq.~\ref{Eq:Sec4:TI} is taken over the sum of all types of (possibly multimeric) complexes along a multidimensional path in the space of the equilibrium constants~\cite{VarillyJCP2012}.

\subsection{Accounting for the elastic deformation of the interfaces}\label{Sec4:Elastic}

The elastic properties of the interfaces play a vital role in regulating linker complexation at all scales. Focusing on the biologically-relevant case of lipid bilayers, different contributions have applied coarse-grained simulations to demonstrate that the molecular details of the membranes can affect the  equilibrium constant of linker dimerisation $K^\mthr{eq,2D}$~\cite{XuWeiklJCP2015,HuWeiklPNAS2013,li2018binding}. These studies showed that $K^\mthr{eq,2D}$ decreases with the molecular roughness of the bilayers and provided portable expressions that could be used to inform multivalent models like those presented in the previous sections.
Elastic properties of the membrane can also mediate phase separation in systems featuring multiple populations of
linkers~\cite{ChenPRE2003,CoombsBJourn2004,WuYiPRE2006,SackmannSoftMatter2014} as relevant, for instance, to the study of synapse
formation~\cite{QiArupPNAS2001,RaychaudhuriKardarPRL2003}.
\\
In the context of trans-membrane transport, elastic deformation of the bilayers is required to wrap particles of different size in processes such as endocytosis and phagocytosis~\cite{alberts2015essential}. Existing literature has extensively studied the effects of size, shape and rigidity of the particles, as reviewed in Ref.~\cite{dasgupta2017nano}. Recent developments described in the Sec.~\ref{Sec4:NonSel} have then seen the integration of this knowledge with the consistent statistical mechanical description of multivalent adhesive free energies reviewed here~\cite{DiMichelePRE2018}.\\
The elastic properties of lipid bilayers can also be regulated by specialised proteins capable of altering, for instance, the local curvature of the surface, as reviewed in Ref.~\cite{2018BiomRem}. Interestingly, Curk {\em et al.}~recently highlighted how membrane--bound components with intrinsic curvature play an important role in regulating passive endocytosis even without direct interaction with the invading particles~\cite{CurkNanoletters2018}.\\
In the context of self-assembly of compliant units, the free energy terms describing particle elastic deformation are sometimes analytically treatable. This is the case, for instance, of oil-in-water emulsion droplets where deformation is regulated by interface tension~\cite{Jasna_PNAS_2012}, or for large GUVs in the limit of strong adhesive forces, where the elastic response is dominated by membrane stretching~\cite{parolini2014thermal} (Sec.~\ref{Sec3Eq}).
In many situations, however, the analytical approach turns out to be unfeasible, for example when attempting to model the reversible, temperature-dependent, self-assembly of relatively small LUVs in which thermal fluctuations have a large influence~\cite{BachmannSoftMatter2016} (Sec.~\ref{Sec3Eq}).
To overcome this issue, Ref.~\cite{BachmannSoftMatter2016} developed a Monte-Carlo simulation platform in which multivalent LUVs are modelled as triangulated
meshes~\cite{Nelson86,Ho_EPL,Ho_PRA,kroll97,vsaric2012fluid} interacting through reversible cross-linking bonds. Density-of-state calculations were then implemented to quantify the free-energy costs associated to the deformation of the LUVs following the formation of a certain number of bonds. The methodology of Ref.~\cite{BachmannSoftMatter2016} enabled the study of the size of the emergent CR as a function of the elastic properties of the membrane~\cite{Nelson86,Ho_EPL,Ho_PRA,kroll97,vsaric2012fluid} as well as the length of the linkers. Importantly, Ref.~\cite{BachmannSoftMatter2016} found that, to an excellent approximation, the overall interaction free energy, accounting for both the multivalent adhesion terms and the vesicle-deformation terms, increases linearly with the number of bridges formed. This results, along with results from MD simulations \cite{VachaNanoletters2011,schubertova2015influence}, could allow parametrising more coarsened models in which the effect of the deformability is implicitly accounted for.

\subsection{Open problems in statistical multivalent models}\label{Sec4:FutureDirections}

In Sec.~\ref{Sec4} we have extensively reviewed equilibrium approaches used to evaluate multivalent adhesion free energies through the estimation of the partition function of statistical mechanical models~\cite{BellBJourn1984,CoombsBJourn2004, KitovJACS2003,Martinez-Veracoechea_PNAS_2011,VarillyJCP2012,XuShawBJour2016,LicataPRL2008}.\\
A different thread of research, focusing specifically on the modelling of multivalent interactions in biological phenomena (\emph{e.g.} endocytosis), adopted a different perspective and attempted to disentangle the effect of kinetic factors, in particular, the finite time required by linkers to diffuse in the CR~\cite{ZhangPRE2008,ShenoyPNAS2005,BihrPRL2012,AtilganBJourn2009,GaoPNAS2005,DecuzziBJourn2008}.
However, most of these contributions are limited to the regime in which bond formation is irreversible, an assumption that is rarely applicable given that, as highlighted throughout this review, functional behaviours often only emerge in the reversible limit (see also Sec.~\ref{Sec3}).
\\
We thus argue that future developments will need to improve current statistical mechanics models by considering how the finite diffusivity of the linkers hampers relaxation towards equilibrium configurations. Moreover, it is likely that complex responsive behaviours arising from the ability of the linkers to probe different states featuring different types of (multimeric) bonds (e.g.~\cite{HalversonJCP2016,TitoArXiv2018,TitoEPJ2016,A-UbertiPRL2017}) will also require careful probing of kinetic bottlenecks related to slow reaction rates between linkers.
\\
At the many--particle level, in Sec.~\ref{Sec6:ReactionRates} we will show how finite reaction rates can drastically alter the collective properties of the system even without multimeric or competing bonds. This has already been discussed in Sec.~\ref{Sec3Kin}, where we showed how small denaturation rates can be exploited to program functionality~\cite{ParoliniACSNano2016,LanfrancoLangmuir2019}. In these situations, neither the currently available kinetic models nor equilibrium thermodynamic approaches are suitable even assuming an infinite diffusivity of the linkers.
%
Moreover, effective modelling capable of tackling the timescale at which biological or biomimetic systems display their functional behaviours will require developing new coarse-grained simulations.\\
Beyond investigating the kinetics of multivalent systems, future developments will require improved treatments for the effects of non-specific interactions: as highlighted in Sec.~\ref{Sec4:NonSel}, very little work has been done in this direction.

\section{Programming self-assembly and collective behaviour } \label{Sec5}

In this section, and particularly in Sec~\ref{Sec6:Many-body}, we generalise the results of Sec.~\ref{Sec4} and discuss the overall multivalent adhesion free energy in ensembles of $N$ finite-size particles or infinite interfaces. In particular, we will show how the mobility of the linkers engender many-body interactions so that $F^\mthr{adh}$ is not a simple sum over particle-particle (interface-interface, or particle-interface) pair-interactions. One should observe that also systems with fixed tethering points can feature many-body interactions. For instance, considering gold nanoparticles decorated by DNA strands of length comparable with their size~\cite{NykypanchukNatMat2008}, Mladek \emph{et al.}~\cite{MladekPRL2012,MladekSoftMatter2013} have quantified three-body interactions arising from the possibility that a given DNA strand binds to a complementary linker tethered to a different nanoparticle. Such an effect is more important and universal in systems with liquid interfaces, given that each linker belonging to a particle can potentially interact with all neighbouring particles. Moreover, as reviewed in Sec.~\ref{Sec6:Many-body}, liquid interfaces can support complex interaction schemes resulting, for instance, in competitions between different types of bonds, multimeric complexes or linkers seeding catalytic reactions (Secs.~\ref{Sec3Kin} and~\ref{Sec6:CollectiveBehav}). These complex interaction pathways are used to leverage many-body interactions resulting in peculiar functional behaviours.\\
Nevertheless, as experimentally demonstrated in Sec.~\ref{Sec3Kin} and discussed in~\ref{Sec4:FutureDirections}, kinetic bottlenecks related to slow denaturation rates can bias the performance of systems with competing bonds. In Sec.~\ref{Sec6:ReactionRates} we review the striking effect of finite reaction rates on multi-body self-assembly and targeting systems. To do so, we review a recently proposed numerical method that explicitly simulates reactions between linkers while evolving particle trajectories using a Brownian Dynamics algorithm. The methods explored in section~\ref{Sec6:ReactionRates} neglect other potentially limiting kinetic factors, such as the finite diffusivity of membrane-bound linkers. It is likely that future developments aiming at modelling biological systems will need considering also the diffusion dynamics of the linkers which has been shown to play an important role~\cite{dasgupta2014membrane}.

\begin{figure}[ht!]
\begin{center}
\includegraphics[width=10.cm,angle=0]{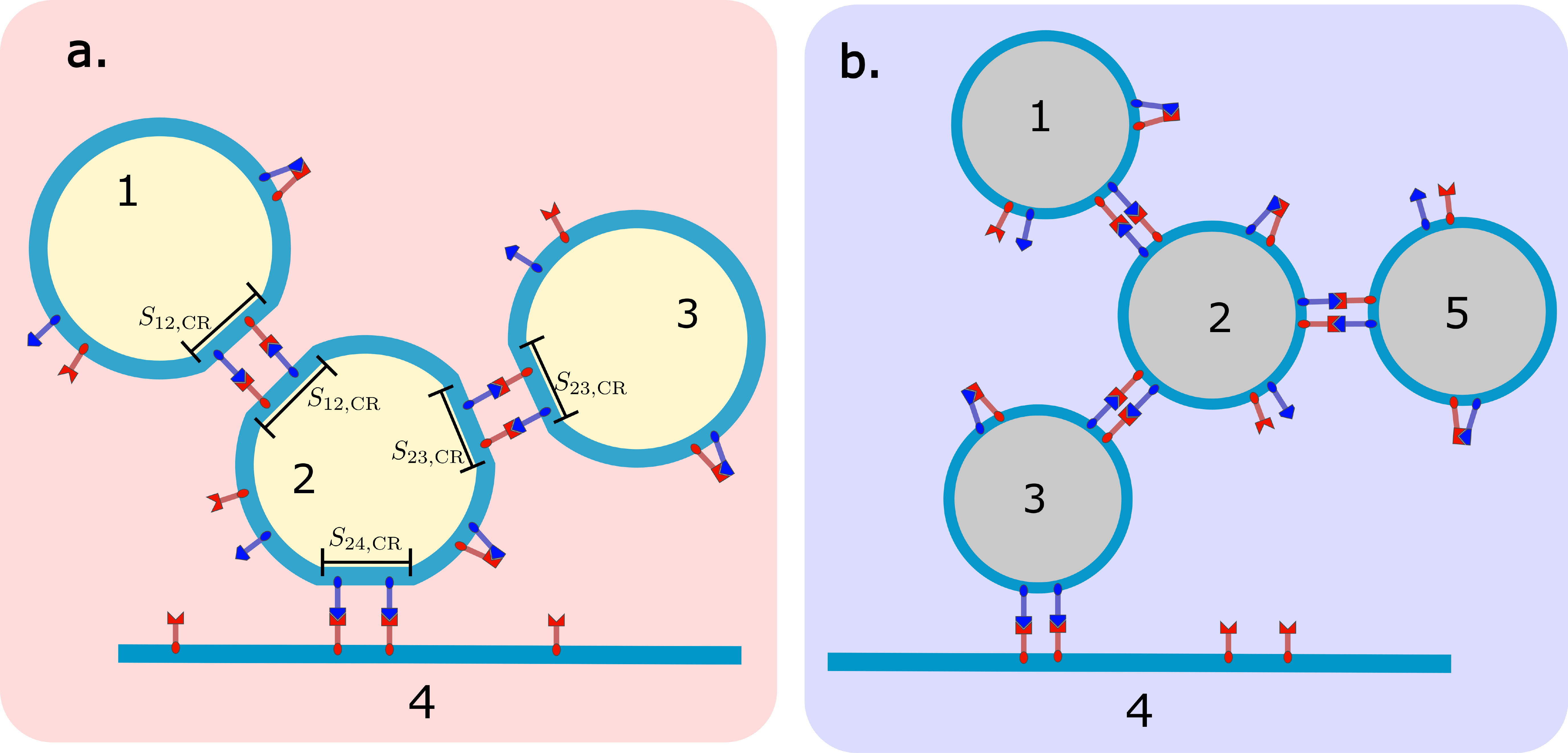}
\caption{{\bf Multivalent free energies in systems featuring more than two interfaces.} {\bf a}, Compliant particles interact through the formation of bonds in the contact regions (CRs). $S_{ij,\mathrm{CR}}$ indicates the area of the CR between particles $i$ and $j$. The area of the outer regions (not shown in panel {\bf a}) is labelled with $S_{i,\mathrm{OR}}$. {\bf b},~In systems of undeformable particle interacting through mobile linkers, as for microspheres coated by supported lipid bilayers, is univocally determined by the positions of the particles  and surfaces $\{ \mathbf{r}\}=\{\mathbf{r}_1, \cdots \mathbf{r}_{N_p}\}$. }\label{Fig:Sec6Fig05}
\end{center}
\end{figure}

\subsection{Many-particle multivalent free energies}\label{Sec6:Many-body}
\subsubsection{General formalism: compliant particles.} \label{Sec6:GeneralFormalism}
First, we generalise the results of Sec.~\ref{Sec4} and calculate the multivalent free energy of a configuration with multiple compliant interfaces, either particles or infinite planes. As done in Fig.~\ref{Fig:Sec4Fig3}, we label with $i$ the different interfaces (for instance, $i=1,\cdots,4$ in Fig.~\ref{Fig:Sec6Fig05}\textbf{a}), while $X$ denotes the type of linkers employed ($X=A,\, B$ in Fig.~\ref{Fig:Sec6Fig05}\textbf{a}). $N_{X_i}$ and $\rho_{X_i,\mthr{id}}$ are the total number and the density of linkers of type $X$ on the interface $i$ when $i$ is, respectively, a particle or a plane. $\on^{\mthr{CR},j}_{X_i}$ is the total number of linkers of type $X$ tethered to particle $i$ and found in the CR between interface $i$ and $j$ at equlibrium, while $\on^\mthr{OR}_{X_i}$ is the total number of $X$-type linkers in the OR of interface $i$. These definitions lead to the constraint $N_{X_i}=\on^\mthr{OR}_{X_i}+\sum_{j\neq i} \on^{\mthr{CR},j}_{X_i}$. Note that in Sec.~\ref{Sec4} the number of linkers of a given type in the CR had been defined simply as $\on_{X}$ because we never considered more than two interfaces and a single CR. $S_{ij,\mthr{CR}}$ is the area of the CR between interfaces $i$ and $j$, both for finite-size particles and infinite planes, while  $S_{i,\mthr{OR}}$ indicates the finite area area of the outer region of particle $i$.
\\
The equilibrium numbers of all possible (possibly multimeric) complexes can be derived from chemical-equilibrium equations as illustrated in Sec.~\ref{Sec4:MobMob}. Knowledge of the numbers of complexes enable the calculation of the numbers of free linkers in each region of each interface. We define as $\on^{\mthr{free,CR},j}_{X_i}$ and $\on^{\mthr{free,OR}}_{X_i}$ the number of free linkers of type $X_i$, respectively, in the CR between interfaces $i$ and $j$ and in the OR of particle $i$. $\on^{\mthr{free,CR},j}_{X_i}$ and $\on^{\mthr{free,OR}}_{X_i}$ are constrained by the requirement that the chemical potential of free linkers in the different regions must remain constant, resulting in relations similar to Eqs~\ref{Eq:Sec4:onA-N} and~\ref{Eq:Sec4:onA-rho}, possibly modified by a compression factor $\chi$ as described in Sec.~\ref{Sec4:CompCon}.\\
Using the equilibrium numbers of linker complexes and free linkers we can derive the overall adhesion free energy of the multi-body system as a sum of binding ($F^\mthr{bind}$) and confining terms ($F^\mthr{conf}$). $F^\mthr{bind}$ follows from Eq.~\ref{Eq:Sec4:Fbind} \cite{A-UbertiJCP2013} or, in the presence of multimeric bonds, Eq.~\ref{Eq:Sec4:FbindMulti}~\cite{DiMicheleJCP2016}. For infinite interfaces, $F^\mthr{conf}$ follows from Eq.~\ref{Eq:Sec4:FconfId}, while for particles featuring more than a single CR it can be derived from a multinomial distribution generalising Eq.~\ref{Eq:Sec4:ZconfN}. Following the discussion of Sec.~\ref{Sec4:ComplexLinkages} we justify the following rules enabling the calculation of $F^\mthr{adh}$ for an arbitrary configuration of many interacting multivalent interfaces (see \emph{e.g.} Fig.~\ref{Fig:Sec6Fig05}\textbf{a}):

\begin{itemize}
\item For each type of bridge-like or loop-like complex moving on a finite-size particle, we add to $F^\mthr{adh}$ a term equal to $k_\mathrm{B}T$ times the equilibrium number of those complexes multiplied by a factor $(m-1)$. $m$ is the number of linkers entering the complex;
\item For each type of linker $X_i$ tethered to a finite-size particle, we add to $F^\mthr{adh}$ a term equal to $k_B T N_{X_i} \log \left( {\on_{X_i}^\mthr{free,OR} \over N_{X_i}} {S_{i,\mthr{Tot}} \over S_{i,\mthr{OR}}}\right)$;
\item For each type of linker $X_i$ tethered to an infinite plane we add to $F^\mthr{adh}$ a term equal to $k_B T  \sum_j \rho_{X_i,\mthr{id}} S_{ij,\mthr{CR}} (1-\chi_{X_i j})$, where $\chi_{X_i j}$ is the compression factor of linker $X_i$ in the CR between interfaces $i$ and $j$ (Sec.~\ref{Sec4:CompCon}), and subtract a term equal to $k_\mathrm{B}T$ times the number of linkers $X_i$ involved in complexes cross-linking two interfaces (bridge-like complexes).
\end{itemize}
The previous rules do not contemplate the possibility of complexes simultaneously connecting more than two interfaces.  As an example, using the previous rules, we can write the multivalent free energy of the configuration of Fig.~\ref{Fig:Sec6Fig05}\textbf{a}, featuring three finite particles and an infinite plane decorated with two types of linkers that can form either loops or bridges
\begin{eqnarray}
{F^\mthr{adh} \over k_B T} &=& \sum_{X=A,B}\sum_{i=1}^3 N_{X_i} \log \left( {\on_{X_i}^\mthr{free,OR} \over N_{X_i}} {S_{i,\mthr{Tot}} \over S_{i,\mthr{OR}}}\right) + \rho_{B_4,\mthr{id}} S_{24,\mthr{CR}}(1-\chi_{B_42})
\nonumber \\
&& +\on^\mthr{loops} + \on^\mthr{bridges}_{12,23} \, .
\label{Eq:Sec6:Ves}
\end{eqnarray}
where $\on^\mthr{loops}$ is the total number of loops and $\on^\mthr{bridges}_{12,23}$ is the total number of bridges between particles 1 and 2 and between particles 2 and 3.
We note that Eq.~\ref{Eq:Sec6:Ves} becomes indeterminate when $S_{i,\mthr{OR}},\, \on^\mthr{OR}_{X_i} \to 0$, a relevant scenario, for instance, in the study of biological tissues featuring close-packed cells. In this case, we can replace the term $\on_{X_i}^\mthr{free,OR}/S_{i,\mthr{OR}}$ in the argument of the logarithm with $\chi_{X_i j} \on_{X_i}^{\mthr{free,CR},j}/S_{ij,\mthr{CR}}$, calculated for an arbitrary $j$.
\\
Note that to guarantee that complexation never depletes linkers and complexes in the OR of infinite planes, we have assumed that each plane is in contact with a finite number of particles. In practical terms, for an interface to be modelled as infinite, the number of linkers forming complexes with all the particles must be a negligible compared to the total number of linkers on the interface. This conditions may not be verified in systems where multivalent interactions lead to the the self-assembly of large particle aggregates or layers at interfaces, in which case, as reviewed in Sec.~\ref{Sec6:CollectiveBehav}, all interfaces must be modelled as finite.

\subsubsection{General formalism: rigid particles with mobile linkers.}
For the general case of compliant particles discussed in Sec.~\ref{Sec6:Many-body} we could derive the multi-body adhesion free energy for a given configuration of the system, \emph{i.e.} given the knowledge of the size and properties of the contact regions and outer regions of all the interfaces involved. However, it is important to realise that the geometry of the such system at equilibrium will ultimately be dictated by the interplay between the multivalent adhesive free energy and the elastic-deformation energy of the interfaces. Although the latter can sometimes be treated analytically or numerically for the case of two interacting surfaces, as discussed in Sec.~\ref{Sec4:Elastic}, the problem becomes prohibitively complex in multi-body compliant systems. In this section we consider the simplified scenario in which particles or interfaces are undeformable as for the case of microspheres coated by fluid lipid bilayers~\cite{VdMeulenJACS2013,RinaldinSoftMatter2018,TroutierAdvCollIntSci20071} (see Secs~\ref{Sec2BilayerParticles} and~\ref{Sec3EmulsAndSolid}). Under the assumption of undeformability, the behaviour of the system is fully determined by the multivalent adhesive free energy, to extent the ensembles of rigid particles decorated by mobile linkers have been studied using Brownian Dynamics simulations in which the forces acting on each particle, ${\bf f}_i$, are derived from $F^\mthr{adh}$. Here we detail the calculation of $F^\mthr{adh}$ and ${\bf f}_i$ for rigid multivalent systems featuring mobile linkers following Refs.~\cite{A-UbertiPRL2014,PetitzonSoftMatter2016,JanaArxiv2018}.
\\
We define by $n_{\alpha\beta\gamma\cdots}$ the number of complexes formed by linkers of type $\alpha$, $\beta$, $\gamma$, $\cdots$, where a given type of linker is specified by a type of molecule, labelled by $X$, and the index of the surface (particle or plane) supporting $X$, \emph{e.g.} $\alpha=X_i$. The number of free (unbound) linkers of type $\alpha$ is then $n_\alpha = N_\alpha -\sum_\beta \left( n_{\alpha\beta} + \sum_{\gamma\leq\beta} n_{\alpha\beta\gamma} + \cdots \cdots \right)$, where we indicate as $N_\alpha$ the total number of linkers of type $\alpha$ (bound and unbound). Equation~\ref{Eq:Sec4:Free}, adapted to the present system, gives the following multivalent adhesion free energy
\begin{eqnarray}
{F^\mthr{adh} \over k_B T} &=& - \log \sum_{\{ n_{\alpha\beta}, n_{\alpha\beta\gamma}, \cdots \}} Z^\mthr{adh}(\{ n_\alpha, n_{\alpha\beta}, n_{\alpha\beta\gamma}, \cdots \})
\nonumber \\
&\approx& - \log Z^\mthr{adh}(\{ \on_\alpha, \on_{\alpha\beta}, \on_{\alpha\beta\gamma}, \cdots \}),
\label{Eq:Sec6:FreeMany}
\end{eqnarray}
where a saddle-point approximation has been applied to obtain the second equality (see Eq.~\ref{Eq:Sec4:FreeSP}) and $\on_{\alpha\beta\gamma\cdots}$ indicate the most probably numbers of free linkers and complexes, found at equilibrium.
Importantly, note that in this section, $n_\alpha$ is not defined as the total number of linkers of type $\alpha$ in the CR as done previously (Secs.~\ref{Sec6:GeneralFormalism} and~\ref{Sec4}). This choice follows from the fact that the concept of contact region (CR) is not a useful one for rigid particles that cannot produce flat interacting surfaces. Indeed, the width of the gap between nearby particles is never uniform nor are, consequently, the configurational costs of forming an inter--particle linker complex. Furthermore these configurational costs are not only a function of the relative distance between tethering points, as assumed in Eq.~\ref{Eq:DGABh}, but also of the curvature of the particles. The model presented here does not rely on the assumption of flat CR, and is thus suitable for systems in which the curvature of the particles is non-negligible as compared to the length of the linkers.\\
The most likely number of complexes, $\on_{\alpha\beta\gamma\cdots}$, satisfy equilibrium conditions $\on_{\alpha\beta\gamma\cdots} = \on_\alpha \on_\beta \on_\gamma \cdots \exp[-\beta \Delta G_{\alpha\beta\gamma\cdots}(\{ {\bf r}\})]$.
In the present modelling, the complexation free energies, $\Delta G_{\alpha\beta\gamma\cdots}(\{ {\bf r}\})$, are derived by considering as reference state that in which all  linkers forming the complex are unbound  free to move over the entire surface to which they are tethered. Therefore different from the approach presented in Sec.~\ref{Sec4} where the cost of translationally-confining the linkers to the CR was computed separately as $F^\mathrm{conf}$, here all the confining costs are accounted for at the single bond level by $\Delta G_{\alpha\beta\gamma\cdots}(\{ {\bf r}\})$. References~\cite{A-UbertiPRL2014,JanaArxiv2018} report explicit expressions of $\Delta G_{\alpha\beta}(\{{\bf r} \})$ for systems of finite-size particles and particles interacting with a surface, respectively, under the assumption of freely-pivoting rigid linkers with negligible mutual steric interactions (see Fig.~\ref{Fig:Sec6Fig05}\textbf{b} and Sec.~\ref{Sec6:CollectiveBehav}). Note that here  $\Delta G_{\alpha\beta\gamma\cdots}$ is a function of the vector identifying the position of the particles and surfaces $\{ {\bf r} \}= \{ {\bf r}_1, \, {\bf r}_2, \cdots , {\bf r}_{N_p}\}$,  given that the relative distances between interfaces control the configurational terms of the hybridisation free energies. $N_p$ is the total number of interfaces. Below we generalise the results of Refs.~\cite{A-UbertiPRL2014,PetitzonSoftMatter2016,JanaArxiv2018} without specifying the nature of each interface (particle or surface) and accounting for the possibility of multimeric complexes formed by more than two linkers. As compared to Secs~\ref{Sec4} and~\ref{Sec6:GeneralFormalism}, planes here are not infinite but carry a depletable number of linkers, as relevant when attempting to model the thermodynamic limit of a system with a finite concentration of particles interacting with a surface.\\
It can be shown that the partition function at a given number of bonds reads as follows
\begin{eqnarray}
 Z^\mthr{adh}(\{ n_\alpha, n_{\alpha\beta}, n_{\alpha\beta\gamma} \cdots \})&=&
 \left[\prod_\alpha {N_\alpha ! \chi_\alpha(\{ {\bf r} \})^{N_\alpha} \over n_\alpha!}\right] \times
\left[ \prod_{\alpha\leq\beta} { e^{- {\Delta G_{\alpha\beta}(\{{\bf r}\}) n_{\alpha\beta} \over k_B T} }\over n_{\alpha\beta}!}\right] \times
\nonumber \\
&& \quad \left[\prod_{\alpha\leq\beta\leq\gamma} { e^{- {\Delta G_{\alpha\beta\gamma}(\{{\bf r}\}) n_{\alpha\beta\gamma} \over k_B T}} \over n_{\alpha\beta\gamma}!}\right] \times \cdots,
\label{Eq:Sec6:Zmany}
\end{eqnarray}
where $\chi_\alpha(\{{\bf r}\})$ is the relative configurational volume of a free linker of type $\alpha$ at a given interfaces’ position $\{{\bf r}\}$ as compared to a configuration in which all the interfaces are far away from each other. Here we have included the possibility of steric interactions between linkers interfaces, hence the first term in the r.h.s.\ of Eq.~\ref{Eq:Sec6:Zmany} emerges and accounts for the resulting repulsive interactions between interfaces as discussed in Sec.~\ref{Sec4:CompCon}~\cite{melting-theory2,melting-theory1}. The combinatorial terms of Eq.~\ref{Eq:Sec6:Zmany} can be obtained by generalising the procedure of Ref.~\cite{JanaArxiv2018}.\\
The stationary conditions enabling the calculation of the equilibrium numbers of free linkers and complexes $\on_{\alpha\beta \dots}$ read as
\begin{eqnarray}
 {\mthr{d} \over \mthr{d} \on_{\alpha\beta \dots}} Z^\mthr{adh}(\{ \on_\alpha, \on_{\alpha\beta}, \on_{\alpha\beta\gamma} \cdots \})=0,
 \label{Eq:Sec6:Zstat}
\end{eqnarray}
and, as discussed in Sec.~\ref{Sec4}, are equivalent to the chemical equilibrium conditions for the number of complexes. The adhesive free energy of the system $F^\mthr{adh}(\{ {\bf r } \})$(Eq.~\ref{Eq:Sec6:FreeMany}) is then given by Eq.~\ref{Eq:Sec4:FbindMulti} augmented by an osmotic term equal to $-k_\mathrm{B}T\sum_\alpha N_\alpha \log \chi_\alpha(\{{\bf r}\})$. Importantly, $F^\mthr{adh}(\{ {\bf r } \})$ is many-body, \emph{i.e.} it is not pair-wise additive, even when the linkers are short as compared to the size of the particles, resulting in short-range direct interactions. The many-body nature of $F^\mthr{adh}(\{ {\bf r } \})$ arises from the mobility of the tethering points such that each linker has the possibility of binding multiple particles. In Sec.~\ref{Sec6:CollectiveBehav}, we discuss how to leverage many-body interactions to engineer peculiar phase behaviours.
\\
Using the partition function, $Z^\mthr{adh}$, we calculate the force acting on particle $i$ as follows
\begin{eqnarray}
{ {\bf f}_i \over k_B T} &=& - { {\bf \nabla}_{{\bf r}_i} F^\mthr{adh}\over k_B T} \approx {{\bf \nabla}_{{\bf r}_i} Z^\mthr{adh}(\{ \on_\alpha, \on_{\alpha\beta}, \on_{\alpha\beta\gamma} \cdots \}) \over Z^\mthr{adh}(\{ \on_\alpha, \on_{\alpha\beta}, \on_{\alpha\beta\gamma} \cdots \})} \, .
\end{eqnarray}
Note that $Z^\mthr{adh}$ is a function of $\{ {\bf r} \}$ through $\on_{\alpha\beta\cdots}$, $\Delta G_{\alpha\beta\cdots}$, and $\chi_\alpha$ (see Eq.~\ref{Eq:Sec6:Zmany}). However, the stationary conditions (Eq.~\ref{Eq:Sec6:Zstat}) imply that we only need to consider variations in  $\Delta G_{\alpha\beta\cdots}$ and $\chi_\alpha$ to calculate ${\bf f}_i$. In particular using Eq.~\ref{Eq:Sec6:Zmany} we find
\begin{eqnarray}
{ {\bf f}_i } &=& k_B T\sum_\alpha N_\alpha {\bf \nabla}_{{\bf r}_i} \chi_\alpha(\{ {\bf r} \}) - \sum_{\alpha\leq\beta} \on_{\alpha\beta}{\bf \nabla}_{{\bf r}_i} \Delta G_{\alpha\beta}(\{{\bf r}\})
\nonumber \\
&& -\sum_{\alpha\leq\beta\leq\gamma}\on_{\alpha\beta\gamma}{\bf \nabla}_{{\bf r}_i} \Delta G_{\alpha\beta\gamma}(\{{\bf r}\})-\cdots \,.
\label{Eq:Sec6:forces}
\end{eqnarray}
Equation~\ref{Eq:Sec6:forces} is easily understandable: Each bond between two interfaces exerts a (usually) attractive force equal to the gradient of the complexation free energy. Instead, each free linker or intra--particle bond compressed between interface $i$ and a neighbour $j$ will contribute to ${\bf f}_i$ and ${\bf f}_j$ with a (usually) repulsive term. In general, the sums in Eq.~\ref{Eq:Sec6:forces} are limited to neighbouring interfaces. For specific expressions in the case of rigid ideal linkers we refer to Refs~\cite{A-UbertiPRL2014,PetitzonSoftMatter2016,JanaArxiv2018}.
Equation~\ref{Eq:Sec6:forces} can be used to implement Brownian Dynamics simulations. Importantly, the evaluation of the forces requires estimating the most likely number of bonds $\on_{\alpha\beta\cdots}$. References~\cite{A-UbertiPRL2014,PetitzonSoftMatter2016,JanaArxiv2018} describe how such estimation can be done on the fly using fixed-point iteration of a set of coupled equations equivalent to the chemical equilibrium conditions (Eq.~\ref{Eq:Sec6:Zstat}). Note that Eq.~\ref{Eq:Sec6:forces} assumes that reactions between linkers happen infinitely fast as compared to the time taken by the interfaces to diffuse away from each other. We discuss the reliability of this approximation in Sec.~\ref{Sec6:ReactionRates}. Finally, we stress that Eq.~\ref{Eq:Sec6:Zmany}, and consequently Eq.~\ref{Eq:Sec6:forces}, do not account for non-specific (\emph{e.g.}, steric or electrostatic) interactions between  the particles and/or surfaces. Those terms are typically pair-wise additive can simply be added to the multivalent forces computed here.\\

\begin{figure}[ht!]
\includegraphics[width=16.cm,angle=0]{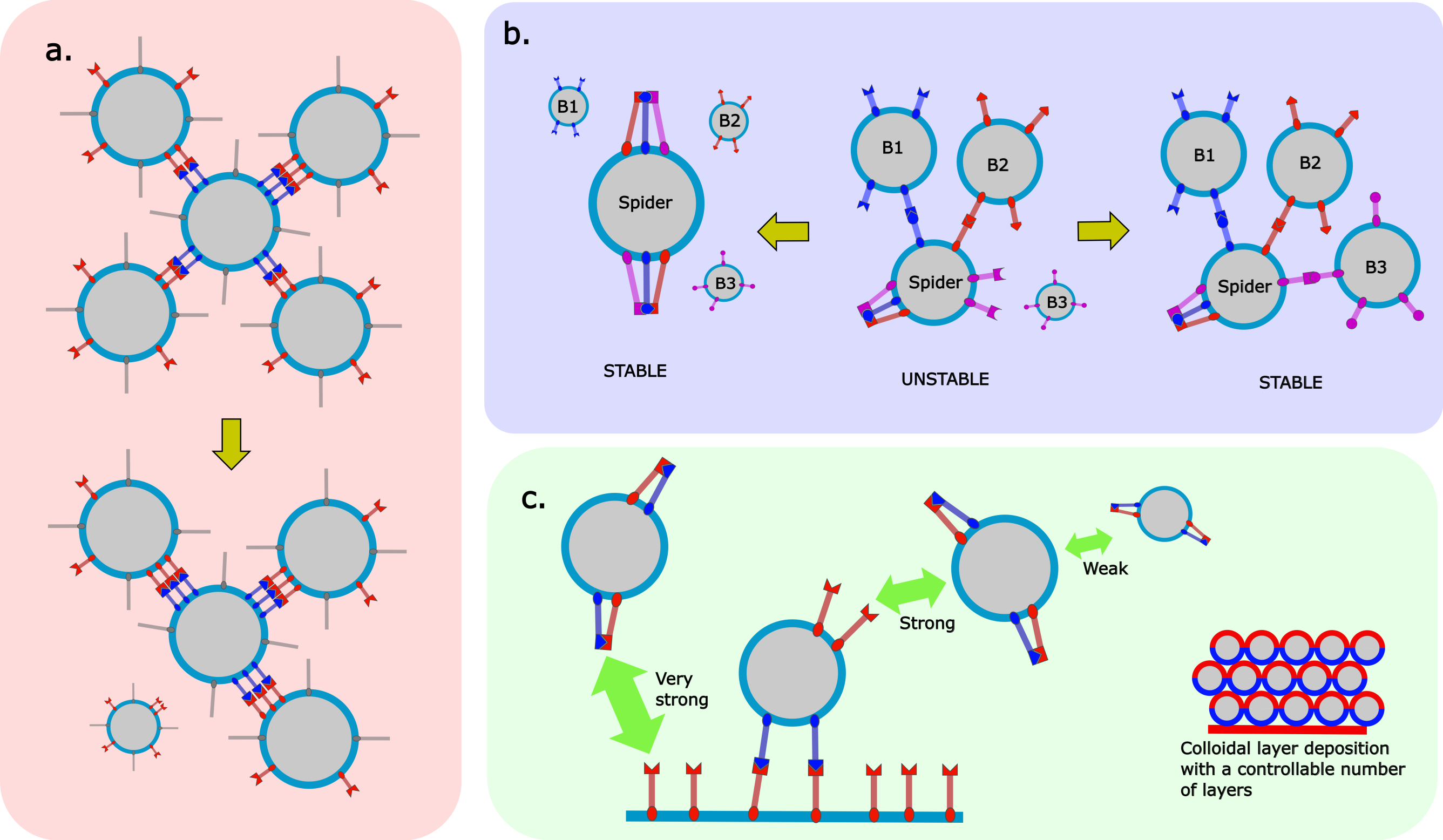}
\caption{ {\bf Linker mobility enables the programming of many-body interactions and emerging functionalities.} {\bf a}, Angioletti-Uberti {\em et al.}~\cite{A-UbertiPRL2014} showed how to combine sticky mobile linkers and inert strands (in grey) to program remote control over the valency of the aggregates. Inert strands do not form bonds but increase particle-particle repulsion (see Sec.~\ref{Sec4:CompCon}). {\bf b}, Halverson {\em et al.}~\cite{HalversonJCP2016} exploited the competition between loops and bridges to prescribe the sequence of binding events leading to the self-assembly of a target structure. In the design of Ref.~\cite{HalversonJCP2016}, particles carry loop-like ``spider'' complexes composed of three linkers. The breakup of a spider in favour of bridges is thermodynamically favourable only in the presence of three neighbouring particles. This cooperative behaviour is used to determine the sequence in which particles attach to the growing aggregate. {\bf c}, Jana {\em et al.}~\cite{JanaArxiv2018} used functionalised interfaces to direct the growth of colloidal crystals with controllable thickness from a flat interface. Linkers on the surface trigger cascade reactions on the particles leading to localised self--assembly. The colour code of the inset shows how the two types of linkers are unevenly distributed on the surfaces of the particles within the aggregate.}\label{Fig:Sec6Fig1}
\end{figure}

\subsubsection{Model-guided engineering new collective behaviours.}\label{Sec6:CollectiveBehav}
In this section, we review three systems where the many-body interactions emerging from linker mobility are exploited to program unique self-assembly behaviours (Fig.~\ref{Fig:Sec6Fig1}).
\\
In a theoretical and numerical contribution, Angioletti-Uberti~{\em et al.\ } have demonstrated control over the valency ($z$) of undeformable colloidal particles functionalised by mobile linkers~\cite{A-UbertiPRL2014}. The authors considered a binary system in which two types of particles are functionalised by complementary linkers that can thus form bridges and lead to adhesion (Fig.~\ref{Fig:Sec6Fig1}\textbf{a}). As the number of particles interacting with a given one increase, the adhesive free energy contribution for each particle pair weakens. This is a simple many-body effect emerging from the fact that, with increasing $z$, the number of bridges each particle can form has to be shared with more neighbours. However, the overall adhesive free energy \emph{per} particle still decreases with increasing $z$, owing to combinatorial and configurational entropic effects. As a result, each particle will always try to maximise its valency. To provide control over $z$ the authors introduced a repulsive free energy term, produced by inert linkers that only contribute with steric repulsion between the particles. The presence of steric repulsion guarantees that a minimum attractive force is required to stabilise a particle-particle contact, effectively limiting $z$. In other words, by combining the monotonic decrease of pair-wise attractive forces with $z$ with the offset introduced by steric repulsion the authors could engineer a \emph{per}-particle total free energy that is minimised for a given finite $z$.
As compared to the experimental droplet-based systems reviewed in Fig.~\ref{fig5}{\bf a}  and discussed Sec.~\ref{Sec3EmulsAndSolid}, the scheme of Ref.~\cite{A-UbertiPRL2014} is based on a free--energy minimisation rather than a control of kinetic factors.
\\
In another theoretical and numerical contribution, Halverson~{\em et al.\ } demonstrated how many-body effects in multivalent interactions can be exploited to engineer the self-assembly pathway of colloidal aggregates~\cite{HalversonJCP2016}. In this design, the target colloidal structure grows one particle at a time, with each new building block sequentially adhering to a prescribed binding site on the aggregate, as illustrated in Fig.~\ref{Fig:Sec6Fig1}. We consider four particles: $S$, $B1$, $B2$, and $B3$. Particle $S$ features intra--particle (loop-like) bonds made of three linkers, dubbed ``spiders'' in Ref.~\cite{HalversonJCP2016}. Each of the three linkers on $S$ is also complementary to linkers tethered to $B1$, $B2$ and $B3$. Binding one or two $B$-type particles to $S$ requires the partial disassembly of the spiders, the free energy penalty of which is not compensated by the formation of just one or two $B$-$S$ bridges. In turn, when all $B$ particles are available for binding to $S$, the free energy gain of forming three bridges is sufficient to overcome the penalty of disassembling the spider. In this way the authors can guarantee that an $S$-type particle can only adhere to a growing aggregate at the locations where it can bind to all 3 $B$-type particles, enabling programmability of the pathway of colloidal self-assembly. This strategy produced an improved yield in the self-assembly of complex colloidal aggregates, as compared to approaches based on conventional selective pair interactions between the particles. Reference~\cite{DiMicheleJCP2016} later applied the statistical mechanical approach discussed in Sec.~\ref{Sec4:Multimeric} to confirm that binding the third $B$-particle to $S$ is is always energetically more favourable than binding the first two. \\
Recently Jana~{\em et al.}~demonstrated by means of computer simulations the use of many-body multivalent effects to direct the growth of colloidal crystals of well defined thickness from a functionalised planar interface~\cite{JanaArxiv2018}. In this implementation, particles are functionalised by two types of linkers forming intra-particle loops when far away from the flat interface, so that inter-particle bridges cannot appear and a colloidal gas-phase is stabilised. The interface is functionalised by a third type of linker that can strongly bind to one of the two on the particles, leading to opening of the loops, adhesion of the particles, and the formation of a first colloidal layer on the interface. Particles in the first layer will feature a number of free linkers from the opening of the initial loops. These will facilitate deposition of a second layer of particles, which will also be ``activated'' by the freeing of some linkers and thus trigger the formation of the next layer. The assembly process proceeds layer after layer, with the caveat that particle-particle adhesion becomes weaker the farther one gets from the interface. As a result, the epitaxial growth of the colloidal crystal does not progresses indefinitely but stabilises when a certain thickness is reached and the solid phase is at equilibrium with the colloidal-gas phase in the bulk.\\
As reviewed in Sec.~\ref{Sec3EmulsAndSolid}, a similar mechanism has been implemented by Zhang~{\em et al.}~\cite{zhang2017sequential} to program the sequential self-assembly of chains of emulsion droplets~(Fig.~\ref{fig5}\textbf{a}). One fundamental difference between the implementation of Ref.~\cite{zhang2017sequential} and that of Ref.~\cite{JanaArxiv2018} is that in the former, to enable sequentiality, particles adhering at different steps feature chemically distinct linkers.


\begin{figure}[ht!]
\begin{center}
\includegraphics[width=15.cm,angle=0]{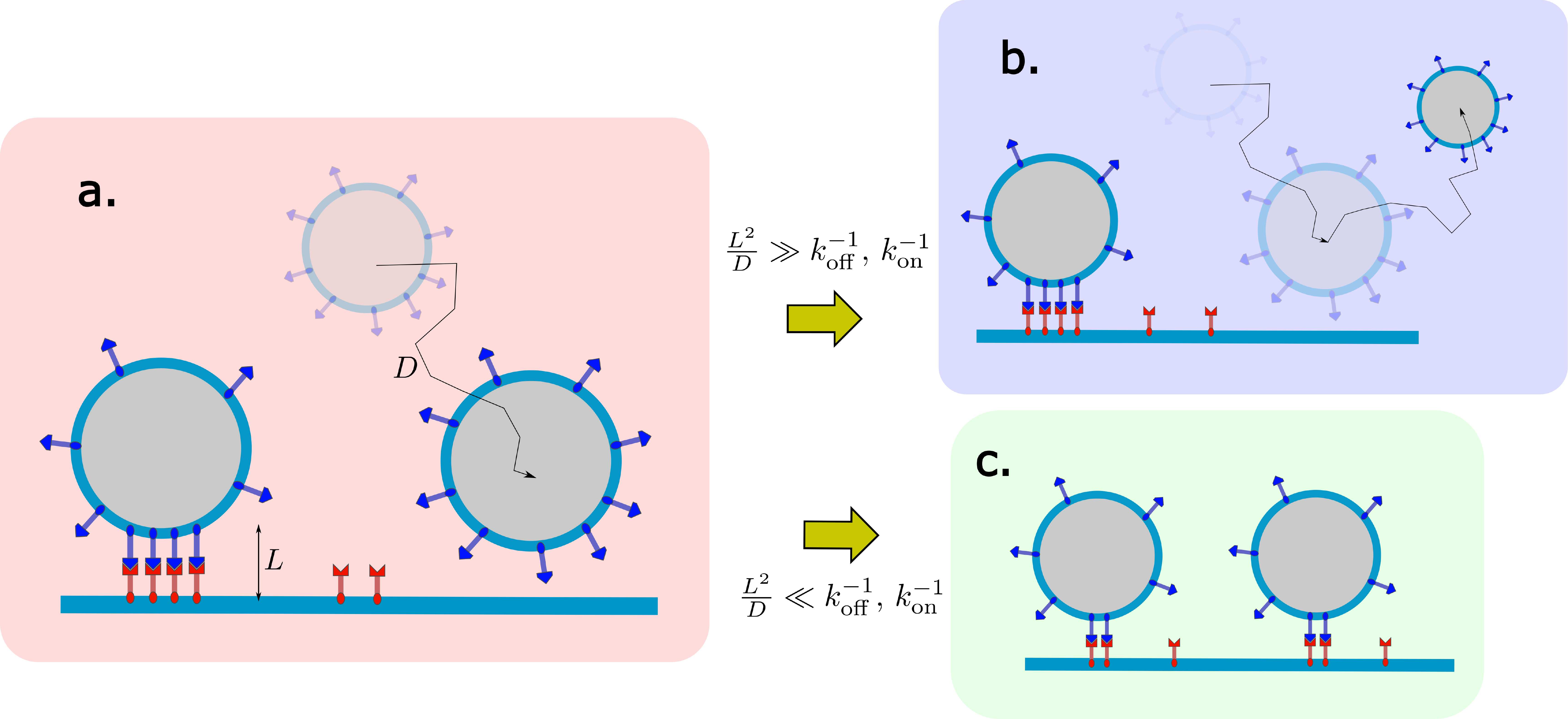}
\caption{ {\bf The reaction rates between complementary linkers control the rate of particle adhesion \emph{via} multivalent interactions.} {\bf a}, We consider particles and a surface decorated by mobile linkers of length $L$, and label as $D$ the diffusion coefficient of the particles. $D/L^2$ is the rate at which particles can diffuse over a distance comparable with the range of the interaction with the surface. {\bf b}, If the rate $k_\mathrm{on}$ at which linkers can react and form bridges is much smaller than $D/L^2$, a particle will need to diffuse in and out the interaction range several times before a bridge can be formed. {\bf c}, In the opposite scenario, the particle and surface are cross-linked at their first encounter. Note that large off-rates are required for surface-tethered linkers to redistribute between the contact regions formed with different particles. Although redistribution of linkers always minimises the avidity term of the multivalent free-energy, small $k_\mathrm{off}$ hampers relaxation towards equilibrium states as leveraged by the system of Fig.~\ref{fig5}{\bf a} to self-assemble low-valency aggregates.}\label{Fig:Sec6Fig2}
\end{center}
\end{figure}

\subsection{Modelling finite reaction rates}\label{Sec6:ReactionRates}
In deriving the models of Sec.~\ref{Sec4} and~\ref{Sec5} we have always assumed chemical equilibrium between the populations of formed complexes and unbound linkers.  The validity of such an assumption requires the reactions of linker complexation and de-complexation to be instantaneous compared to all other relevant timescales, including the lateral-diffusion times (see Sec.~\ref{Sec4:FutureDirections}) of mobile linkers and those of the functionalised particles diffusing in bulk.
However, while reviewing experimental studies on the self-assembly of multivalent liposomes~\cite{parolini2014thermal} and liposome-particle systems~\cite{LanfrancoLangmuir2019} in Sec~\ref{Sec3Kin}, we have encountered multiple instances in which the chemical-equilibrium assumption is not justified, and an interplay between the timescales of linker reactions and other processes produce an interesting and programmable phenomenology.\\
Bachmann, Petitzon~{\em et al.\ } addressed the question of how finite linker-reaction rates affect the self-assembly process of multivalent particles in Ref.~\cite{PetitzonSoftMatter2016} through numerical simulations explicitly accounting for the reaction kinetics of the linkers.\\ The method of Ref.~\cite{PetitzonSoftMatter2016} uses Brownian Dynamics simulations to upgrade the position of the particles using Eq.~\ref{Eq:Sec6:forces}. However, instead of calculating the most likely number of bonds using chemical equilibrium equations, corresponding to infinitely fast reaction kinetics, the method of Ref.~\cite{PetitzonSoftMatter2016} explicitly simulates reactions between linkers. In particular, at each step of the algorithm, while keeping the position of the particles $\{{\bf r}\}$ fixed, the Gillespie algorithm is used to sample one (de)complexation reaction at a time, along with the time for it to happen. After simulating reactions for a time $\Delta t$ corresponding to the integration step of the algorithm, forces are calculated using Eq.~\ref{Eq:Sec6:forces} and the particles’ configuration $\{ {\bf r} \}$ is upgraded through a standard Brownian Dynamics scheme~\cite{allen2017computer,frenkel2001understanding}. As compared to existing methods also simulating
reactions~\cite{DemboPRSL1998,chang1996influence,chang2000state,shah2011modeling}, the scheme of Ref.~\cite{PetitzonSoftMatter2016} can deal with competing reactions, \emph{i.e.}, cases in which a multiple linkers can bind the same complementary linker.
\\
The reaction kinetics is specified by $on$ and $off$ rates of forming a possible type of bond (\emph{e.g.}\ $\alpha\beta$) that are calculated as
follows~\cite{PetitzonSoftMatter2016,ParoliniACSNano2016,JanaArxiv2018,LanfrancoLangmuir2019,C8SM01430B}
\begin{eqnarray}
k_\mthr{on}= k^0_\mthr{on}  \exp[-\beta \Delta G^\mthr{conf}_{\alpha\beta}(\{ {\bf r} \})] &\qquad& k_\mthr{off} = \rho_0 k^0_\mthr{on} \exp[\beta \Delta G^0_{\alpha\beta}],
\label{Eq:Sec6:ReactionRates}
\end{eqnarray}
where $\rho_0$ is the standard concentration, $\rho_0=1\,$M$=N_A/\mthr{liter}$, and $N_A$ is the Avogadro's number. In Eq.~\ref{Eq:Sec6:ReactionRates} we decompose the hybridisation free energy into a configurational term (see Eqs~\ref{Sec3EqDeltaG}-\ref{Eq:DGABh}) and  the contribution of the reactive complexes  when free in solution, $\Delta G_{\alpha\beta} (\{ {\bf r} \}) = \Delta G^\mthr{conf}_{\alpha\beta} (\{ {\bf r} \}) + \Delta G^0_{\alpha\beta}$.
Eq.~\ref{Eq:Sec6:ReactionRates} assumes that the $off$ rates of linkers free in solution and tethered are equal. This is a fair approximation when considering linkers carrying DNA sticky ends given that the structure of the double helix is not affected by loads of strength comparable with the thermal
scale~\cite{Ho_BiophJ_2009}. However, in general, such approximation is not used~\cite{DemboPRSL1998,chang1996influence,chang2000state,shah2011modeling}.
In Eq.~1, $k^0_\mthr{on}$ is a reaction constant that has been used in Ref.~\cite{PetitzonSoftMatter2016} as a model parameter to study the effects of changing the speed of the reactions without affecting the thermodynamics of the system (for short DNA sticky ends free in solution, $k^0_\mthr{on}=10^6$M$^{-1}$s$^{-1}$). A useful non-dimensional parameter is given by $k^{0*}_\mthr{on}=k^0_\mthr{on}/(DL)$.  $k^{0*}_\mthr{on}$ can be interpreted as the ratio between the time taken to one particle to diffuse over a distance comparable with the size of the linker ($L$) and the time to react two linkers~\cite{PetitzonSoftMatter2016}.
\\
Reference~\cite{PetitzonSoftMatter2016} shows how finite reaction rates (small $k^{0*}_\mthr{on}$) drastically affect the self-assembly of a binary colloidal system decorated with mobile linkers.  Particularly at low temperature, the system forms aggregates resembling branched chains rather than compact aggregates as obtained when considering infinite reaction rates. Indeed, the $off$ rates become exponentially small at low temperature (see Eq.~\ref{Eq:Sec6:ReactionRates}) making it harder for the growing aggregates to reorient existing bonds and compactify the clusters. This result nicely shows how reaction rates can affect the organisation of functionalised particles at a larger scale.
\\
As already reviewed in Sec.~\ref{Sec3Kin}, finite reaction rates between linkers significantly affect adsorption of DNA functionalised gold nanoparticles (GNPs) onto LUVs carrying complementary linkers~\cite{LanfrancoLangmuir2019}. Using numerical simulations based on the method developed in Refs.~\cite{PetitzonSoftMatter2016,JanaArxiv2018}, Ref.~\cite{LanfrancoLangmuir2019} corroborated the molecular model presented in Sec.~\ref{Sec3Kin} in which linkers tethered to the LUV are sequestrated by the GNPs binding the vesicles at the early stages of the adsorption process, hindering later adhesion events. Similarly to what reported by Ref.~\cite{PetitzonSoftMatter2016}, numerical simulations also clarified how the number of GNPs adhering to each LUV in steady conditions is smaller than what expected at equilibrium.


\section{Outlook and applications}~\label{Sec6}
We believe that the comprehensive understanding of multivalent compliant systems enabled by the body of theoretical work summarised in Secs.~\ref{Sec4} and \ref{Sec5}, along with the know-how demonstrated by the numerous experimental implementations reviewed in Secs.~\ref{Sec2} and~\ref{Sec3}, will guarantee a thriving future for the field.\\
Broadly speaking, we envisage two main directions:  biology, including both interactions with living matter and \emph{in vitro} mimics of biological complexity, and technology \emph{per se}, as in sensors for biomolecules, responsive materials or even materials with basic information processing power.\\

Last-generation experimental tools reviewed in Secs.~\ref{Sec2} and~\ref{Sec3} have brought the level of control and programmability that we have on artificial multivalent systems to unprecedented heights, which paves the way to the design of new and improved \emph{in vitro} biophysical assays. As discussed, many of the open questions on compliant multivalent systems concern the dynamic coupling of processes taking place at different timescales, including linker diffusion, bond formation and release, and deformation modes of the interfaces. These issues could be addressed with dedicated assays, where a combination of micro-manipulation and microscopy could enable monitoring the dynamics of individual adhesion events between cell-like liposomes, particles, droplets \emph{etc.}, keeping track of interface deformation, linker complexation, and their distribution. Thanks to the tools of DNA nanotechnology one should be able to precisely control of both the thermodynamic properties and the kinetics of linker complexation, while the size and structure of the linkers could be prescribed to tune non-specific interactions. Guided by theoretical and numerical modelling, experiments performed with these fully controllable platforms could shed light on a whole class of biological processes that span all the way from basic physiology of cells (cargo uptake through endocytosis, pinocytosis) to specific engulfment of particles as in phagocytosis, and processes where the ``cargo'' plays a more active role, as in many host/pathogen interactions.   Examples of the latter are very widespread in infectious disease, and still quite poorly understood  in viral~\cite{Delgusteeaat1273,vahey2019influenza} bacterial~\cite{bryant12,paul13} and other species such as malaria~\cite{cicuta2019,Boulant2015,Pizarro-Cerda:2006aa}.\\

Biological cells sense their environment through receptor proteins located on the plasma membrane. Environmental sensing is crucial for a variety of processes in biology, from embryonic development~\cite{sharpe2015} to the regulation of inflammatory response~\cite{sedger2014} and the maintenance of stem cell niches~\cite{Inaba2016}.   When bound to specific ligands (neurotransmitters, drugs, hormones, toxins \emph{etc.}), most receptors dimerise or change structure, enabling signal transduction across the membrane. Each signal generally triggers the activation of a specific biochemical pathway within the cell, leading to regulation of specific gene expression.\\
We argue that the recent developments in DNA nanotechnology, along with our understanding of cooperative effects in multivalent systems and their coupling with membrane deformation, will enable the design and production of biomimetic nanostructures capable of replicating more and more of the complex functionalities of biological membrane-receptors, besides that of inducing adhesive forces.\\
For example, one could imagine interacting networks of synthetic membrane receptors capable of sensing a variety of environmental and biochemical stimuli, including the presence of specific analytes, light, temperature, pH or even hydrostatic pressure. Mechanisms to achieve sensitivity to all these stimuli are already available if one considers the use of DNA nanotechnology to produce the receptors and exploit chemical-functionalisations~\cite{Hernandez-Ainsa:2016aa}, DNA aptamers~\cite{Milam:2016aa,Del-Grosso:2019aa}, pH-responsive DNA motifs~\cite{Idili:2014aa}, as well as the coupling of the nanostructure with stimuli-dependent features of lipid membranes such as their fluidity and phase-separation state. The presence of the target signal could induce a change in the steady state populations of receptor complexes, which could then be made to produce a detectable fluorescent signal.
Probes with these capabilities will enable \emph{in-situ} studies of the physico-chemical environment of biological cells, both \emph{in vivo} and \emph{in vitro}, and may help to unravel compelling biological effects at the single-cell and population level, along the lines of some early attempts discussed in Sec.~\ref{Sec3Cells}.\\
Even more interestingly, the ``activation'' of artificial receptors could be coupled to downstream signalling pathways within the cell. If this endeavour were successful, we would then be able to equip biological cells with a formidable nanotechnological arsenal to temporarily influence and even enhance their sensing and communication capabilities, which could have a great impact on healthcare. Indeed, most modern drugs are small molecules acting directly as artificial ligands on naturally occurring membrane receptors, either by activating them, or by blocking their binding sites, or by enhancing their sensitivity to naturally occurring molecules. In turn, synthetic membrane receptors could represent a new class of nanotechnological  drugs that instead of targeting existing receptors will simply establish new ones. Artificial receptors could for instance be designed to sensitise the cells to chemical signals that would otherwise be ignored, enabling a response to new drugs developed \emph{ad hoc} for the artificial receptors, or setting up cells to act as reporters of specific processes (inflammation, cancer, \emph{etc.}) in nearby tissues~\cite{floss2019}. In analogy to recently proposed synthetic protein networks in cells~\cite{elowitz2018}, one can even imagine an information-processing circuitry based around DNA materials, in cross talk to genetic and protein networks (both natural and  synthetic/edited).\\
Synthetic membrane-receptors could also be applied to artificial-cellular technologies in the context of bottom-up synthetic biology, enabling an unprecedented level of control over environmental sensing and communications in synthetic cell mimics that, to date, are mostly restricted by the use of reconstituted biological machinery with limited room for the bottom-up design of new functionalities.\\

The quantitative understanding we have of the class soft materials reviewed here makes them also valuable for direct technological applications. An aspect that we argue should be pursued is the potential of compliant multivalent units in biomolecular sensing, which was partially explored in Ref.~\cite{Amjad:2017aa}.   Systems can be designed in which  lipid membranes, (\emph{e.g.} vesicles), adhere to each other, linked by bridges in which one element is the biomolecule to be sensed, and the other components of the bridge are DNA linkers suitably functionalised to interact with the molecules of interest.  Such platforms could be designed to be very sensitive to the concentration of ligands, such that adhesion sets in sharply when a given concentration is reached and liposomes compactify into a tissue-like material. As discussed in Sec.~\ref{Sec3}, adhesion produces large-scale deformation in vesicles that, interestingly, could at the same time amplify the detected signal and form the basis of a convenient readout mechanism, based for instance on changes on the mechanical properties or the electrical impedance of the formed material. The simultaneous use of different types of linkers could enable multiplexed detection of different analytes.\\

{\bf Acknowledgments:} The authors would like to thank Lucia Parolini, Jurij Kotar, Omar Amjad, Shunsuke Shimobayashi, Pritam Kumar Jana, Stephan Bachmann, Marius Petitzon, Stefano Angioletti-Uberti, and Daan Frenkel for collaborating with us on the topics of this review. BMM was supported by an ARC (ULB) grant of the {\em F\'ed\'eration Wallonie-Bruxelles} and by the F.R.S.-FNRS under grant n$^\circ$ MIS F.4534.17. LDM and PC acknowledge support from the EPSRC Programme Grant CAPITALS, number EP/J017566/1. LDM acknowledges support from the Royal Society through a University Research Fellowship (UF160152).

\section*{References}
\bibliography{biblio}{}
\bibliographystyle{unsrt}

\end{document}